\documentclass[preprint,aps,tightenlines,byrevtex,showpacs,nofootinbib]{revtex4-1}
\usepackage{graphicx}
\begin{document}

\title{
Sensitivity of T2KK to the non-standard
interaction in propagation
}

\author{Haruna Oki}
\altaffiliation{Present address: Toppan Printing Co., Ltd., Japan.}
\author{Osamu Yasuda}
\email{yasuda_at_phys.metro-u.ac.jp}
\affiliation{Department of Physics, Tokyo Metropolitan University,
Minami-Osawa, Hachioji, Tokyo 192-0397, Japan}

\date{\today}

\begin{abstract}
Assuming only the non-zero electron
and tau neutrino components $\epsilon_{ee}$,
$\epsilon_{e\tau}$, $\epsilon_{\tau\tau}$
of the non-standard matter effect and
postulating the atmospheric neutrino constraint
$\epsilon_{\tau\tau}=|\epsilon_{e\tau}|^2/(1+\epsilon_{ee})$,
we study the sensitivity to the non-standard interaction
in neutrino propagation of the T2KK neutrino
long-baseline experiment.
It is shown that T2KK can constrain the parameters
$|\epsilon_{ee}|\lesssim 1$,
$|\epsilon_{e\tau}|\lesssim 0.2$.
It is also shown that
if $|\epsilon_{e\tau}|$ and $\theta_{13}$
are large, then
T2KK can determine
the Dirac phase and the phase of
$\epsilon_{e\tau}$ separately,
due to the information at the two baselines.
We also provide an argument that the components
$|\epsilon_{\alpha\mu}|$ $(\alpha=e,\mu,\tau)$
must be small for the disappearance oscillation probability
to be consistent with high-energy
atmospheric neutrino data,
which justifies our premise that these quantities are negligible.
\end{abstract}
\vskip 0.1cm
\pacs{14.60.Pq, 14.60.St}
\maketitle

\section{introduction}

It has been shown by experiments with solar and atmospheric
neutrinos \cite{Amsler:2008zz} that neutrinos have masses and mixings.
In the standard three-flavor framework,
neutrino oscillations are described by three mixing angles,
$\theta_{12}$, $\theta_{13}$, $\theta_{23}$, one CP phase $\delta$,
and two independent mass-squared differences, $\Delta m^2_{21}$ and
$\Delta m^2_{31}$.  The values of the set of the parameters
($\Delta m^2_{21}$, $\theta_{12}$) and ($\Delta m^2_{31}$, $\theta_{23}$)
have been determined to a certain precision by solar and atmospheric
neutrino experiments, respectively.
On the other hand, only the upper bound on $\theta_{13}$
is known\footnote{
In Refs.\,\cite{Fogli:2008jx,Schwetz:2008er,Fogli:2008cx,Ge:2008sj,Fogli:2009ce,GonzalezGarcia:2010er}, a
global analysis of the neutrino oscillation data has been performed, in
which a non-vanishing best-fit value for $\theta_{13}$ is obtained. This result,
however, is compatible with $\theta_{13} = 0$ at less than $2 \sigma$,
and it is not yet statistically significant enough to be taken
seriously.}, $\sin^2\theta_{13} < 0.04$ at 90\%CL,
and there is an absence of information on $\delta$.
In future neutrino long-baseline experiments (see, e.g.,
Ref.\,\cite{Bandyopadhyay:2007kx}),
the values of $\theta_{13}$ and
$\delta$ are expected to be determined precisely.
As in the case of B factories\,\cite{belle:0000,babar:0000},
such highly precise measurements will enable us to
search for deviation from the standard three-flavor oscillations.
One such possibility,
which will be discussed in this paper,
is the effective non-standard neutral current-neutrino interaction with
matter\,\cite{Wolfenstein:1977ue,Guzzo:1991hi,Roulet:1991sm}, given by
\begin{eqnarray}
{\cal L}_{\mbox{\rm\scriptsize eff}}^{\mbox{\tiny{\rm NSI}}} 
=-2\sqrt{2}\, \epsilon_{\alpha\beta}^{fP} G_F
(\overline{\nu}_\alpha \gamma_\mu P_L \nu_\beta)\,
(\overline{f} \gamma^\mu P f'),
\label{NSIop}
\end{eqnarray}
where $f$ and $f'$ stand for fermions (the only relevant
ones are electrons, u, and d quarks),
$G_F$ is the Fermi coupling constant, and $P$ stands for
a projection operator that is either
$P_L\equiv (1-\gamma_5)/2$ or $P_R\equiv (1+\gamma_5)/2$.
In the presence of the interaction Eq.\,(\ref{NSIop}),
the standard matter effect\,\cite{Wolfenstein:1977ue,Mikheev:1986gs}
is modified.
Since we discuss the long-baseline experiments
on the Earth, we make an approximation
that the number densities of electrons ($N_e$),
protons, and neutrons are equal.\footnote{
This assumption is not valid in other environments, e.g., in the Sun.}
By introducing the notation
$\epsilon_{\alpha\beta}
\equiv \sum_{P}
\left(
\epsilon_{\alpha\beta}^{eP}
+ 3 \epsilon_{\alpha\beta}^{uP}
+ 3 \epsilon_{\alpha\beta}^{dP}
\right)$,
the hermitian $3 \times3 $ matrix of the matter potential becomes
\begin{eqnarray}
{\cal A}\equiv A\left(
\begin{array}{ccc}
1+ \epsilon_{ee} & \epsilon_{e\mu} & \epsilon_{e\tau}\\
\epsilon_{\mu e} & \epsilon_{\mu\mu} & \epsilon_{\mu\tau}\\
\epsilon_{\tau e} & \epsilon_{\tau\mu} & \epsilon_{\tau\tau}
\end{array}
\right),
\label{matter-np}
\end{eqnarray}
where $A\equiv\sqrt{2}G_FN_e$.

Constraints on $\epsilon_{\alpha\beta}$
from various neutrino experiments have been discussed
in Refs.\,\cite{Fornengo:2001pm,Berezhiani:2001rs,Davidson:2003ha,GonzalezGarcia:2004wg,Miranda:2004nb,Barranco:2005ps,Barranco:2007ej,Bolanos:2008km,Escrihuela:2009up}.
Since the coefficients $\epsilon_{\alpha\beta}$ in Eq.\,(\ref{matter-np})
are given by
$\epsilon_{\alpha\beta}\sim
\epsilon^e_{\alpha\beta}
+3\epsilon^u_{\alpha\beta}
+3\epsilon^d_{\alpha\beta}$,
considering the constraints by 
Refs.\,\cite{Fornengo:2001pm,Berezhiani:2001rs,Davidson:2003ha,GonzalezGarcia:2004wg,Miranda:2004nb,Barranco:2005ps,Barranco:2007ej,Bolanos:2008km,Escrihuela:2009up},
we have the following constraints\,\cite{Biggio:2009nt} at 90\%CL\footnote{
Here, we adopt the conservative bounds on $\epsilon_{\alpha\beta}$
which were derived without using the one-loop arguments,
because the bounds obtained by the loop-contributions
are known to be model-dependent\,\cite{Biggio:2009kv}.
However, if we accept results based on one-loop arguments, then
we get the following bounds at 90\%CL:
$|\epsilon_{e\mu}|<
[\sum_P(|\epsilon_{e\mu}^{eP}|^2+|3\epsilon_{e\mu}^{uP}|^2
+|3\epsilon_{e\mu}^{dP}|^2)]^{1/2}\sim
5\times 10^{-3}$\,\cite{Kitazawa:2006iq,Davidson:2008,Yasuda:2008zza},
where $|\epsilon_{e\mu}^{eP}|<5\times10^{-4}$\,\cite{Davidson:2003ha},
$|\epsilon_{e\mu}^{uP}|<3.1\times10^{-4}$\,\cite{Davidson:2008,Yasuda:2008zza}
and
$|\epsilon_{e\mu}^{dP}|<3.1\times10^{-4}$\,\cite{Davidson:2008,Yasuda:2008zza}
are used;
$|\epsilon_{e\tau}|<
[\sum_P(|\epsilon_{e\tau}^{eP}|^2+|3\epsilon_{e\tau}^{uP}|^2
+|3\epsilon_{e\tau}^{dP}|^2)]^{1/2}\sim
1.7$\,\cite{Davidson:2008,Yasuda:2008zza},
where $|\epsilon_{e\tau}^{eP}|<0.32$\,\cite{Davidson:2008,Yasuda:2008zza},
$|\epsilon_{e\tau}^{uP}|<0.28$\,\cite{Davidson:2008,Yasuda:2008zza}
and
$|\epsilon_{e\tau}^{dP}|<0.28$\,\cite{Davidson:2008,Yasuda:2008zza}
are used.
}:
\begin{eqnarray}
\left(
\begin{array}{ccc}
|\epsilon_{ee}| < 4\times 10^0 & \quad|\epsilon_{e\mu}| < 3\times 10^{-1}
& \quad|\epsilon_{e\tau}| < 3\times 10^{0\ }\\
& \quad |\epsilon_{\mu\mu}| < 7\times 10^{-2}
& \quad|\epsilon_{\mu\tau}| < 3\times 10^{-1}\\
& & \quad|\epsilon_{\tau\tau}| < 2\times 10^{1\ }
\end{array}
\right).
\label{epsilon-m}
\end{eqnarray}
From this prior study, it is known that the bounds on 
$\epsilon_{ee}$, $\epsilon_{e\tau}$ and $\epsilon_{\tau\tau}$ are
much weaker than $\epsilon_{\alpha\mu}~(\alpha=e,\mu,\tau)$,
and typically
$\epsilon_{\alpha\beta}\sim {\cal O}(1)~(\alpha, \beta = e, \tau)$
is allowed.

On the other hand,
the new physics with components $\epsilon_{\alpha\beta}~(\alpha,\beta=e,\tau)$
should be consistent with
the high-energy atmospheric neutrino data, which suggest the
behavior of the disappearance oscillation probability
\begin{eqnarray}
1-P(\nu_\mu\rightarrow\nu_\mu)\sim
\sin^22\theta_{\text{atm}}
\sin^2\left(\frac{\Delta m^2_{\text{atm}}L}{4E}\right)\,
\propto\, \frac{1}{E^2}\,,
\label{atm-he}
\end{eqnarray}
where $\sin^22\theta_{\text{atm}}$ and
$\Delta m^2_{\text{atm}}$ are the oscillation
parameters in the two-flavor formalism.
Note that the terms of ${\cal O}(E^0)$
and ${\cal O}(E^{-1})$ are absent in Eq.\,(\ref{atm-he}).
As shown later,
the elements $\epsilon_{\alpha\mu}~(\alpha=e,\mu,\tau)$
should be small so as to produce no term of ${\cal O}(E^0)$
in Eq.\,(\ref{atm-he}).
Furthermore, absence of terms of ${\cal O}(E^{-1})$ in Eq.\,(\ref{atm-he}) implies
\begin{eqnarray}
|\epsilon_{e\tau}|^2
\simeq \epsilon_{\tau\tau} \left( 1 + \epsilon_{ee} \right),
\label{atm}
\end{eqnarray}
as pointed out in Ref.\,\cite{Friedland:2004ah,Friedland:2005vy}.
When Eq.\,(\ref{atm}) is satisfied,
two of the three eigenvalues of the matrix
(\ref{matter-np}) with $\epsilon_{\alpha\mu}=0~(\alpha=e,\mu,\tau)$
become zero.  Only in this case, one of the frequencies
of the three oscillation modes
at high energy coincides with the one in the vacuum oscillation,
and the disappearance oscillation
probability of $\nu_\mu$ behaves as in Eq.\,(\ref{atm-he}).
The effect of the non-standard interaction in propagation for solar neutrinos
has also been discussed in Refs.\,\cite{Friedland:2004pp,Miranda:2004nb,Bolanos:2008km,Escrihuela:2009up,Palazzo:2009rb}, and
Refs.\,\cite{Escrihuela:2009up} and \cite{Palazzo:2009rb}
give a constraint $-0.06<\epsilon_{e\tau}^{dV}\sin\theta_{23}< 0.41$
(at 90\%CL) and 
$|\epsilon_{e\tau}^{dV}|\lesssim 0.4$ (at $\Delta\chi^2=4$ for 2 d.o.f.),
respectively.

The sensitivity of the ongoing long-baseline experiments to the non-standard interaction in
propagation has been studied for MINOS\,\cite{minos:0000} in
Refs.\,\cite{Friedland:2006pi,Yasuda:2007tx,Sugiyama:2007ub,Blennow:2007pu}, and for
OPERA\,\cite{opera:0000} in
Refs.\,\cite{EstebanPretel:2008qi,Blennow:2008ym}.  On the other hand,
for the sensitivity
of future long-baseline experiments, 
Ref.\,\cite{Ribeiro:2007jq} provided the sensitivity of the T2KK
experiment\,\cite{Ishitsuka:2005qi,Hagiwara:2005pe},
an extension of the T2K neutrino oscillation
experiment\,\cite{Itow:2001ee} with a far detector in Korea,
in the two-flavor analysis with $\epsilon_{e\alpha}=0~(\alpha = e, \mu,
\tau)$.  The sensitivity of the reactor and super-beam experiments was
discussed in Ref.\,\cite{Kopp:2007ne}, and the sensitivity of neutrino
factories\,\cite{Geer:1997iz,Bandyopadhyay:2007kx} has been discussed
by many
authors\,\cite{Gago:2001xg,Ota:2001pw,Huber:2001de,Campanelli:2002cc,Ribeiro:2007ud,Kopp:2008ds,Gago:2009ij,Meloni:2009cg}.

In the present paper we analyze the sensitivity to
the parameters $\epsilon_{\alpha\beta}$
of the T2KK experiment,
assuming 
$\epsilon_{\mu\alpha}=0~(\alpha = e, \mu, \tau)$
and $\epsilon_{\tau\tau}=|\epsilon_{e\tau}|^2/( 1 + \epsilon_{ee})$.
We do not discuss the so-called parameter
degeneracy\,\cite{BurguetCastell:2001ez,Minakata:2001qm,Fogli:1996pv,Barger:2001yr},
since little is known about parameter degeneracy in the presence of the new physics,
and the study of the subject is beyond the scope of this paper.

The paper is organized as follows.
In sect.\ref{atmospheric}, we discuss the constraints
on the $\epsilon_{\alpha\beta}$ parameters
from the high-energy atmospheric neutrino data.
In sect.\ref{sensitivity}, we analyze
the sensitivity of T2KK to the new physics parameters.
In sect.\ref{conclusions}, we draw our conclusions.
In the appendices \ref{appendixa}--\ref{appendixg} we provide details
of the derivation of the analytic formulae for the
oscillation probabilities and their high-energy behaviors.

\section{Atmospheric neutrinos and the non-standard interaction in propagation
\label{atmospheric}}
In Refs.\,\cite{Friedland:2004ah,Friedland:2005vy}, it was pointed out that
the new physics should be consistent with the constraint imposed by
the atmospheric neutrino data, which suggest
that the disappearance oscillation probability at high-energy satisfies
Eq.\,(\ref{atm-he}).
In the standard three-flavor scheme, the
high-energy behavior is
\begin{eqnarray}
%&{\ }&
\lim_{E\to\infty}
\frac{1-P(\nu_\mu\rightarrow\nu_\mu)}{(\Delta E_{31}/A)^2}
%\nonumber\\
&\simeq&
\lim_{E\to\infty}
\frac{1}{(\Delta E_{31}/A)^2}
\left[
4\frac{|U_{\mu3}|^2}{c^2_{13}}
\left(1-\frac{|U_{\mu3}|^2}{c^2_{13}}\right)
\sin^2\left(\frac{c^2_{13}\Delta E_{31}L}{2}\right)
\right.\nonumber\\
&{\ }&\left.
\qquad\qquad~~~~
+4\left(\frac{\Delta E_{31}}{A}\right)^2
|U_{\mu3}|^2s^2_{13}\sin^2\left(
\frac{AL}{2}\right)
\right]
\nonumber\\
&\simeq&
\sin^22\theta_{23}
\left(\frac{c^2_{13}AL}{2}\right)^2
+s^2_{23}\sin^22\theta_{13}\sin^2\left(
\frac{AL}{2}\right),
\label{he-std}
\end{eqnarray}
where $\Delta E_{jk}\equiv E_j-E_k\simeq \Delta m^2_{jk}/2E$.
(See the appendix \ref{appendixa}
for derivation of the oscillation probability
in constant density matter and the appendix \ref{appendixb}
for derivation of its high-energy behavior (\ref{he-std})).
In the following discussions,
we assume the generic matter potential
(\ref{matter-np}), and derive the high-energy behavior
\begin{eqnarray}
1-P(\nu_\mu\rightarrow\nu_\mu)\simeq
c_0 + \frac{c_1}{E} + {\cal O}\left(\frac{1}{E^2}\right),
\label{expansion}
\end{eqnarray}
and postulate $|c_0|\ll 1$ and $|c_1|\ll 1$.
Note that the term of ${\cal O}(E^{-2})$
corresponds to the standard results
(\ref{atm-he}) in the two-flavor scheme
or (\ref{he-std}) in the three-flavor case,
where the information on the atmospheric neutrino
oscillation parameters appears.
In the presence of the matter potential (\ref{matter-np}),
as we discussed in detail in the appendix \ref{appendixc},
assuming the constant density of matter,
we can obtain the analytic
form for the behavior (\ref{atm-he})
of the disappearance oscillation probability
at high energy, in which
$|A/\Delta E_{31}|\equiv \sqrt{2}G_FN_e/(|\Delta m^2_{31}|/2E)\ll 1$
is satisfied.
The leading term $c_0$ in Eq.\,(\ref{expansion}) is given by
\begin{eqnarray}
c_0& \simeq&
\left[
1-\frac{1+\epsilon_{ee}+\epsilon_{\tau\tau}}
{(1+\epsilon_{ee})|\epsilon_{\mu\tau}|^2+\epsilon_{\tau\tau}|\epsilon_{e\mu}|^2
-2\mbox{\rm Re}(\epsilon_{e\mu}\epsilon_{\mu\tau}
(\epsilon_{e\tau})^\ast)}
\left(\frac{|\epsilon_{e\mu}|^2+|\epsilon_{\mu\mu}|^2+|\epsilon_{\mu\tau}|^2}
{1+\epsilon_{ee}+\epsilon_{\tau\tau}}-\epsilon_{\mu\mu}
\right)^2\right]
\nonumber\\
&{\ }&~~\times
\sin^2\left[AL
\frac{(1+\epsilon_{ee})|\epsilon_{\mu\tau}|^2
+\epsilon_{\tau\tau}|\epsilon_{e\mu}|^2
-2\mbox{\rm Re}(\epsilon_{e\mu}\epsilon_{\mu\tau}
(\epsilon_{e\tau})^\ast)}
{1+\epsilon_{ee}+\epsilon_{\tau\tau}}
\right]
\nonumber\\
&{\ }&+4\left[
\frac{|\epsilon_{e\mu}|^2+|\epsilon_{\mu\mu}|^2+|\epsilon_{\mu\tau}|^2}
{(1+\epsilon_{ee}+\epsilon_{\tau\tau})^2}
-\frac{(1+\epsilon_{ee})|\epsilon_{\mu\tau}|^2
+\epsilon_{\tau\tau}|\epsilon_{e\mu}|^2
-2\mbox{\rm Re}(\epsilon_{e\mu}\epsilon_{\mu\tau}
(\epsilon_{e\tau})^\ast)}
{(1+\epsilon_{ee}+\epsilon_{\tau\tau})^3}\right]
\nonumber\\
&{\ }&~~\times
\sin^2\left(\frac{(1+\epsilon_{ee}+\epsilon_{\tau\tau})AL}{2}\right).
\label{he1}
\end{eqnarray}
For Eq.\,(\ref{he1}) to be consistent with Eq.\,(\ref{atm-he}),
we should have
$\epsilon_{e\mu}\simeq\epsilon_{\mu\mu}\simeq\epsilon_{\mu\tau}\simeq0$
(see appendix \ref{appendixc}).

In Refs.\,\cite{Fornengo:2001pm,GonzalezGarcia:2004wg,Mitsuka:2008zz},
the two-flavor analysis of the atmospheric neutrino data
with the matter effect
\begin{eqnarray}
A\left(
\begin{array}{cc}
\epsilon_{\mu\mu} & \epsilon_{\mu\tau}\\
\epsilon_{\mu\tau} & \epsilon_{\tau\tau}
\end{array}
\right)
\nonumber
\end{eqnarray}
was performed.  In this case, in the limit $E\to\infty$
the disappearance oscillation probability results in
\begin{eqnarray}
1-P(\nu_\mu\rightarrow\nu_\mu)
\simeq
\frac{|\epsilon_{\mu\tau}|^2}
{(\epsilon_{\mu\mu}-\epsilon_{\tau\tau})^2/4
+|\epsilon_{\mu\tau}|^2}
\sin^2\left(AL\sqrt{
(\epsilon_{\mu\mu}
-\epsilon_{\tau\tau})^2/4
+|\epsilon_{\mu\tau}|^2}\right).
\nonumber
\end{eqnarray}
This suggests that the two parameters
$|\epsilon_{\mu\tau}|$ and $|\epsilon_{\mu\mu}
-\epsilon_{\tau\tau}|$
should be small
so as to be consistent with the high-energy behavior (\ref{atm-he}),
and in fact, the authors of Ref.\,\cite{Mitsuka:2008zz}
obtained the bounds $|\epsilon_{\mu\tau}|<1.5\times10^{-2}$
and $|\epsilon_{\mu\mu}-\epsilon_{\tau\tau}|<4.9\times10^{-2}$ at 90\%CL.
In the three flavor framework, as seen below,
$\epsilon_{\alpha\beta}\sim{\cal O}(1)~(\alpha, \beta=e, \tau)$
can be consistent with
the atmospheric neutrino data as long as 
$\epsilon_{\tau\tau}-|\epsilon_{e\tau}|^2/(1+\epsilon_{ee})
\simeq0$ is satisfied\,\cite{Friedland:2004ah,Friedland:2005vy}.
We see that the two flavor constraints
in Refs.\,\cite{Fornengo:2001pm,GonzalezGarcia:2004wg,Mitsuka:2008zz}
are consistent with three flavor ones as follows:
As for $\epsilon_{\tau\tau}$, since $|\epsilon_{\mu\mu}|\ll1$,
the bound $|\epsilon_{\mu\mu}-\epsilon_{\tau\tau}|\ll1$
in the two flavor framework implies
$|\epsilon_{\tau\tau}|\ll1$.
The two flavor framework can be regarded as a subset of
the three flavor case in the limiting case
$\epsilon_{ee}=\epsilon_{e\tau}=\theta_{13}=\Delta m_{21}^2=0$, so
the constraint Eq.\,(\ref{atm}) in the two flavor case leads to
$|\epsilon_{\tau\tau}|\simeq0$.
On the other hand, the bound on $|\epsilon_{\mu\tau}|$
in the three flavor case is independent of other components
$\epsilon_{\alpha\beta}$, so
the bound $|\epsilon_{\mu\tau}|\lesssim {\cal O}(10^{-2})$
is expected to be valid both in the two and three flavor cases.
While $|\epsilon_{\mu\mu}|\ll1$ and $|\epsilon_{\mu\tau}|\ll1$ were
first shown in Refs.\,\cite{Davidson:2003ha} and \cite{Fornengo:2001pm},
respectively, if we do not accept the one-loop
arguments\,\cite{Davidson:2003ha} to constrain $\epsilon_{e\mu}$
as in Ref.\,\cite{Biggio:2009kv}, then
the observation in this paper that $|\epsilon_{e\mu}|\ll1$ follows
from the the atmospheric neutrino constraint is new,
although our discussions are based only on an analytical
treatment.\footnote{Before Ref.\,\cite{Biggio:2009kv} appeared,
the bounds $|\epsilon_{e\mu}^{fP}|<{\cal O}(10^{-4})$ (90\%) in 
Ref.\,\cite{Davidson:2003ha} based on
the one-loop arguments were widely accepted, and this was
used to justify the assumption $\epsilon_{e\mu}\simeq0$
in Refs.\,\cite{Friedland:2004ah,Friedland:2005vy}.
}

In the following discussions, therefore, let us assume that 
$\epsilon_{e\mu}=\epsilon_{\mu\mu}=\epsilon_{\mu\tau}=0$.
Then we obtain the coefficient $c_1$ of the
term of the next-to-leading order in $1/E$ in Eq.\,(\ref{expansion})
(see appendix \ref{appendixd}):
\begin{eqnarray}
c_1
&\simeq&
-\frac{2s^2_{23}\Delta m^2_{31}}{(1+\epsilon_{ee})A\Delta t_{32}}
\left[\frac{t_3^2}{\zeta'}
\sin^2\left\{\frac{(1+\epsilon_{ee})\zeta' AL}{2t_3}\right\}
+\frac{\zeta'}{t_3^2}
\sin^2\left\{\frac{(1+\epsilon_{ee})t_3AL}{2}\right\}
\right],
\label{he2}
\end{eqnarray}
where
\begin{eqnarray}
\zeta'&\equiv&\frac{1}{1+\epsilon_{ee}}\left(
\epsilon_{\tau\tau}-\frac{|\epsilon_{e\tau}|^2}{1+\epsilon_{ee}}
\right)
\nonumber\\
t_3&\equiv&
\frac{1}{2(1+\epsilon_{ee})}\left[
1+\epsilon_{ee}+\epsilon_{\tau\tau}
+\sqrt{(1+\epsilon_{ee}+\epsilon_{\tau\tau})^2-4(1+\epsilon_{ee})^2\zeta'}\,
\right]
\nonumber\\
\Delta t_{32}&\equiv&
\frac{1}{2(1+\epsilon_{ee})}
\sqrt{(1+\epsilon_{ee}+\epsilon_{\tau\tau})^2-4(1+\epsilon_{ee})^2\zeta'}\,.
\nonumber
\end{eqnarray}
Again, for Eq.\,(\ref{he2}) with Eq.\,(\ref{expansion}) to be consistent with
the high-energy behavior (\ref{atm-he}), we should have
\begin{eqnarray}
\epsilon_{\tau\tau}-\frac{|\epsilon_{e\tau}|^2}{1+\epsilon_{ee}}
\simeq0.
\nonumber
\end{eqnarray}
This agrees with the conclusion
$\epsilon_{\tau\tau}\simeq|\epsilon_{e\tau}|^2/(1+\epsilon_{ee})$
in Refs.\,\cite{Friedland:2004ah,Friedland:2005vy}.

Thus, let us assume
$\epsilon_{\tau\tau}-|\epsilon_{e\tau}|^2/(1+\epsilon_{ee})=0$.
In this case, it is convenient to introduce the new variable
\begin{eqnarray}
t_\beta\equiv\tan\beta\equiv
\frac{|\epsilon_{e\tau}|}{1+\epsilon_{ee}}.
\label{tanbeta}
\end{eqnarray}
Then, we have the following high-energy behavior
(see appendix \ref{appendixe}):
\begin{eqnarray}
\frac{1-P(\nu_\mu\rightarrow\nu_\mu)}{(\Delta E_{31}/A)^2}
&\simeq&
4\frac{s^2_{23}}
{(c''_{13})^2}
\left\{1-\frac{s^2_{23}}
{(c''_{13})^2}
\right\}
\left\{\frac{(c''_{13})^2AL}{2}
\right\}^2
\nonumber\\
&{\ }&+\frac{s^2_{23}}{(c''_{13})^2}
\sin^22\theta''_{13}
\left(\frac{c^2_\beta}{1+\epsilon_{ee}}\right)^2
\sin^2\left(\frac{(1+\epsilon_{ee})AL}{2c^2_\beta}\right).
\label{he3}
\end{eqnarray}
where $c''_{13}\equiv \cos\theta''_{13}$,
$c_\beta\equiv\cos\beta$, $s_\beta\equiv\sin\beta$, and
the new angle $\theta''_{13}$,
which is introduced in the appendix \ref{appendixf}
to diagonalize the mass matrix in the presence of the new physics,
is defined by Eq.\,(\ref{doubleprime13}).

Comparing Eqs.\,(\ref{he-std}) and (\ref{he3}),
we see that if
\begin{eqnarray}
1+\epsilon_{ee}=c^2_\beta
\label{circle}
\end{eqnarray}
is satisfied, then by introducing the two effective
mixing angle
\begin{eqnarray}
\sin^2\theta^{\text{eff}}_{23}&\equiv&
\frac{s^2_{23}}{(c''_{13})^2}
\label{eff1}\\
\theta^{\text{eff}}_{13}&\equiv&\theta''_{13}\,,
\label{eff2}
\end{eqnarray}
Eq.\,(\ref{he3}) shows almost the identical
behavior as that of the standard scheme (\ref{he-std}).

A few remarks are in order.

Firstly, although Eq.\,(\ref{circle}) is satisfied only in a narrow
region,\footnote{
In fact, Eq.\,(\ref{circle}) stands for an upper half circle
around a center (-1/2,0) with
a radius 1/2 in the $(\epsilon_{ee},|\epsilon_{e\tau}| )$ plane.}
as long as Eq.\,(\ref{circle}) holds, the high-energy behavior of
the disappearance oscillation probability coincides with that
of the standard three-flavor scheme.  Off this upper half circle,
equivalence between the behaviors of Eqs.\,(\ref{he3}) and
(\ref{he-std}) are lost, but it is expected that due to
the experimental errors around this
upper half circle, there exist
some regions in which the behaviors of Eq.\,(\ref{he3})
and (\ref{he-std}) are similar.

Secondly, Eq.\,(\ref{eff2}) indicates that the angle
$\theta''_{13}$
plays a role similar to that of $\theta_{13}$ in the standard scheme.
Note that the corrections in Eq.\,(\ref{he-std}) due to
$\theta_{13}$ were not discussed in Refs.\,\cite{Friedland:2004ah,Friedland:2005vy},
where it was suggested that the quantities that appear
in Eq.\,(\ref{he3}) imply the effective
two-flavor mixing angle,
$\sin^2\theta_{\text{atm}}=(1+t^2_\beta)s^2_{23}/
(1+s^2_{23}t^2_\beta)$ and the effective mass-squared difference
$\Delta m^2_{\text{atm}}=\Delta m^2_{32}(1+s^2_{23}t^2_\beta)
/(1+t^2_\beta)$.  While the former is exactly the same
as Eq.\,(\ref{eff1}) in the limit $\theta_{13}\to0$,
the latter does not appear in our result.
This is because the correction factor
$(1+s^2_{23}t^2_\beta)/(1+t^2_\beta)$
naturally arises from the three-flavor
contributions, i.e., from the $\theta_{13}$ dependent terms,
and we need not normalize $\Delta m^2$.
If we postulate the effective
mixing angle to be $\theta^{\text{eff}}_{23}=\pi/4$
in Eq.\,(\ref{eff1}), then $c_{23}\equiv\cos\theta_{23}>0$ can be
expressed by $\beta$ and $\theta_{13}$ as
\begin{eqnarray}
c_{23}=
\frac{s_\beta c_\beta s_{13} c_{13}\cos\Phi}
{2-s^2_\beta c^2_{13}}
+\left\{
\frac{1+c^2_\beta s^2_{13}}
{2-s^2_\beta c^2_{13}}
+\left(
\frac{s_\beta c_\beta s_{13} c_{13}\cos\Phi}
{2-s^2_\beta c^2_{13}}
\right)^2
\right\}^{1/2},
\label{c23}
\end{eqnarray}
where we have introduced
\begin{eqnarray}
\Phi\equiv\delta+\mbox{\rm arg}(\epsilon_{e\tau}).
\nonumber
\end{eqnarray}
In the limit $\theta_{13}\to0$,
Eq.\,(\ref{c23}) agrees with the expression
$c^2_{23}=1/(1+c^2_\beta)$ obtained
in Refs.\,\cite{Friedland:2004ah,Friedland:2005vy}.

Thirdly, Ref.\,\cite{Hosaka:2006zd} performed
a three-flavor analysis of the atmospheric
neutrino data, and the authors concluded that
the atmospheric neutrino data alone
gives $s^2_{13}<0.14~(0.27)$ at 90\%CL
for a normal (inverted) mass hierarchy.
This implies that the range $s^2_{13}<0.14$
is consistent at 90\%CL with the high-energy atmospheric
neutrino data, i.e., the upward going $\mu$ events.
In the present case, we found that
the value of $(s''_{13})^2$
can be made smaller than 0.14 in almost all the region
for $0\le\sin^22\theta_{13}<0.15$
and $0\le s^2_\beta<0.5$, by adjusting the value of $\Phi$.
With the conditions (\ref{circle}), (\ref{eff1}), (\ref{eff2}), therefore,
the region around the upper half circle for
$|\epsilon_{e\tau}|\lesssim 0.5$ and
$-1/2\lesssim 1+\epsilon_{ee}\lesssim 0$
is expected to be consistent with the atmospheric
neutrino data.

Thus, taking into account the various constraints described above,
we will work with the ansatz
\begin{eqnarray}
{\cal A}= A\left(
\begin{array}{ccc}
1+ \epsilon_{ee}~~ & 0 & \epsilon_{e\tau}\\
0 & 0 & 0\\
\epsilon_{e\tau}^\ast & 0 & ~~|\epsilon_{e\tau}|^2/(1 + \epsilon_{ee})
\end{array}
\right)
\label{ansatz}
\end{eqnarray}
in the following discussions.

\section{Sensitivity of T2KK to $\epsilon_{ee}$ and $\epsilon_{e\tau}$
\label{sensitivity}}

In this section we discuss the sensitivity of the T2KK
experiment to the non-standard interaction in propagation with the ansatz
(\ref{ansatz}).
Since $\epsilon_{\tau\tau}$ is expressed in terms of
$\epsilon_{e\tau}$ and $\epsilon_{ee}$,
the only new degrees of freedom are
$\epsilon_{ee}$, $|\epsilon_{e\tau}|$
and $\mbox{\rm arg}(\epsilon_{e\tau})$.
Firstly, in sect.\,\ref{t2kk},
we briefly describe the setup of the
T2KK experiment.
Secondly, in sect.\,\ref{discrimination},
we consider the $(\epsilon_{ee}, |\epsilon_{e\tau}|)$ plane
and discuss the region in which
T2KK can discriminate the non-standard interaction in propagation
from the standard three-flavor scenario.
Thirdly, in sect.\,\ref{precision}, we study
the case in which new physics can be
discriminated and discuss
how precisely T2KK can determine
$\epsilon_{ee}$ and $|\epsilon_{e\tau}|$.
Then, in sect.\,\ref{cp},
we consider whether
the two complex phases
$\delta$ and $\mbox{\rm arg}(\epsilon_{e\tau})$
can be determined separately.

\subsection{The T2KK experiment
\label{t2kk}}
The T2KK experiment\,\cite{Ishitsuka:2005qi,Hagiwara:2005pe}
is a proposal for the future extension
of the T2K experiment\,\cite{Itow:2001ee}.
In this proposal, a water
Cherenkov detector
is placed not only in Kamioka (at a baseline length $L$ = 295 km) but also
in Korea (at $L$ = 1050 km), whereas the power of the beam
at J-PARC in Tokai Village is upgraded to 4 MW.
As in the T2K experiment, it is assumed that
T2KK uses an off-axis beam with a 2.5$^\circ$
angle between the directions
of the charged pions and neutrinos,
and the neutrino energy spectrum has a peak approximately
at 0.7 GeV.
Because the two detectors are assumed to be identical,
some of the systematic errors cancel.
Also, because the distances of the two detectors
from the source are different, parameter
degeneracy in the three-flavor oscillation
scenario\,\cite{BurguetCastell:2001ez,Minakata:2001qm,Fogli:1996pv,Barger:2001yr}
is expected to be resolved with this
complex\,\cite{Ishitsuka:2005qi,Hagiwara:2005pe,Hagiwara:2006vn,Kajita:2006bt,Hagiwara:2006nn,Hagiwara:2009bb}.

In this paper, we assume the same setup
as that in Refs.\,\cite{Kajita:2006bt,Ribeiro:2007jq}.
In our analysis, we use the disappearance channel
$\nu_\mu\to\nu_\mu$ and $\bar{\nu}_\mu\to\bar{\nu}_\mu$,
the appearance one $\nu_\mu\to\nu_e$ and
$\bar{\nu}_\mu\to\bar{\nu}_e$, and data from
single-Cherenkov-ring electron and muon events.
We assume that the measurement will run for 8 years in total,
4 years each for the neutrino and anti--neutrino beams.
The fiducial volume of each detector is 0.27 Mton.
The density of the Earth is assumed to be
$\rho=2.3\,[{\rm g/cm^3}]$ in the case of
Tokai--Kamioka, and $\rho=2.8\,[{\rm g/cm^3}]$
in the case of Tokai--Korea.
The electron fraction $Y_e$ is assumed to be 0.5.
The energy resolution is considered to be 80 MeV.
We use various information such as the
neutrino flux from Ref.\,\cite{Kaneyuki:2008}.

\subsection{Bounds on $\epsilon_{ee}$ and $\epsilon_{e\tau}$
\label{discrimination}}

Firstly, we discuss the case of the
region ($\epsilon_{ee}$, $|\epsilon_{e\tau}|$), in which
we can distinguish the new physics with ansatz
(\ref{ansatz}) from the standard three-flavor scheme.
To perform such a test, we introduce
the following quantity:

\begin{eqnarray}
\nonumber \Delta\chi^2  &=& \min_{\text{param},\epsilon_\ell}
 \bigg [
 \sum_{k=1}^4 \bigg \{ \sum_{i=1}^5 \frac{1}{\sigma_i^2(e)}
\{ N^0_i(e) +
 B^0_i(e) - N_i(e) \sum_{l=3,7}(1+ f(e)_l^i \epsilon_l )
\nonumber\\
&{\ }&\qquad\qquad\qquad - B_i(e) \sum_{l=1,2,7} (1+ f(e)_l^i \epsilon_l ) \}^2
 \nonumber \\
&{\ }&\qquad\qquad\quad +  \sum_{i=1}^{20}  \frac{1}{\sigma_i^2(\mu)}
 \{ N^0_i(\mu) +
 B^0_i(\mu) - N_i(\mu) \sum_{l=4,5,7} (1+ f(\mu)_l^i \epsilon_l )
\nonumber\\
&{\ }&\qquad\qquad\qquad - B_i(\mu) \sum_{l=4,6,7} (1+ f(\mu)_l^i \epsilon_l ) \}^2
 \bigg \} \nonumber\\
&{\ }&\qquad\qquad + \sum_{l=1}^7 (\frac{\epsilon_l}{\tilde{\sigma}_l})^2 
+ \Delta\chi^2_{\text{prior}}\bigg ],
\label{chi1}
\end{eqnarray}
where the prior $\Delta\chi^2_{\text{prior}}$ is given by
\begin{eqnarray}
\Delta\chi^2_{\text{prior}}\equiv
\frac{2.7 \times (\sin^22\theta_{23}-1.0)^2}{(0.06)^2}
 +\frac{(\sin^2\theta_{13}-0.02)^2}{(0.01)^2} 
 +\frac{(|\Delta
 m^2_{31}|-2.4\times10^{-3}[{\rm eV}^2])^2}{(1.5\times10^{-4}[{\rm eV}^2])^2}.
\nonumber
\end{eqnarray}
In principle we could perform an analysis
without the prior $\Delta\chi^2_{\text{prior}}$,
but in that case it would take more computation time
by minimizing $\Delta\chi^2$ for the parameter region which is
already excluded by the present data of the
atmospheric and reactor experiments.
So we have included the prior in our analysis
to save computation time.
In Eq.\,(\ref{chi1}),
$N^0_i(e)$, $N^0_i(\mu)$ ($B^0_i(e)$, and $B^0_i(\mu)$)
are the expected signal (background) numbers of events in the presence of
the new physics (\ref{ansatz}),
while $N_i(e)$, $N_i(\mu)$ ($B_i(e)$, and $B_i(\mu)$)
are the expected signal (background) numbers of events in the three-flavor
framework with the standard matter effect.
All these numbers except $B^0_i(e)$ and $B_i(e)$
depend on the neutrino oscillation parameters.
The indices $i$ and $k=1,\cdots,4$
stand for the number of the neutrino energy bin 
for electrons and muons and the four combinations 
of detectors in Kamioka and Korea with the 
neutrino and anti-neutrino beams, respectively.
For the electron events, there are five energy bins
(400-500~MeV, 500-600~MeV, 600-700~MeV, 700-800~MeV, and 
800-1200~MeV), whereas for the muon events,
there are twenty bins from 200 to 1200~MeV with 50~MeV width.
$\sigma_i(\ell)~(\ell=e, \mu)$ stands for the statistical
uncertainties, whereas $\epsilon_\ell~(\ell=1,\cdots,7)$ stands for the systematic
uncertainties in the expected number of signals and backgrounds.
$\Delta\chi^2$ is defined by minimizing the quantity
inside the square bracket in Eq.\,(\ref{chi1}) with respect
to the uncertainties $\epsilon_\ell$ as well as the
oscillation parameters ($|\Delta m^2_{31}|$, sign($\Delta m^2_{31}$),
$\theta_{23}$,
$\delta$) of the standard three-flavor scheme, on which the numbers of events
$N_i(e)$, $N_i(\mu)$, $B_i(e)$, and $B_i(\mu)$ depend.
The uncertainties in $B_i(e)$ and
$N_i(e)$
are represented by 4 parameters $\epsilon_j$ ($j=1,2,3,7$).
The backgrounds in the muon events are referred to as
non-quasi-elastic events in Refs.\,\cite{Kajita:2006bt,Ribeiro:2007jq}.
The uncertainties in  $B_i(\mu)$ and $N_i(\mu)$  
are represented by 4 parameters $\epsilon_j$ ($j=4,\cdots,7$). 
The parameter $f(e~{\rm or}~\mu)^i_j$ indicates the
possible dependence of the parameter $\epsilon_j$ on the $i$-th energy bin.
$\epsilon_1$ stands for the uncertainty
in the overall background normalization
for electron events with $\tilde{\sigma}_{1}$ = 0.05.
$\epsilon_2$ is the energy-dependent uncertainty
for the background electron-like events with a function
$f(e)_2^i=((E_\nu-800~\text{MeV}) / 400~\text{MeV})$ and
$\tilde{\sigma}_{2}$ = 0.05.
$\epsilon_3$ is the uncertainty in the detection efficiency
for the electron signal events with $\tilde{\sigma}_{3}=0.05$.
$\epsilon_4$ is the energy-dependent uncertainty for
both the muon signal and background events with
the function $f(\mu)_4^i = (E_\nu-800~\text{MeV}) / 800~\text{MeV}$
and $\tilde{\sigma}_{4} = 0.05$.
$\epsilon_5$ is the uncertainty in the 
signal detection efficiency
for the muon signal events with $\tilde{\sigma}_{5} = 0.05$.
$\epsilon_6$ is the uncertainty in the separation of quasi-elastic and 
non-quasi-elastic interactions in the muon events
and $\tilde{\sigma}_{6}$ = 0.20.
$\epsilon_7$ stands for the uncertainty in the neutrino flux
in Korea, and $\tilde{\sigma}_{7}$ is assumed to be
the predicted flux difference between those in Kamioka and in 
Korea, given in Ref.\,\cite{Meregaglia}.

In Eq.\,(\ref{chi1}),
the numbers of events $N^0_i(e)$, $N^0_i(\mu)$ ($B^0_i(e)$, and $B^0_i(\mu)$)
depend not only on the new physics parameters $\epsilon_{ee}$,
$|\epsilon_{e\tau}|$, arg($\epsilon_{e\tau}$) but also on
the standard oscillation parameters, which we denote as
$\bar{\theta}_{12}$,
$\bar{\theta}_{13}$, $\bar{\theta}_{23}$, $\Delta \bar{m}^2_{21}$,
$\Delta \bar{m}^2_{31}$, and $\bar{\delta}$.
Here, we take the best-fit values
for most of the standard oscillation parameters
as the reference values:
\begin{eqnarray}
\sin^2 (2\bar{\theta}_{12}) &=& 0.87 
\nonumber\\
\sin^2 (2\bar{\theta}_{23}) &=& 1.0
\nonumber\\
\Delta \bar{m}_{21}^2 &=& 7.9 \times 10^{-5} {\rm eV}^2
\nonumber\\
\Delta \bar{m}_{32}^2 &=& 2.4 \times 10^{-3} {\rm eV}^2
\label{central-values}
\end{eqnarray}
On the other hand, since we have no information
on $\bar{\theta}_{13}$ and $\bar{\delta}$,
we will take several reference values
for these parameters.

\begin{figure}
    \includegraphics[width=6.5cm]{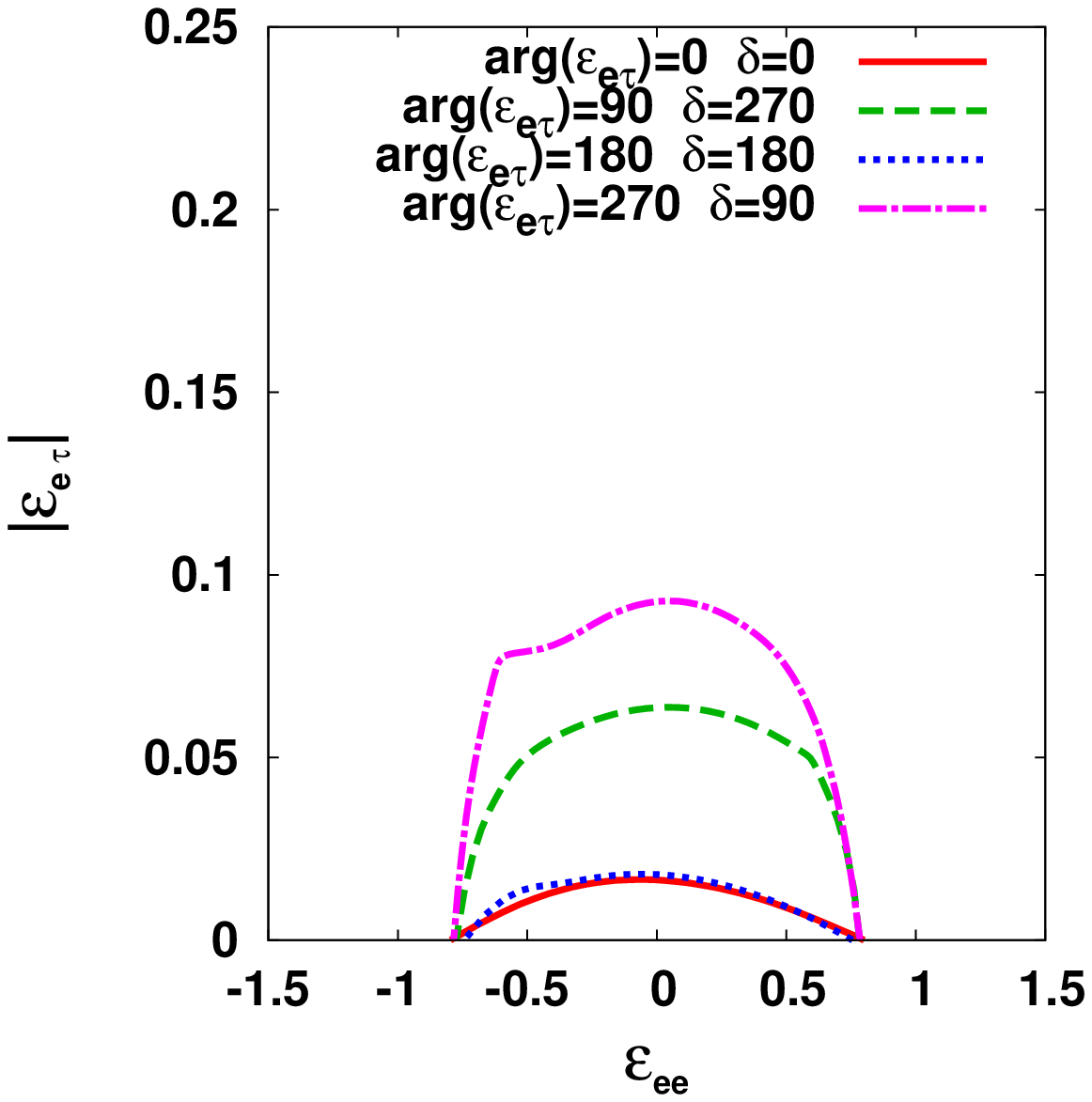}
    \includegraphics[width=6.5cm]{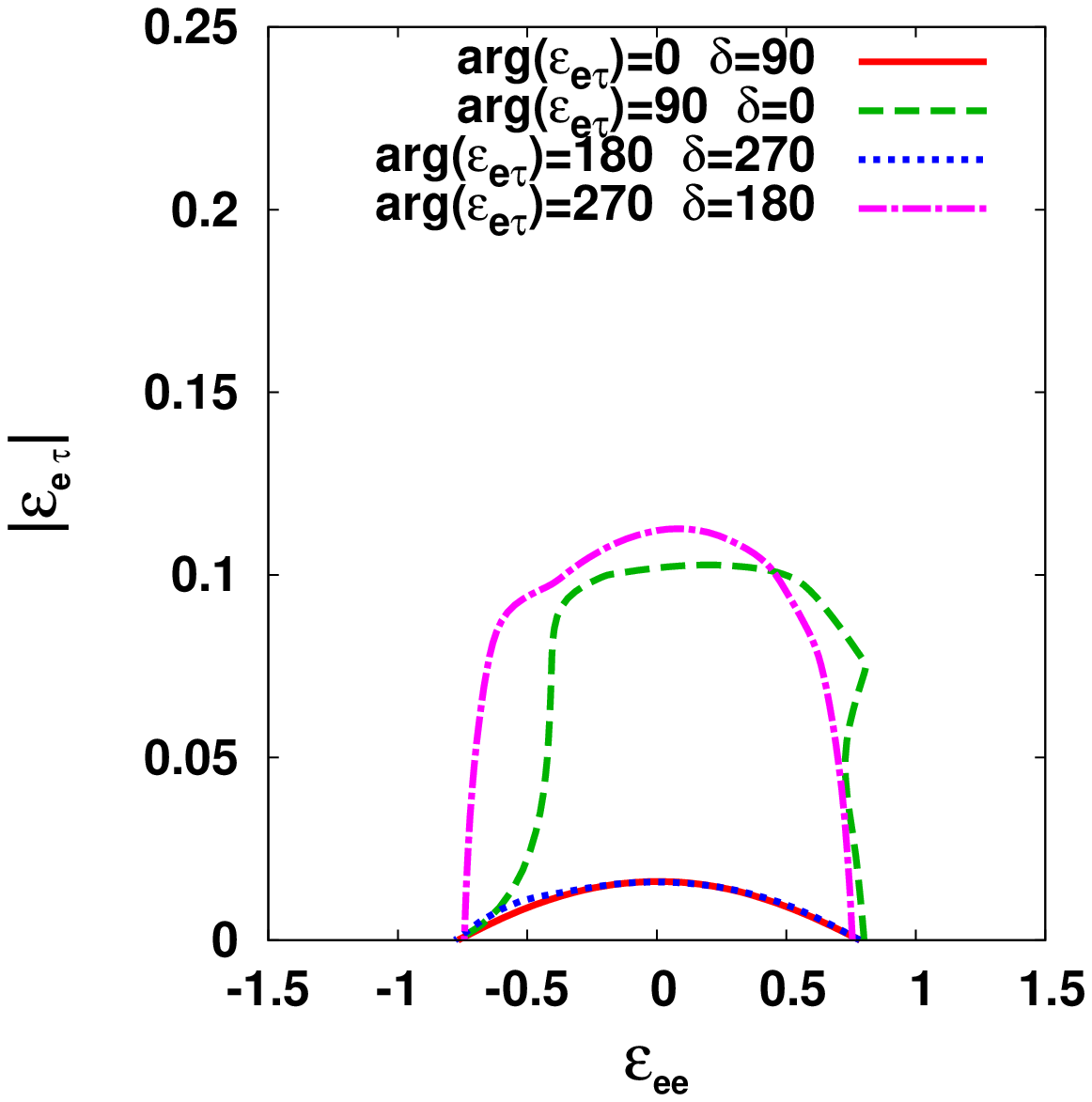}
    \includegraphics[width=6.5cm]{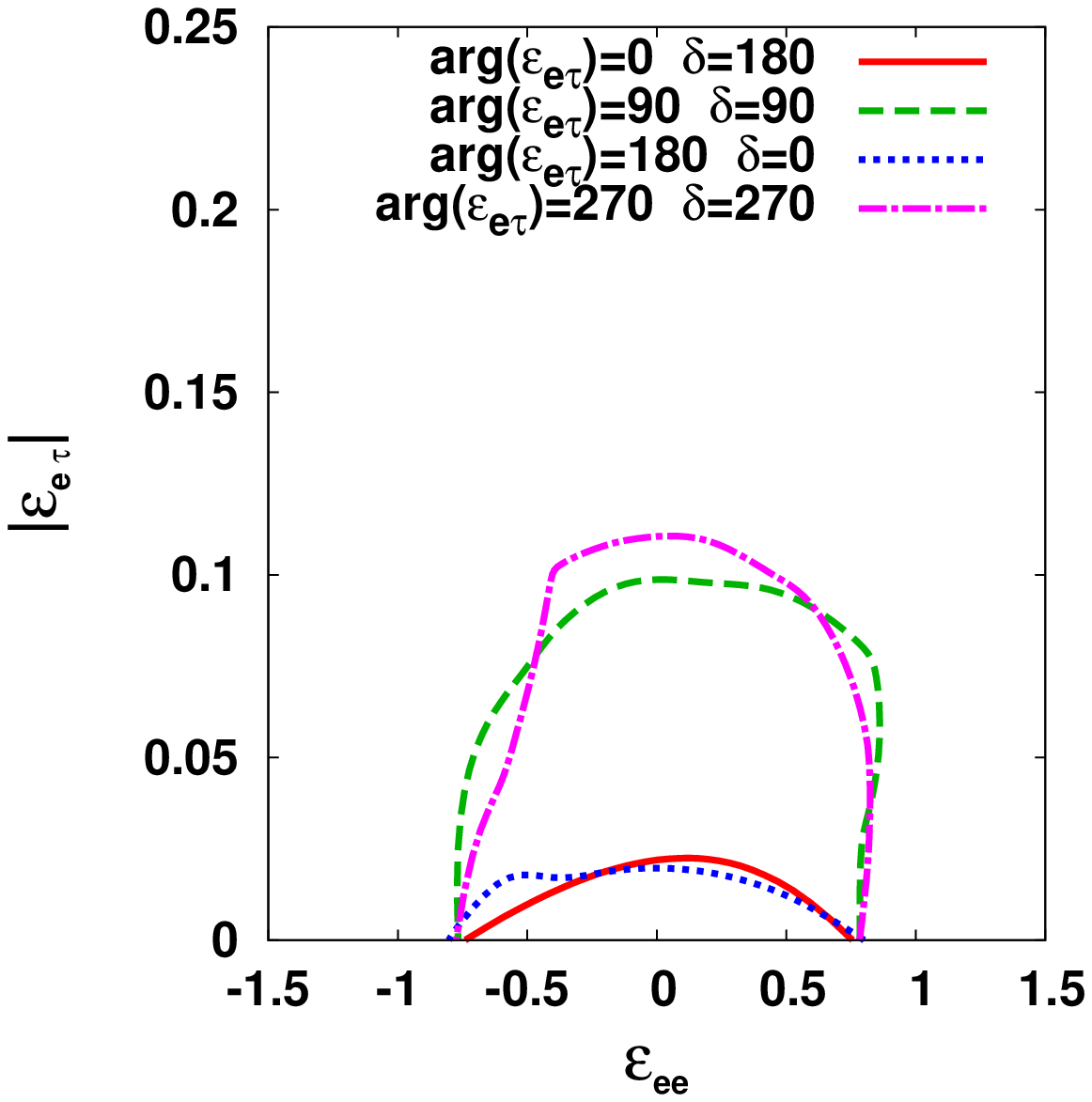}
    \includegraphics[width=6.5cm]{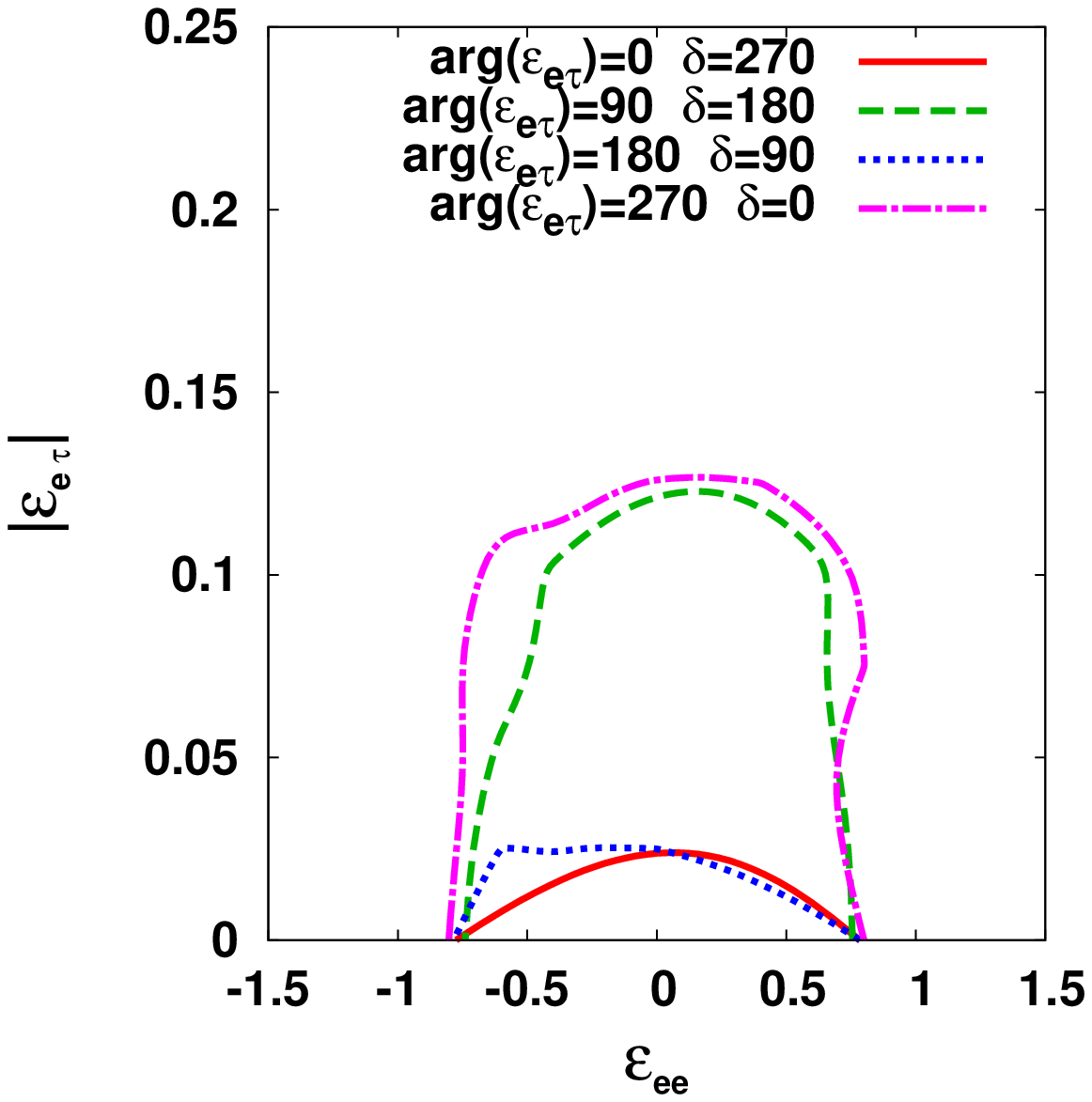}
\caption{Region in which the new physics is discriminated at 90\%CL
from the standard three-flavor scheme for $\sin^22\bar{\theta}_{13}=10^{-4}$.}
\label{fig1}
\end{figure}

\begin{figure}
    \includegraphics[width=6.5cm]{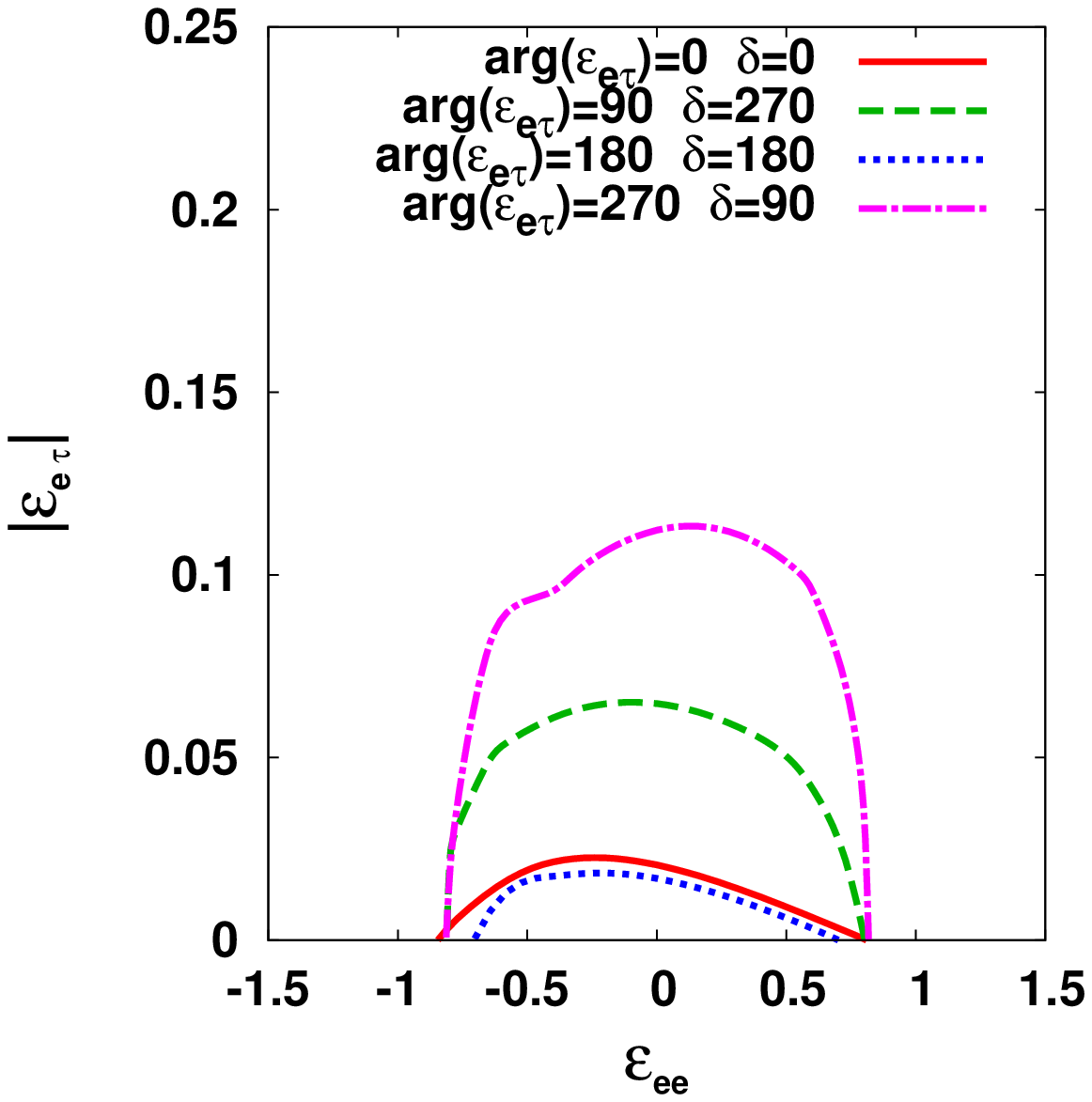}
    \includegraphics[width=6.5cm]{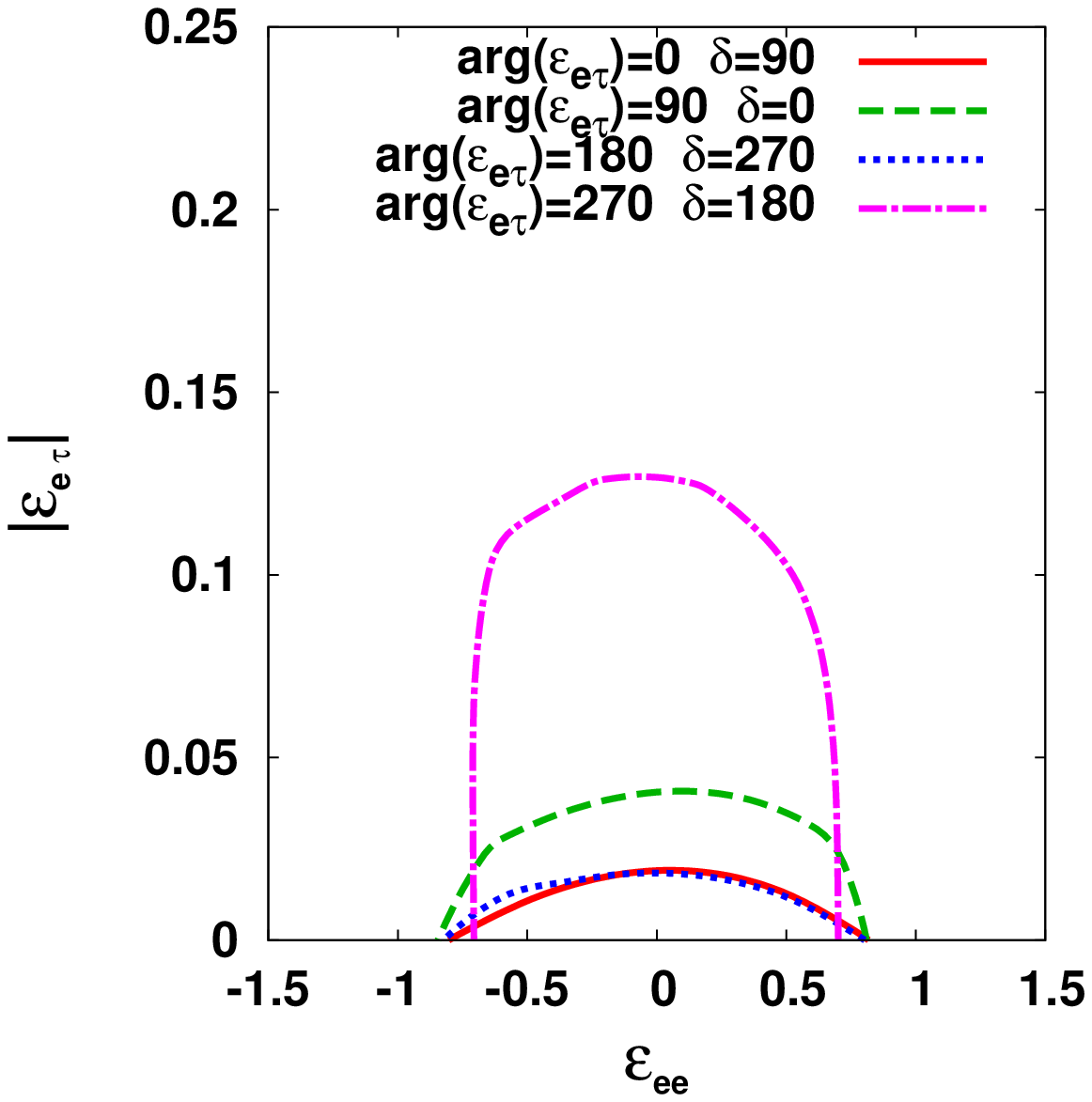}
    \includegraphics[width=6.5cm]{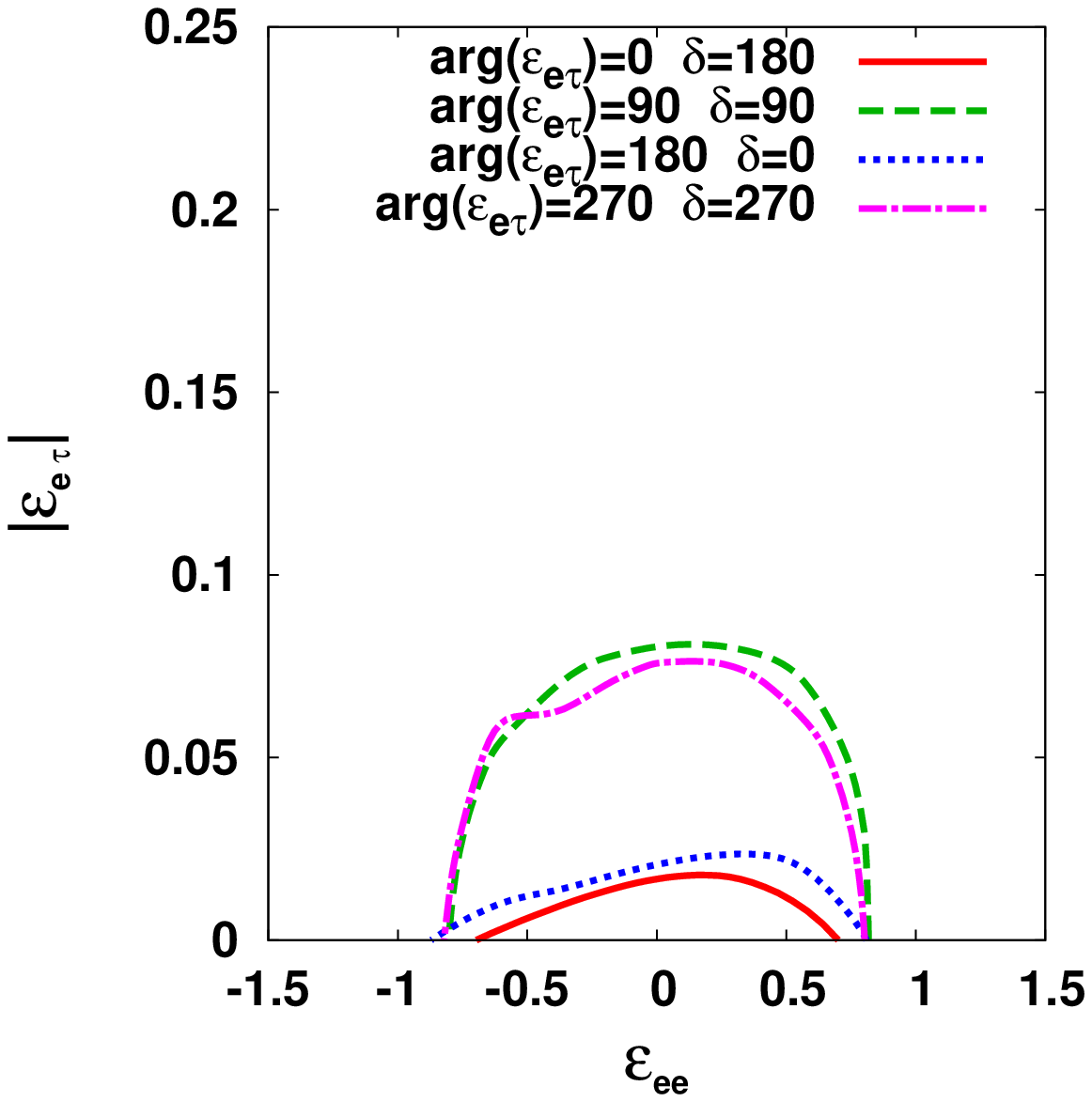}
    \includegraphics[width=6.5cm]{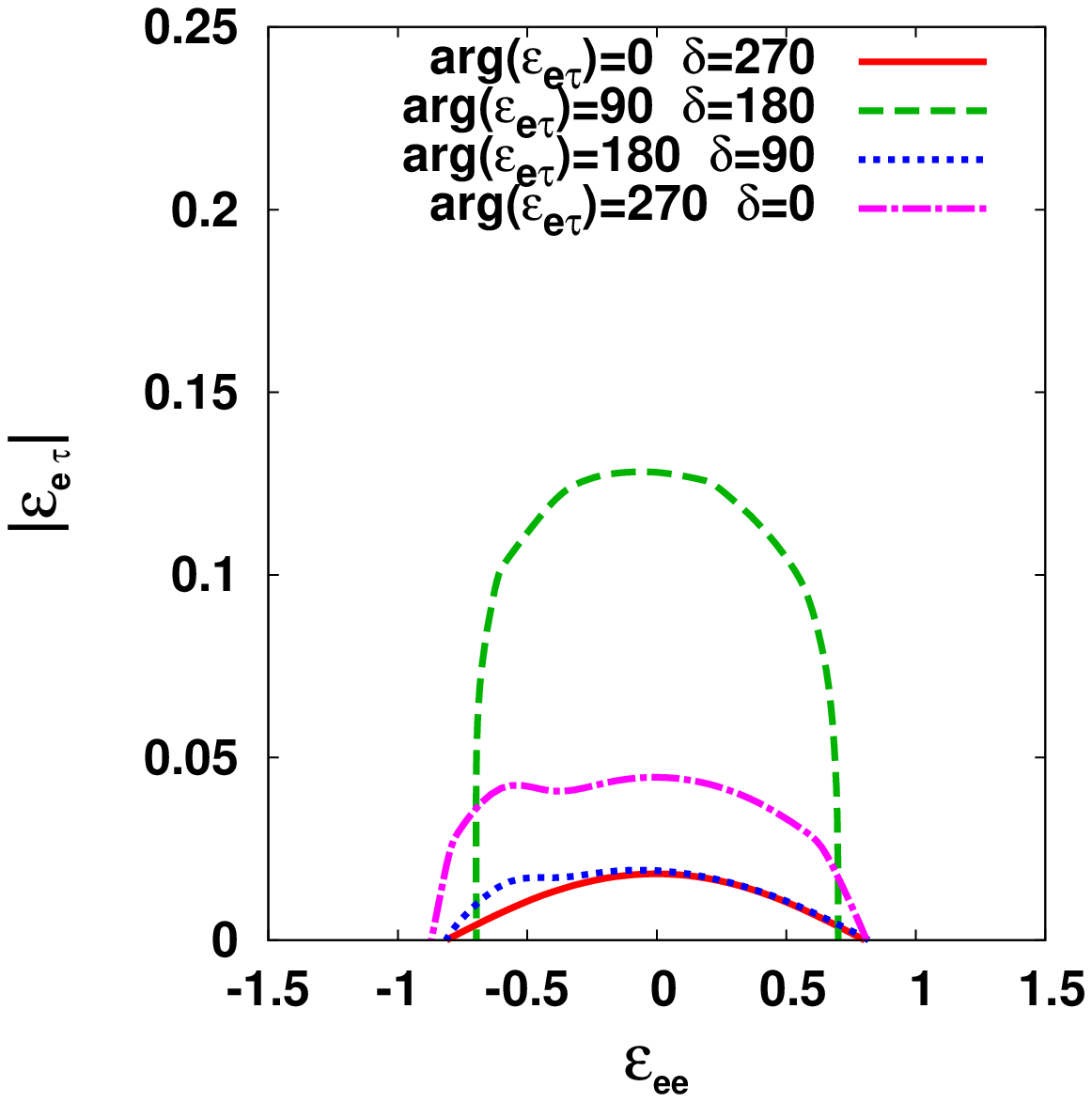}
\caption{Region in which the new physics is discriminated at 90\%CL
from the standard three-flavor scheme for $\sin^22\bar{\theta}_{13}=10^{-3}$.}
\label{fig2}
\end{figure}

\begin{figure}
    \includegraphics[width=6.5cm]{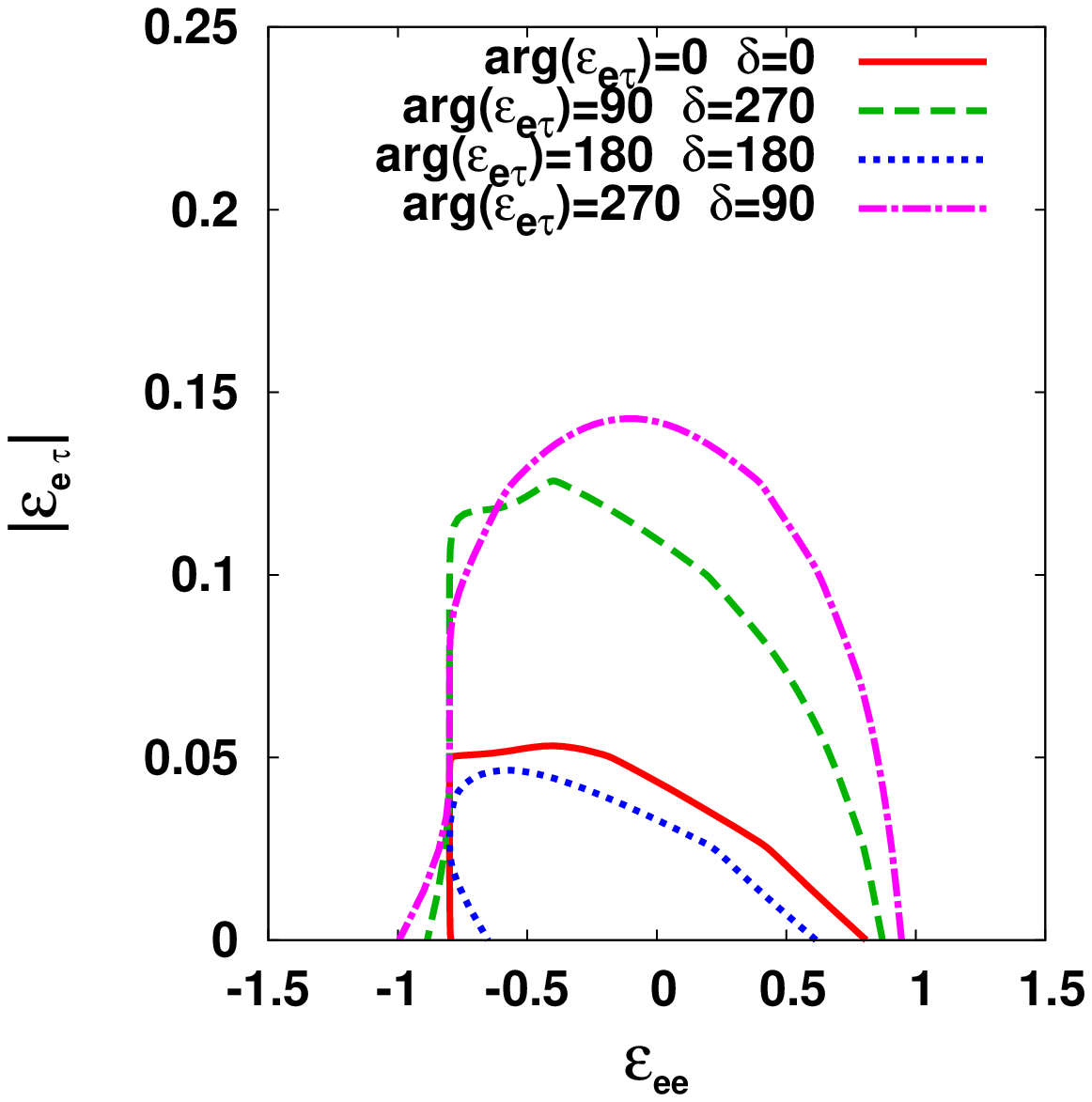}
    \includegraphics[width=6.5cm]{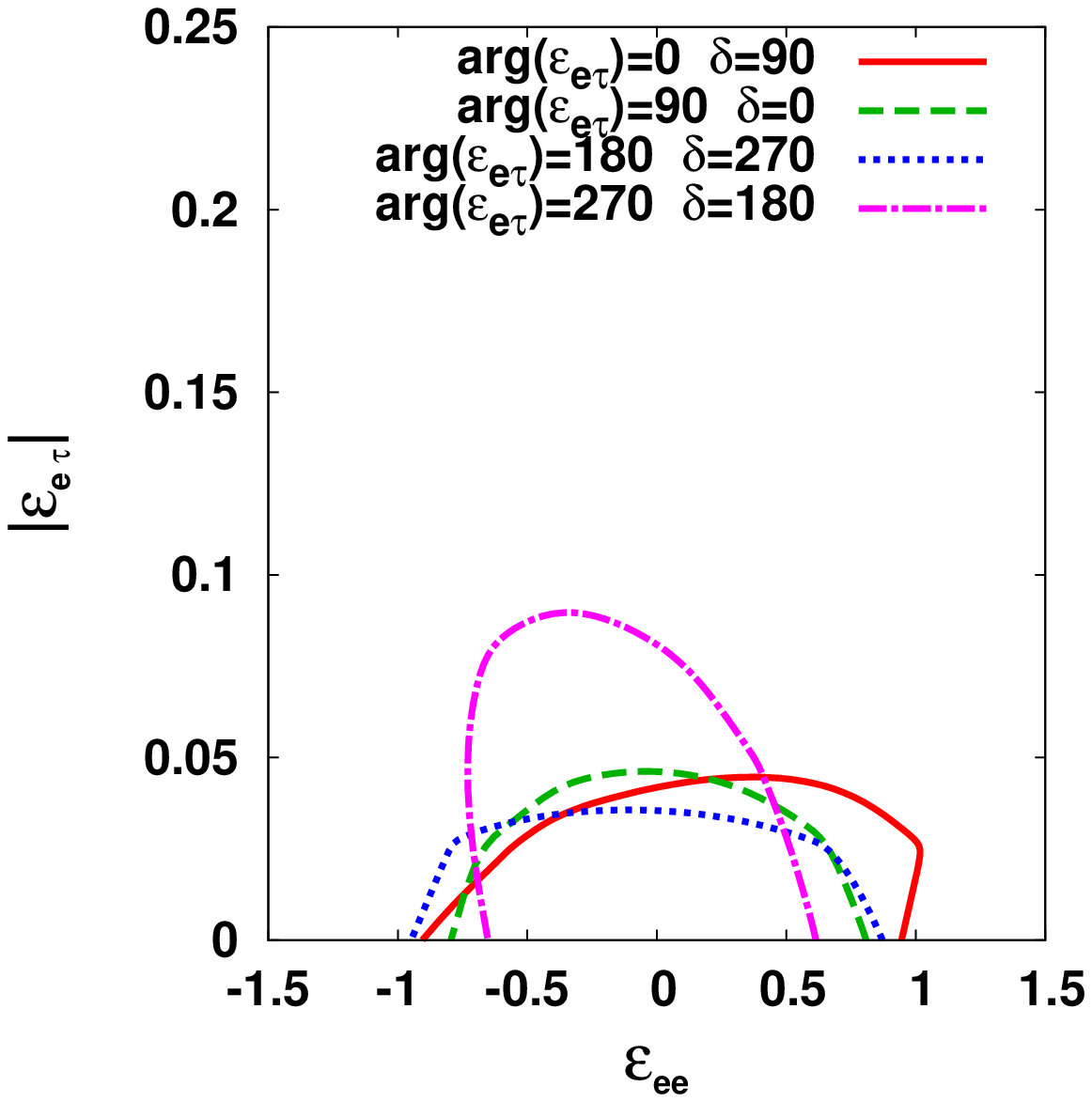}
    \includegraphics[width=6.5cm]{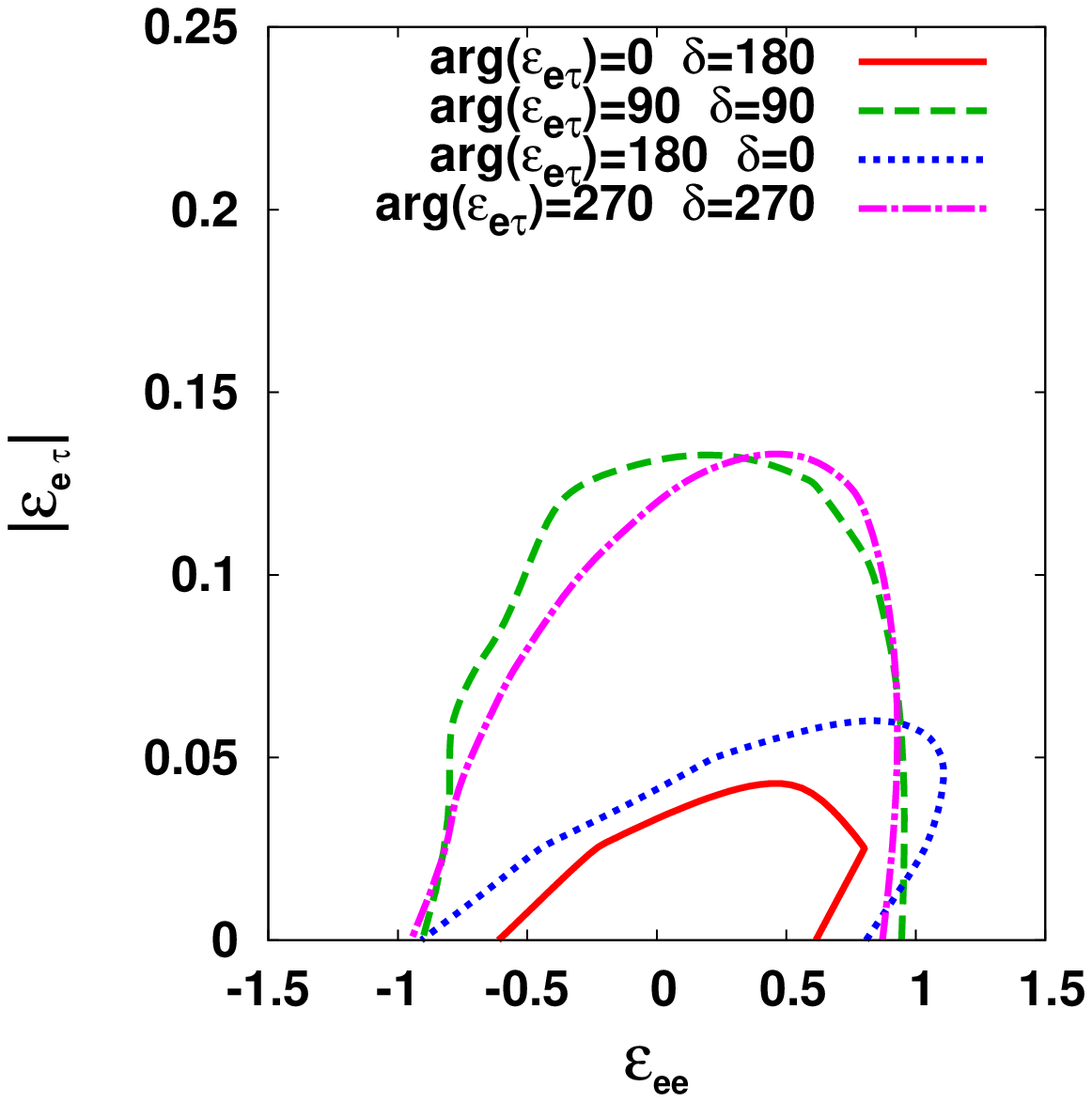}
    \includegraphics[width=6.5cm]{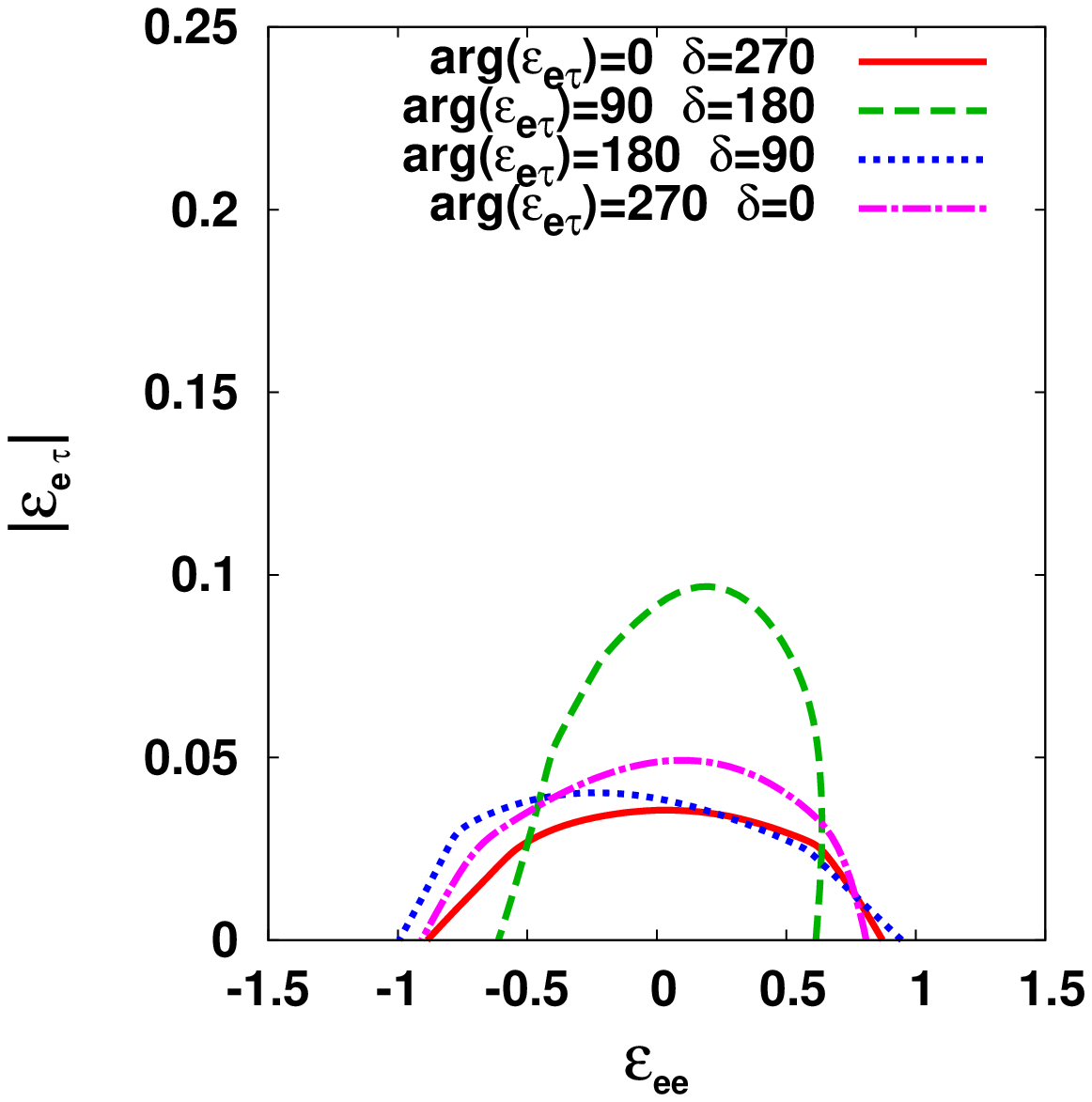}
\caption{Region in which the new physics is discriminated at 90\%CL
from the standard three-flavor scheme for $\sin^22\bar{\theta}_{13}=10^{-2}$.}
\label{fig3}
\end{figure}

\begin{figure}
    \includegraphics[width=6.5cm]{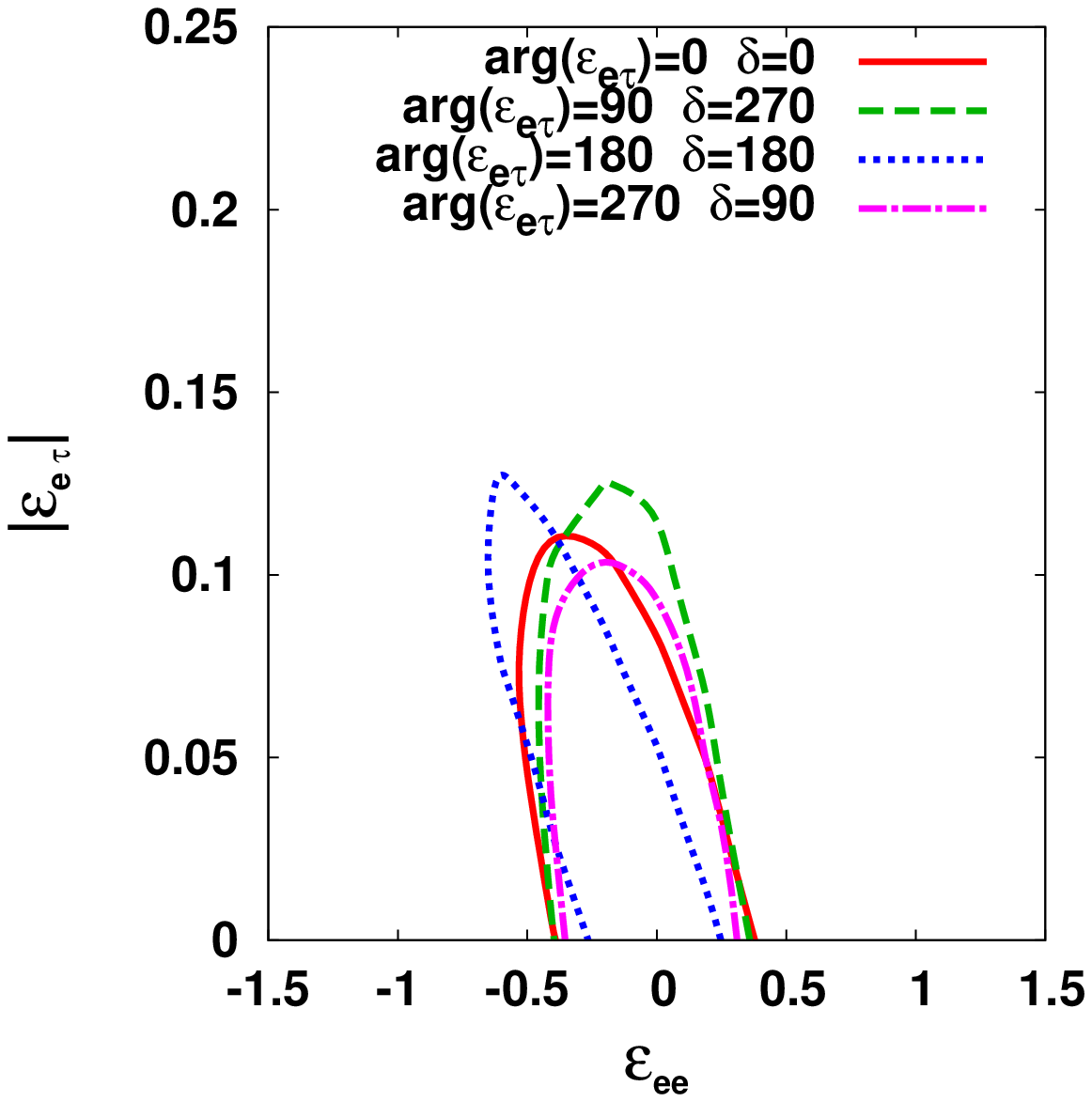}
    \includegraphics[width=6.5cm]{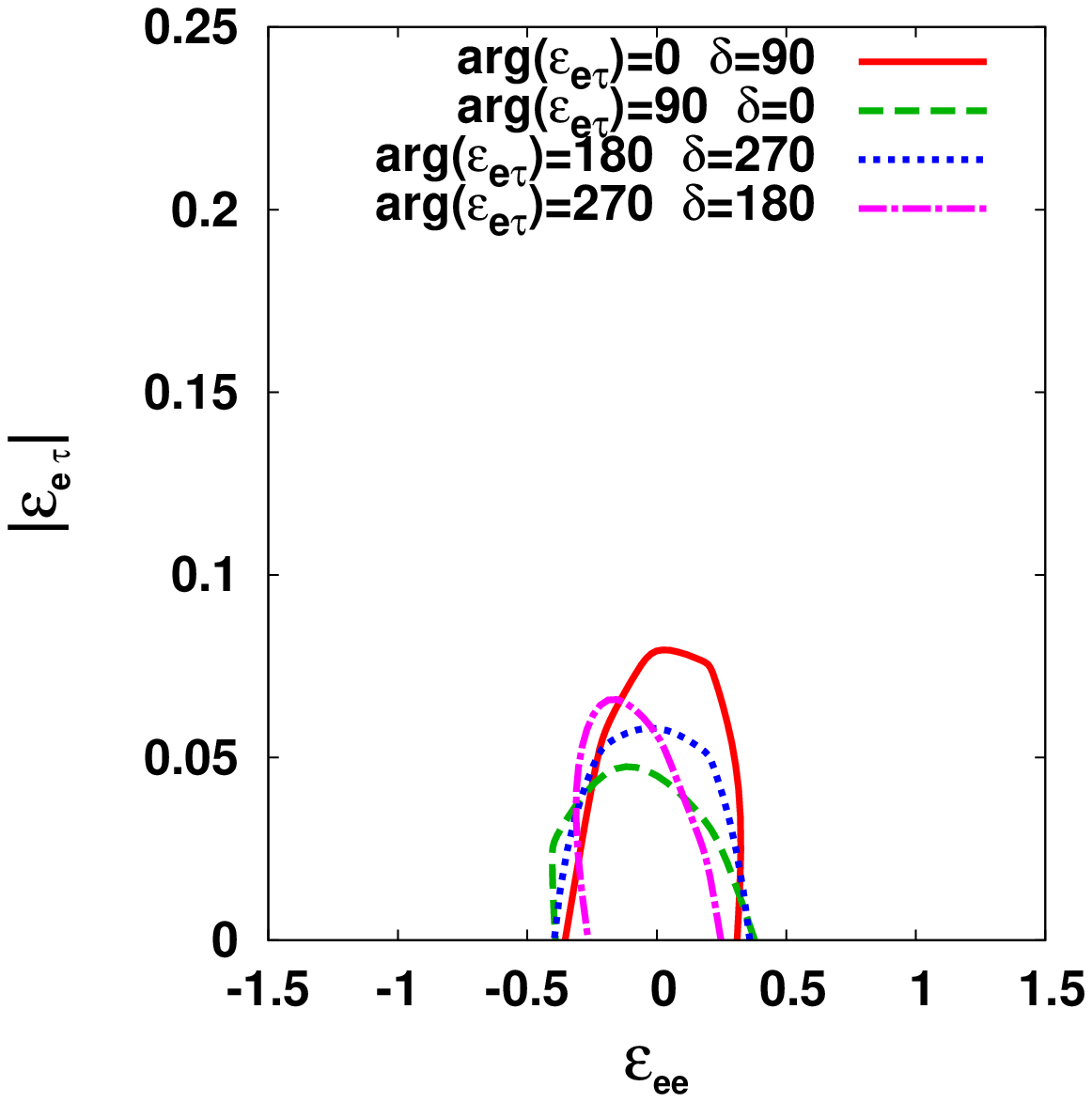}
    \includegraphics[width=6.5cm]{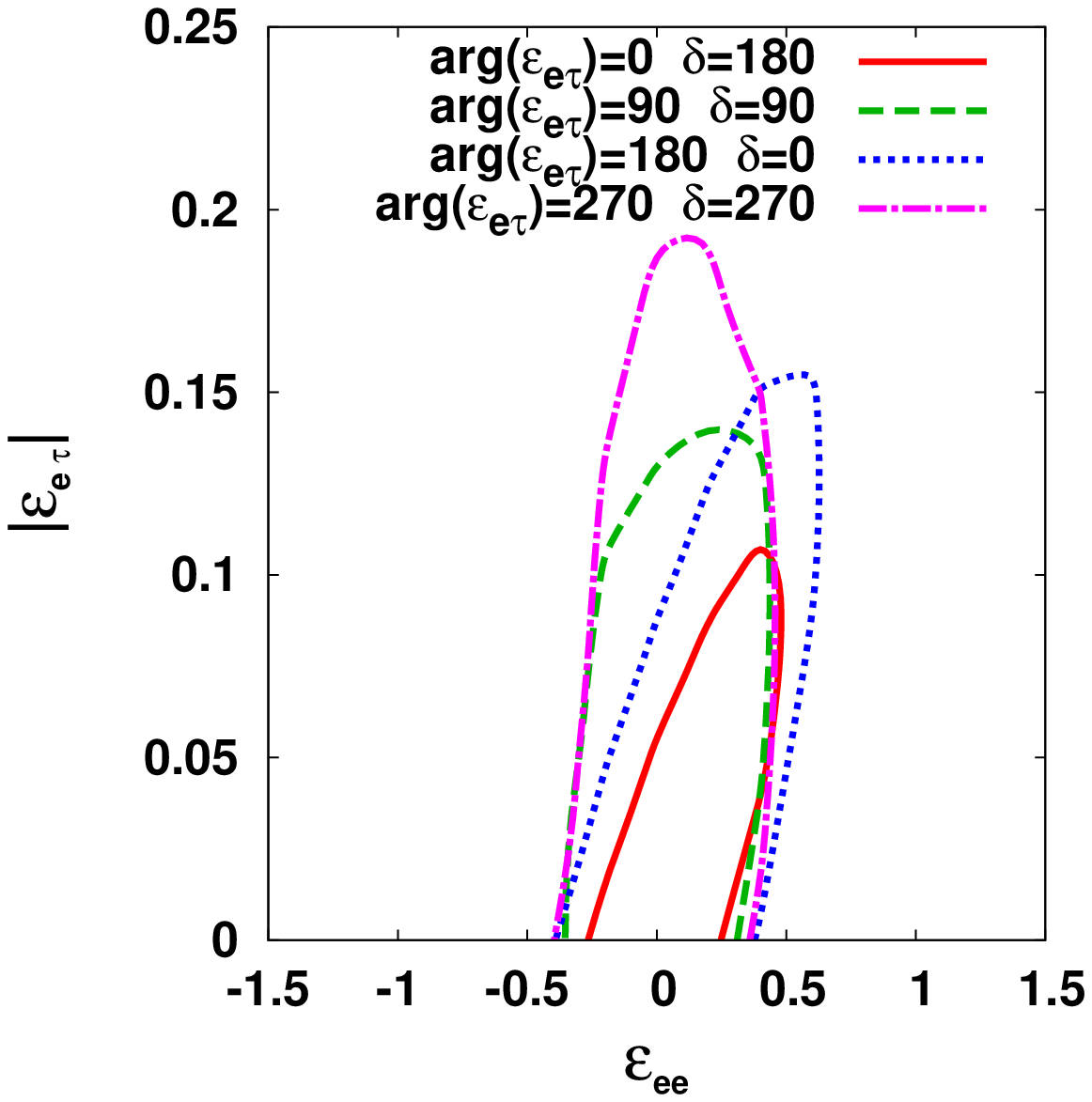}
    \includegraphics[width=6.5cm]{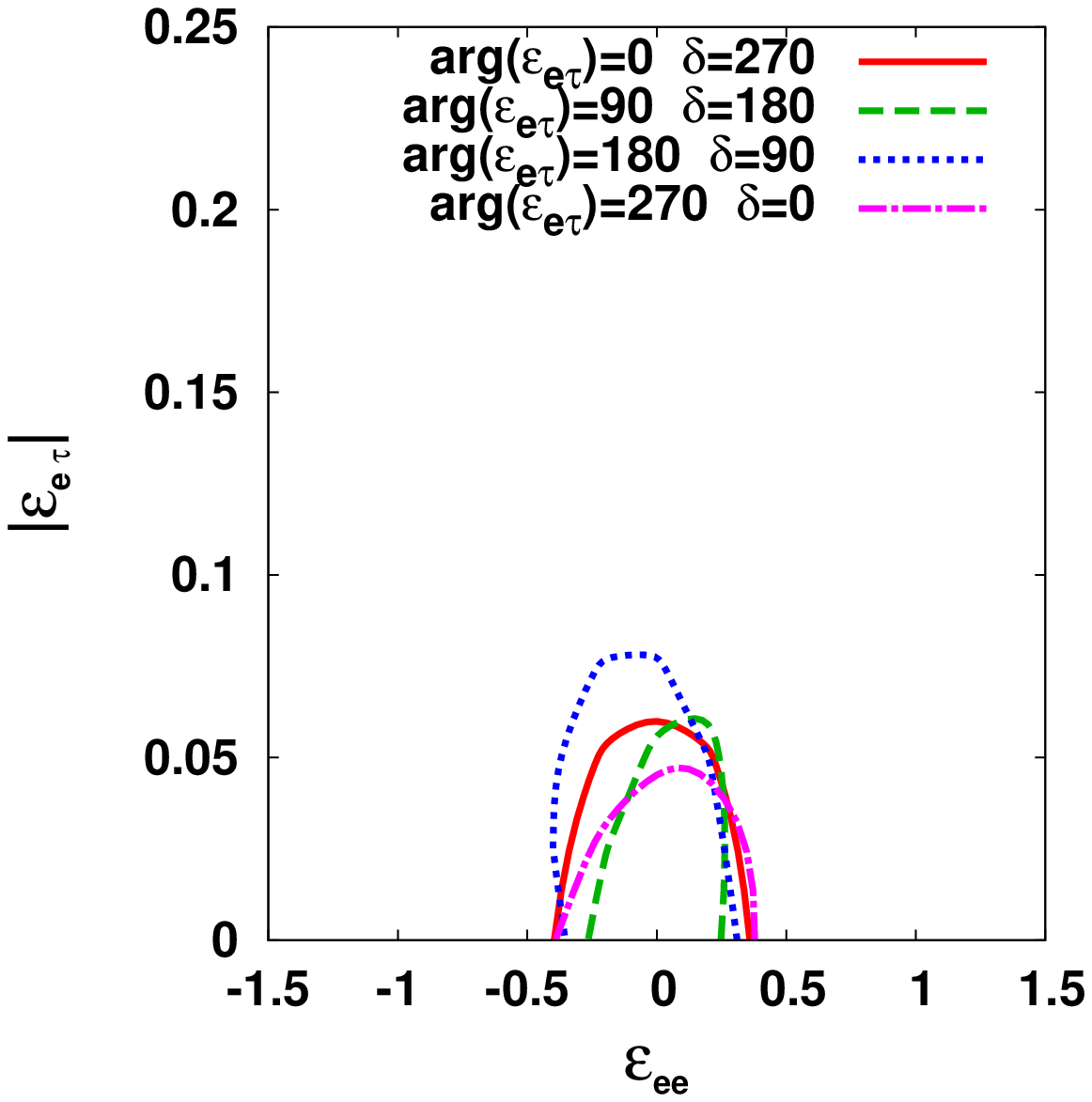}
\caption{Region in which the new physics is discriminated at 90\%CL
from the standard three-flavor scheme for $\sin^22\bar{\theta}_{13}=0.12$.}
\label{fig4}
\end{figure}

The results are shown in Figs.\ref{fig1}--\ref{fig4},
where the curves are drawn at 90\%CL
($\Delta\chi^2=4.6$ for 2 degrees of freedom).
The new physics with the ansatz (\ref{ansatz}) can be
distinguished from the standard three-flavor scheme
outside the curves.  Four different choices
for the phases $\delta$ and arg($\epsilon_{e\tau}$)
are taken, where the sum of the two phases is
the same in each figure.  It has been known\,\cite{Ota:2001pw,Yasuda:2007jp}
that in the limit of
$\Delta m^2_{21}\to0$, the oscillation probability
$P(\nu_\mu\to\nu_e)$ depends only on the
combination $\delta+\mbox{\rm arg}(\epsilon_{e\tau})$
of the phases.  If the four curves in Figs.\ref{fig1}--\ref{fig4}
coincided with each other, then it would mean that
the contribution of the solar mass-squared difference
were small.  From these figures, however,
the behaviors of the four curves are different
even if $\delta+\mbox{\rm arg}(\epsilon_{e\tau})$ = constant,
so the contribution of $\Delta m^2_{21}$ is not
negligible.  This is because we are considering
the oscillation probabilities in Korea,
where $\Delta m^2_{21}L/4E\sim0.3$.
The analytic form of the
oscillation probability $P(\nu_\mu\to\nu_e)$
is given in the appendix \ref{appendixf},
where the correction to $P(\nu_\mu\to\nu_e)$ to the first order in
$\Delta m^2_{21}$ is also given.
The corrections to the energy eigenvalues $\tilde{E}_j~(j=1,2,3)$
are proportional to either
$\sin^2\theta''_{12}$ or $\cos^2\theta''_{12}$,
where $\theta''_{12}$ is defined in Eq.\,(\ref{doubleprime12}),
and $\theta''_{12}$ depends mainly on $\mbox{\rm arg}(\epsilon_{e\tau})$.
From these figures we conclude that the T2KK experiment can
distinguish the new physics with ansatz (\ref{ansatz})
at 90\%CL approximately for $|\epsilon_{ee}|\gtrsim 1$ and
$|\epsilon_{e\tau}|\gtrsim 0.2$.
In other words, if T2KK lacks evidence
of a new physics, then with the ansatz (\ref{ansatz}),
we can put bounds on the two parameters:
$|\epsilon_{ee}|\lesssim 1$ and $|\epsilon_{e\tau}|\lesssim 0.2$.
While the bound on $\epsilon_{ee}$ is modest, the one on
$|\epsilon_{e\tau}|$ is impressive compared with
the present bound (\ref{epsilon-m}).

\subsection{Precision in determination of
$\epsilon_{ee}$, $|\epsilon_{e\tau}|$
\label{precision}}
Let us now turn to the case with affirmative results
in the T2KK experiment, i.e., we will discuss the
points that lie {\it outside} the curves
in the ($\epsilon_{ee}$, $|\epsilon_{e\tau}|$) plane
in Figs.\ref{fig1}--\ref{fig4}.\footnote{
Since we have the fixed value of $\theta_{13}$,
we do not have the $\theta_{13}-\epsilon_{e\tau}$ confusion
in Ref.\,\cite{Huber:2001de}, in which
it was shown that the same neutrino survival probability is
produced by the different pairs of ($\theta_{13}$, $\epsilon_{e\tau}$).}

First, we discuss the experimental errors
in $\epsilon_{ee}$ and $|\epsilon_{e\tau}|$, i.e.,
the correlation of errors for these two variables.
In this case, we introduce the following quantity,
which is similar to Eq.\,(\ref{chi1}):
\begin{eqnarray}
\nonumber \Delta\chi^2  &=& \min_{\text{param},\epsilon_\ell}
 \bigg [
 \sum_{k=1}^4 \bigg \{ \sum_{i=1}^5 \frac{1}{\sigma_i^2(e)}
\{ N^0_i(e) +
 B^0_i(e) - N_i(e) \sum_{l=3,7}(1+ f(e)_l^i \epsilon_l )
\nonumber\\
&{\ }&\qquad\qquad\qquad - B_i(e) \sum_{l=1,2,7} (1+ f(e)_l^i \epsilon_l ) \}^2
 \nonumber \\
&{\ }&\qquad\qquad\quad +  \sum_{i=1}^{20}  \frac{1}{\sigma_i^2(\mu)}
 \{ N^0_i(\mu) +
 B^0_i(\mu) - N_i(\mu) \sum_{l=4,5,7} (1+ f(\mu)_l^i \epsilon_l )
\nonumber\\
&{\ }&\qquad\qquad\qquad - B_i(\mu) \sum_{l=4,6,7} (1+ f(\mu)_l^i \epsilon_l ) \}^2
 \bigg \} \nonumber\\
&{\ }&\qquad\qquad + \sum_{l=1}^7 (\frac{\epsilon_l}{\tilde{\sigma}_l})^2 
\bigg ],
\label{chi2}
\end{eqnarray}
where most of the definitions are the same as those
in Eq.\,(\ref{chi1}).  The only differences between
(\ref{chi1}) and (\ref{chi2}) are that
the prior is absent in the latter, and that
$N_i(\ell)~(\ell=e, \mu)$ and $B_i(\ell)~(\ell=e, \mu)$
in the latter (in the former)
are the expected numbers of events in the presence of
new physics with ansatz (\ref{ansatz})
(in the standard scheme), respectively.
Namely, both $N^0_i(\ell)~(\ell=e, \mu)$ and 
($B^0_i(\ell)~(\ell=e, \mu)$ and 
are the expected number of events in the presence of a
new physics with the ansatz (\ref{ansatz}) in Eq.\,(\ref{chi2}).
The number of events $N^0_i(\ell)~(\ell=e, \mu)$
depends on the parameters of the new physics
($\bar{\epsilon}_{ee}$, $|\bar{\epsilon}_{e\tau}|$,
and arg($\bar{\epsilon}_{e\tau}$)) as well as
the oscillation parameters of the standard scheme
$\bar{\theta}_{12}$, $\bar{\theta}_{13}$,
$\bar{\theta}_{23}$, $\Delta \bar{m}^2_{21}$,
$\Delta \bar{m}^2_{31}$, and $\bar{\delta}$.
We fix $\bar{\epsilon}_{ee}$, $|\bar{\epsilon}_{e\tau}|$
at some points outside the curves in Figs.\ref{fig1}--\ref{fig4}
and evaluate $\Delta\chi^2$ as a function of
$\epsilon_{ee}$ and $|\epsilon_{e\tau}|$, which appear in
the argument of $N_i(\ell)~(\ell=e, \mu)$.
For simplicity,
we assume the central values given in Eqs.\,(\ref{central-values})
for $\bar{\theta}_{12}$, $\bar{\theta}_{23}$,
$\Delta \bar{m}^2_{21}$, and $\Delta \bar{m}^2_{31}$.
For $\bar{\theta}_{13}$,
we take a few representative values
$\sin^22\bar{\theta}_{13}=10^{-4}, 10^{-2}, 0.12$.
We assume normal hierarchy and fix the value
of the phases as
$\bar{\delta}=\pi$ and arg$(\bar{\epsilon}_{e\tau})=\pi$
for simplicity.
As for the variables in $N_i(\ell)~(\ell=e, \mu)$,
for simplicity we equate the variables
$\theta_{12}$, $\theta_{13}$, $\theta_{23}$,
$\Delta m^2_{21}$, $\Delta m^2_{31}$,
$\delta$, and arg$(\epsilon_{e\tau})$
to $\bar{\theta}_{12}$, $\bar{\theta}_{13}$, $\bar{\theta}_{23}$,
$\Delta \bar{m}^2_{21}$, $\Delta \bar{m}^2_{31}$,
$\bar{\delta}$, and arg$(\bar{\epsilon}_{e\tau})$
in $N^0_i(\ell)~(\ell=e, \mu)$, respectively.
In this analysis, we do not introduce any
prior because it will be difficult to estimate it
in the presence of the new physics.
Thus, in Eq.\,(\ref{chi2}),
we only minimize the quantity in the
square bracket with respect to the parameters $\epsilon_\ell$ ($\ell=1,\cdots,7$),
and we evaluate $\Delta\chi^2$ as a function of the variables
$\epsilon_{ee}$ and $|\epsilon_{e\tau}|$.

\begin{figure}
    \includegraphics[width=.75\textwidth]{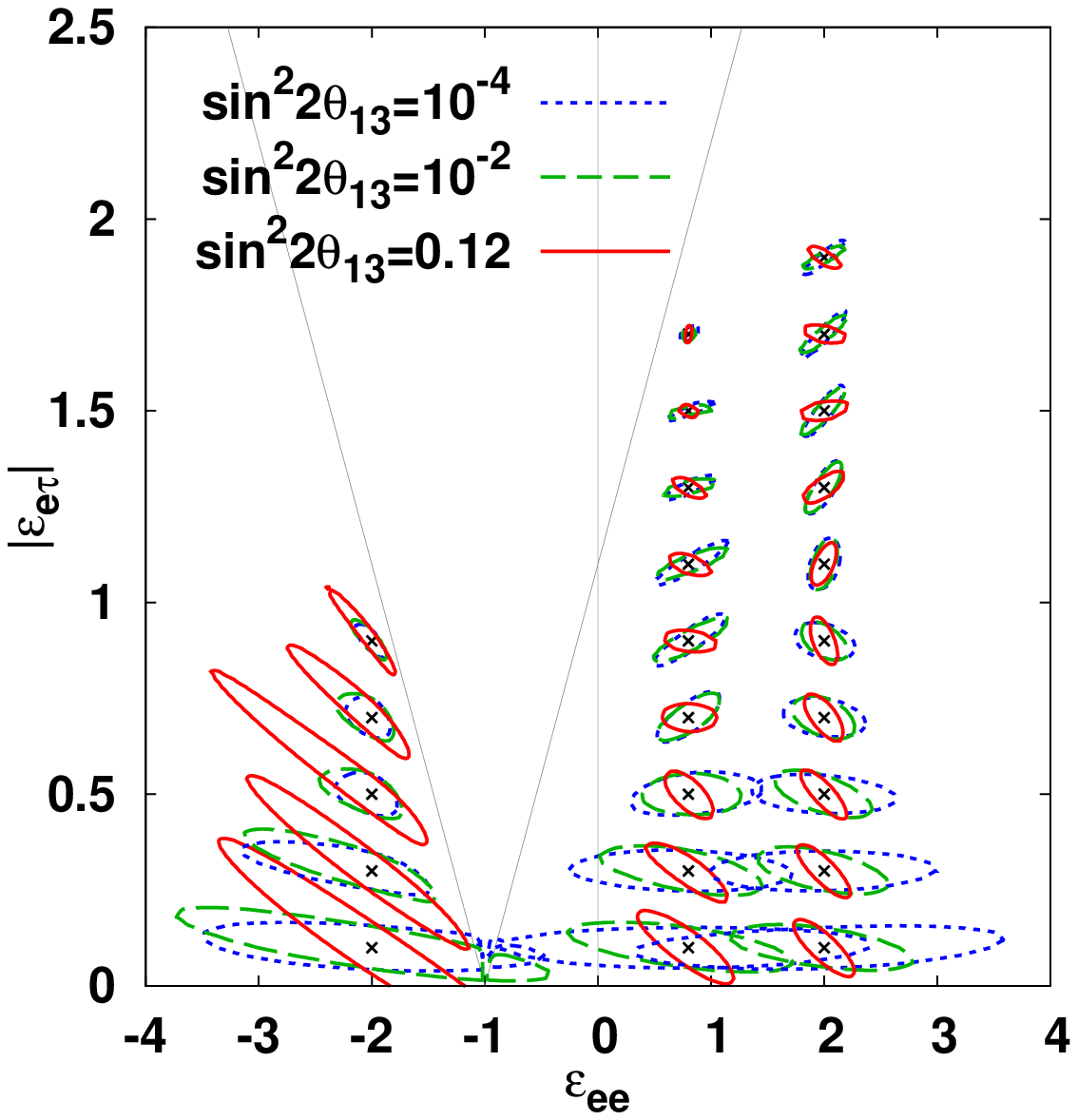}
\caption{Correlation in ($\epsilon_{ee}$, $|\epsilon_{e\tau}|$):
The contours around the true points are
depicted at 90\%CL.
$\bar{\delta}=\mbox{\rm arg}(\bar{\epsilon}_{e\tau})=\pi$ is assumed.}
\label{fig5}
\end{figure}

The results are shown in Fig.\ref{fig5}, where
the contours around the true points are
depicted at 90\%CL
($\Delta\chi^2=4.6$ for 2 degrees of freedom)
for three different values of $\theta_{13}$.
The straight lines $|\epsilon_{e\tau}|=1.1\times|1+\epsilon_{ee}|$
in Fig.\ref{fig5} stand for the approximate bound
from the atmospheric neutrinos, which lead to
$\cos^2\beta>0.45$ or
$|\tan\beta|\lesssim 1.1$\,\cite{Friedland:2005vy},
and we have examined only the points below
these straight lines.
As seen from Fig.\ref{fig5}, the error in
$|\epsilon_{e\tau}|$ is relatively smaller compared
with that in $\epsilon_{ee}$ for all the values of
$\theta_{13}$.  The experimental
error in $\epsilon_{ee}$ increases
for smaller values of $|\epsilon_{e\tau}|$
and $\theta_{13}$.  For $|\epsilon_{e\tau}|\lesssim0.3$,
therefore,
the possibility of $\epsilon_{ee}=0$ cannot be excluded
by the experiment, while $|\epsilon_{e\tau}|=0$
can be for $|\epsilon_{e\tau}|\gtrsim0.2$.
For $1+\epsilon_{ee}<0$, the correlation
in $\epsilon_{ee}$ and $|\epsilon_{e\tau}|$
increases, and in this case, even if T2KK
can discriminate the new physics from the standard scheme,
the determination of these parameters is
difficult.
We have also analyzed
other combinations of the phases $\delta$ and arg($\epsilon_{e\tau}$)
and they share the same features as in Fig.\ref{fig5}.

\subsection{CP violating phases
\label{cp}}
In the ansatz (\ref{ansatz}), there are two
phases $\delta$ and arg($\bar{\epsilon}_{e\tau}$),
and if a new physics exists at all, then it becomes
important whether we can determine these two
phases separately.  Thus, we address this question next.
In this case, we can use the same $\Delta\chi^2$ in Eq.\,(\ref{chi2}),
but there are two differences between this and sect.\ref{precision}.
Firstly, here we vary the variables $\epsilon_{ee}$ and $|\epsilon_{e\tau}|$
in $N_i(\ell)~(\ell=e, \mu)$
and minimize the quantity in the square bracket
in Eq.\,(\ref{chi2}) with respect to these two
parameters as well as the parameters $\epsilon_\ell$.
Secondly, $\Delta\chi^2$ is plotted as a function of
the two variables $\delta$ and arg($\bar{\epsilon}_{e\tau}$)
in $N_i(\ell)~(\ell=e, \mu)$ here, while
it is plotted as a function of
the two variables $\epsilon_{ee}$ and $|\epsilon_{e\tau}|$
in $N_i(\ell)~(\ell=e, \mu)$ in sect.\ref{precision}.

\begin{figure}
    \includegraphics[width=4.5cm]{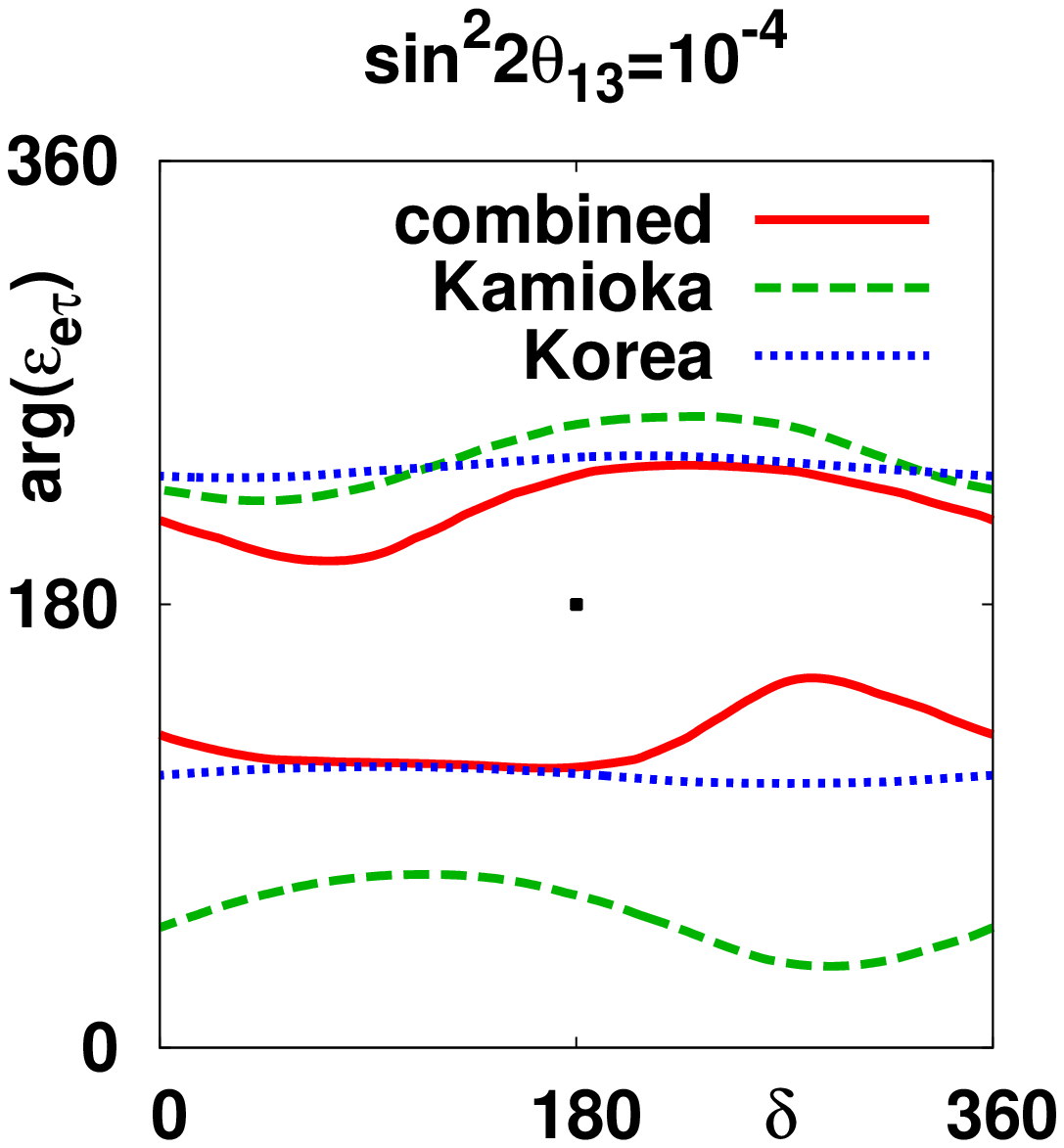}
\hglue -0.9cm
    \includegraphics[width=4.5cm]{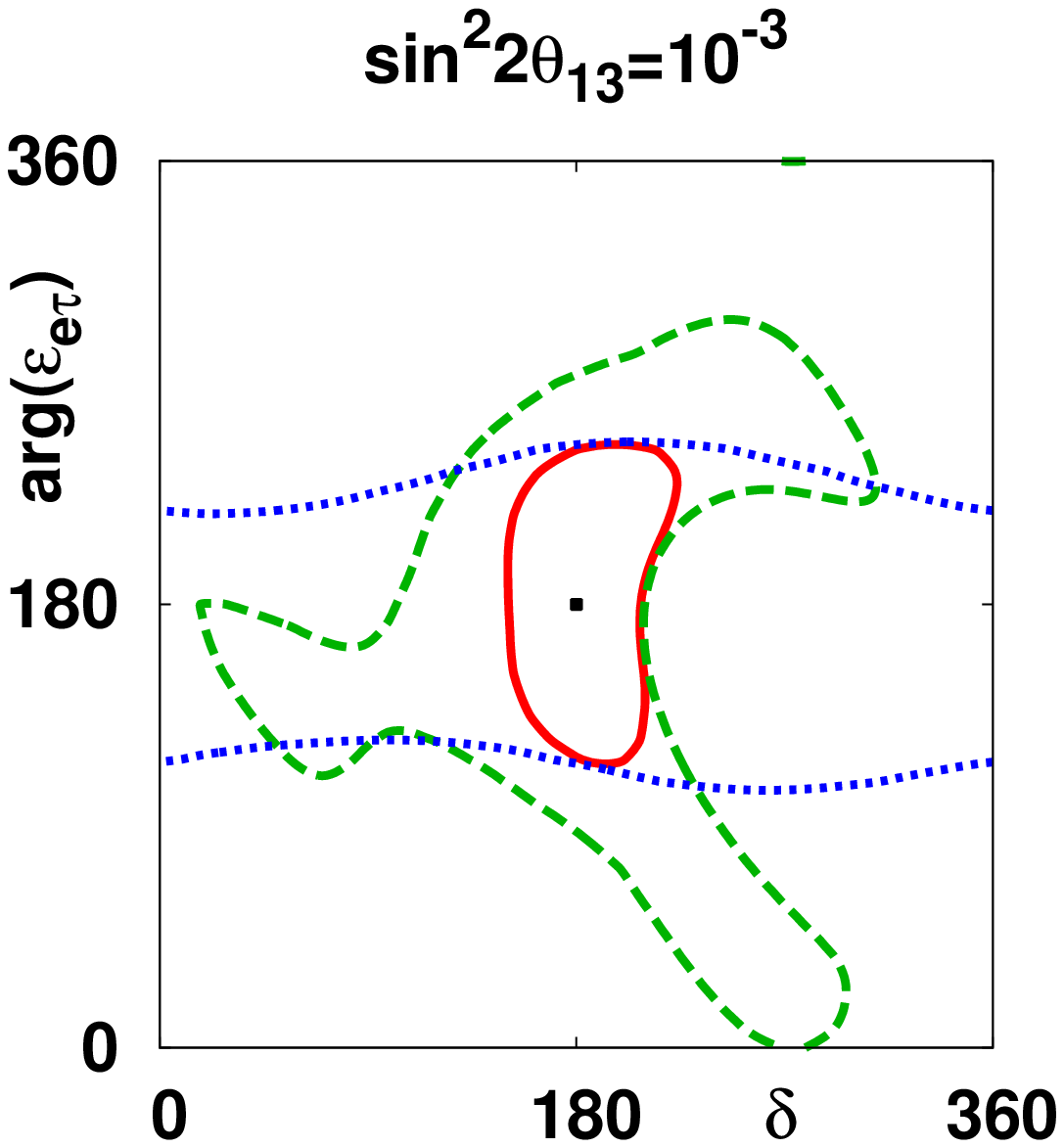}
\hglue -0.9cm
    \includegraphics[width=4.5cm]{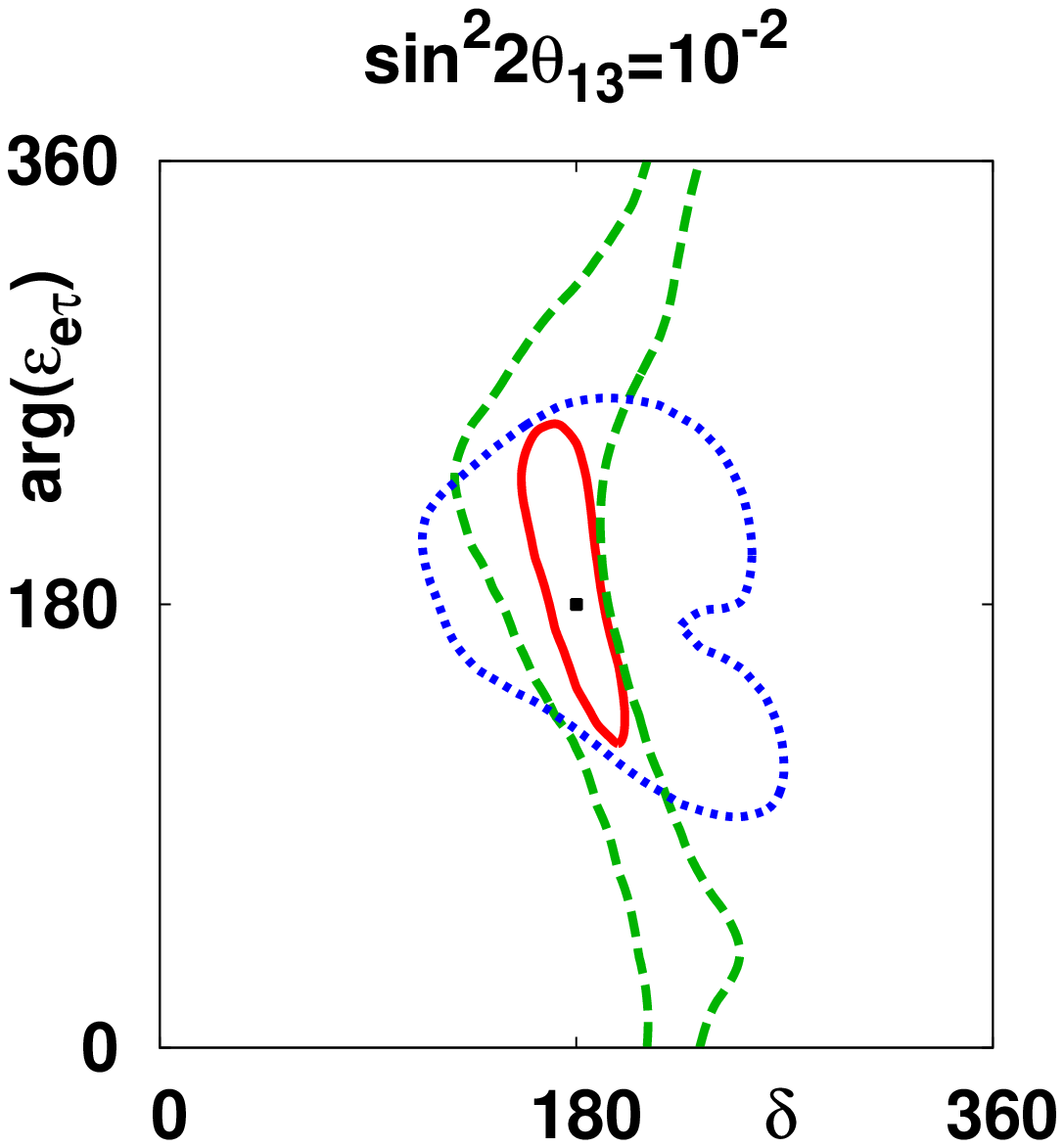}
\hglue -0.9cm
    \includegraphics[width=4.5cm]{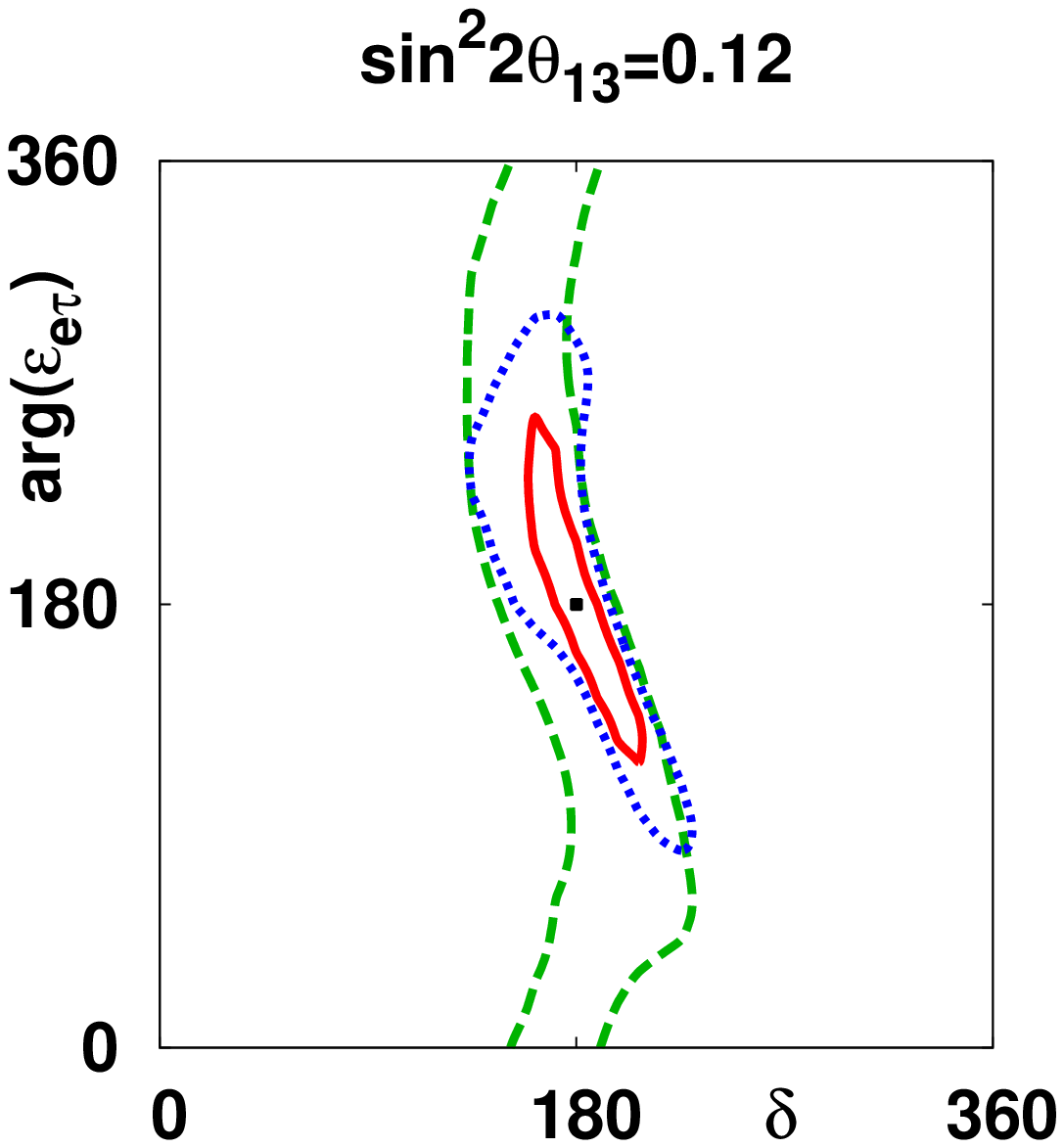}
\caption{Correlation in ($\delta$, arg($\epsilon_{e\tau}$))
for $\epsilon_{ee}=0.8$, $|\epsilon_{e\tau}|=0.2$.
The contours around the true points are
depicted at 90\%CL.
}
\label{fig6}
\end{figure}

\begin{figure}
    \includegraphics[width=4.5cm]{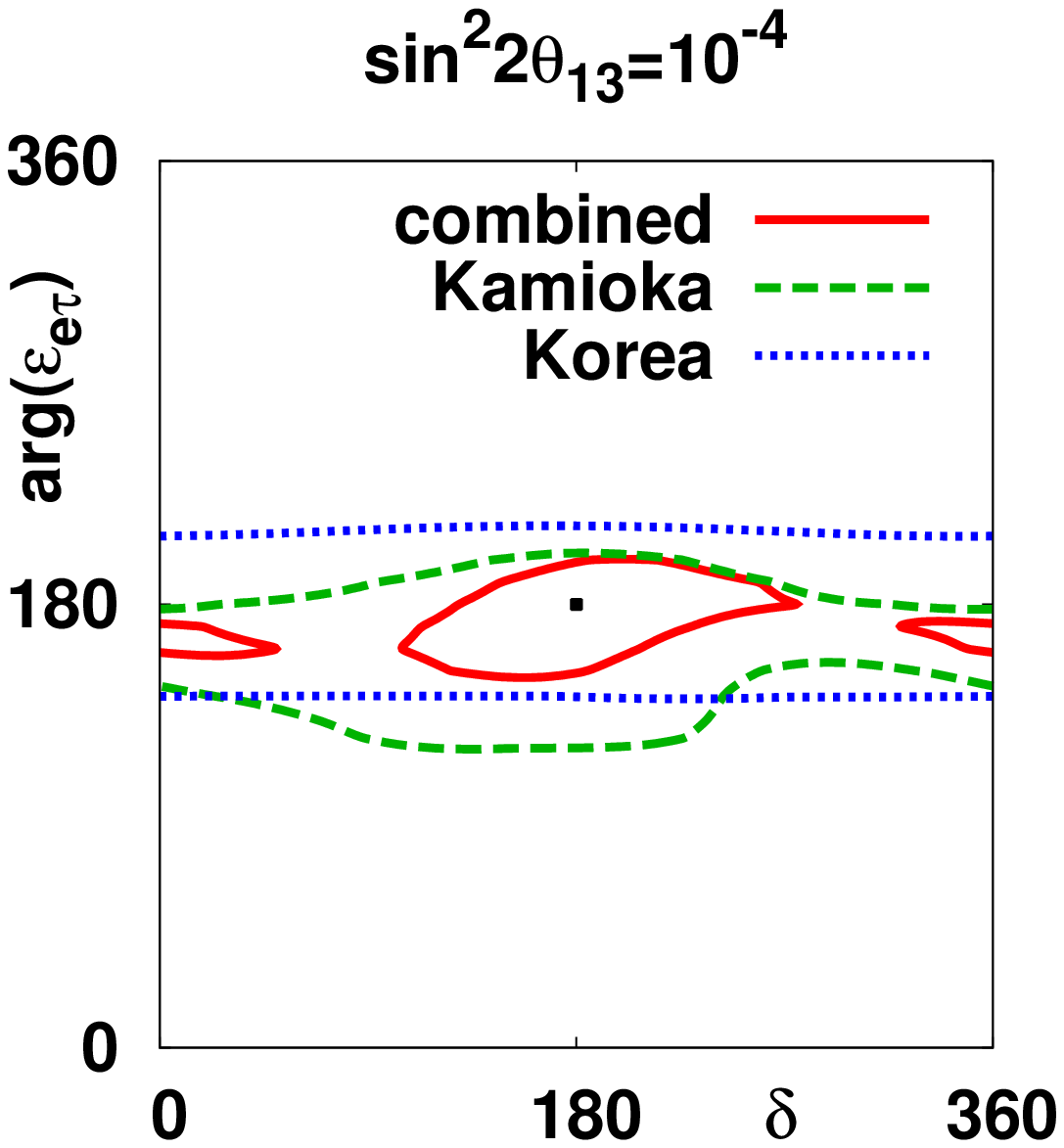}
\hglue -0.9cm
    \includegraphics[width=4.5cm]{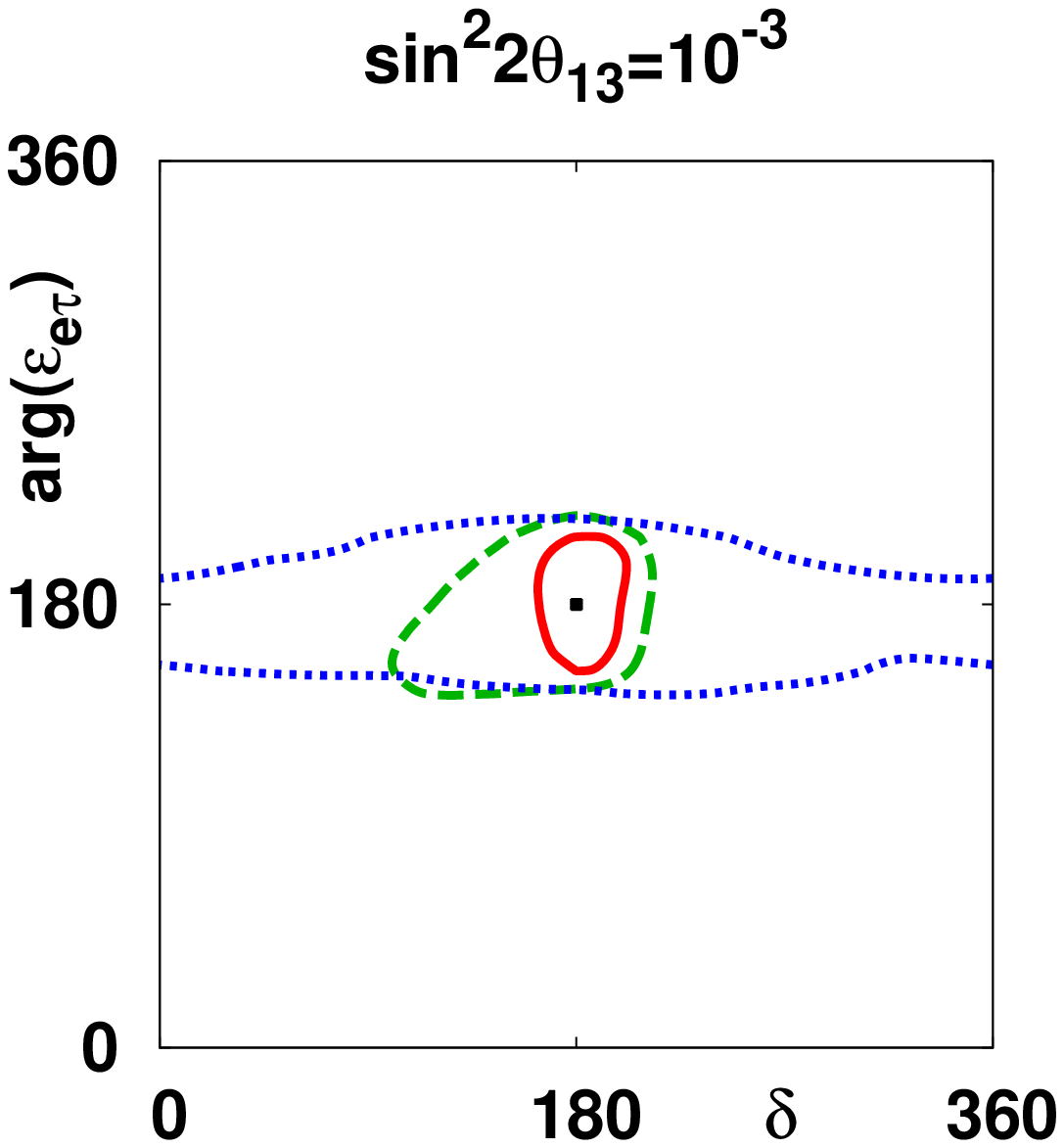}
\hglue -0.9cm
    \includegraphics[width=4.5cm]{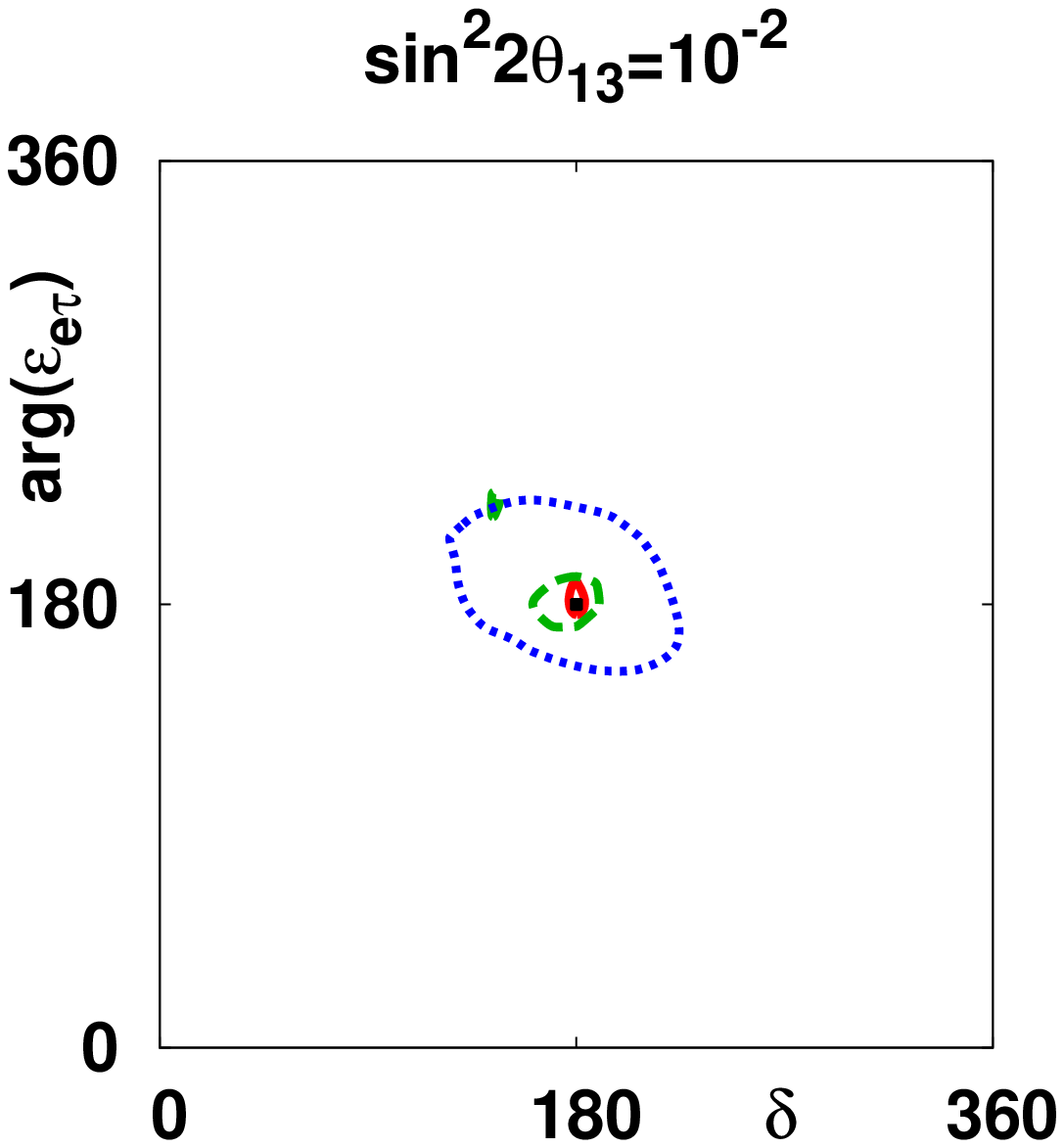}
\hglue -0.9cm
    \includegraphics[width=4.5cm]{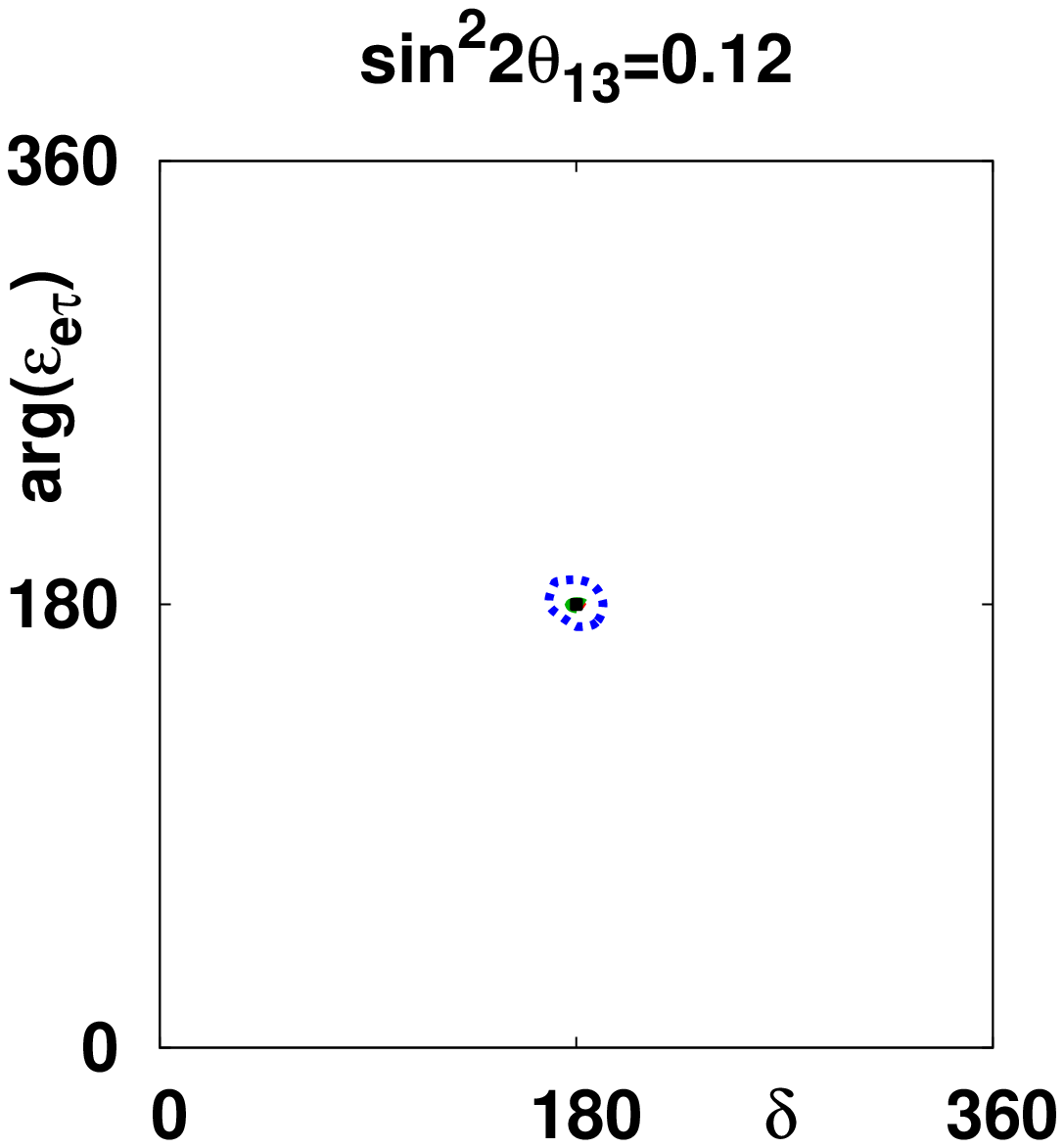}
\caption{Correlation in ($\delta$, arg($\epsilon_{e\tau}$))
for $\epsilon_{ee}=2.0$, $|\epsilon_{e\tau}|=2.0$.
The contours around the true points are
depicted at 90\%CL.
}
\label{fig7}
\end{figure}

The results at 90\%CL are shown in Figs.~\ref{fig6} and \ref{fig7}
for ($\bar{\epsilon}_{ee}$, $|\bar{\epsilon}_{e\tau}|$) = (0.8, 0.2)
and (2.0, 2.0), respectively.
Since we are discussing the cases that can be distinguished
from the standard scheme, relatively large values of $|\bar{\epsilon}_{e\tau}|$
are chosen in both examples.  Again 
$\bar{\delta}=\mbox{\rm arg}(\bar{\epsilon}_{e\tau})=\pi$ is assumed
for the true values of the phases.
To clarify the roles of the detectors at the two baseline lengths,
separate contours are given for the result from the detector in Kamioka,
for that from the detector in Korea, and for that from the combination of the two.
As in the standard three-flavor case, if $\theta_{13}$ is very small,
neither the detector in Kamioka nor the one in Korea can
provide any information on $\delta$.
As the value of $\theta_{13}$ increases,
the sensitivity to $\delta$ of the detectors in Kamioka
and Korea increases.  For larger values of $\theta_{13}$,
the sensitivity to arg($\epsilon_{e\tau}$) depends on
the value of $|\epsilon_{e\tau}|$.
For larger (smaller) values of $|\epsilon_{e\tau}|$,
sensitivity to arg($\epsilon_{e\tau}$) is good (poor).
These features can be understood qualitatively
by looking at the T violating term
in the oscillation probability $P(\nu_\mu\to\nu_e)$
(see appendix \ref{appendixg} for details).
In the present case, there are two sources
for T violation, the standard one
Im$\{Y^{\mu e}_3(Y^{\mu e}_2)^\ast\}_{\text{std}}$
and the extra one
Im$\{Y^{\mu e}_3(Y^{\mu e}_2)^\ast\}_{\text{NP}}$
because of the new physics.  From the explicit forms
(\ref{jarlgkog-std}) and (\ref{jarlgkog-np}),
the ratio of the two terms is roughly given by
(std)/(NP)$\sim 
|\Delta E_{31}||U_{e3}|/(A|\epsilon_{e\tau}||U_{\tau3}|)
\sim 10 s_{13}/|\epsilon_{e\tau}|$.
This implies that if
$\sin^22\theta_{13}\gtrsim (\lesssim)~{\cal O} (10^{-3})$
and if $|\epsilon_{e\tau}|=0.2$,
then the contribution of $\theta_{13}$ is large (small).
If $|\epsilon_{e\tau}|$ is very large,
then the oscillation probability
and the number of events becomes so large that the
sensitivity of both the detectors to the two phases
increases, as shown in Fig.~\ref{fig7}.

\section{Conclusions
\label{conclusions}}
In this paper we have studied the sensitivity of the T2KK
experiment to the non-standard
interaction in propagation with the
ansatz (\ref{ansatz}).

To justify our premise, we have provided an argument that
$\epsilon_{\mu\alpha}~(\alpha=e, \mu, \tau)$
must be small for the behavior of the disappearance probability
to be consistent with the high-energy atmospheric neutrino data.
Using the analytical form of the disappearance probability,
we showed that
$\epsilon_{\mu\alpha}~(\alpha=e, \mu, \tau)\simeq0$
as well as $\epsilon_{\tau\tau}\simeq
|\epsilon_{e\tau}|^2/(1+\epsilon_{ee})$ are
necessary for the disappearance probability to
be consistent with the high-energy behavior
$1-P(\nu_\mu\rightarrow\nu_\mu)\propto 1/E^2$ in
the high-energy atmospheric neutrino data.
This speculation should be verified explicitly by numerical
calculations in the future.

With the ansatz (\ref{ansatz}), we have looked for
the region in the ($\epsilon_{ee}$, $|\epsilon_{e\tau}|$)
plane in which T2KK can distinguish the new physics
from the standard three-flavor scheme.
At 90\%CL T2KK can discriminate the new physics from
the standard case for approximately $|\epsilon_{ee}|\gtrsim 1$ and
$|\epsilon_{e\tau}|\gtrsim 0.2$.
These values can be interpreted as bounds
for these parameters, if T2KK has negative evidence
for a new physics.
While the bound on $\epsilon_{ee}$ by T2KK is modest,
the bound on $|\epsilon_{e\tau}|$ by T2KK is much stronger
than the present one $|\epsilon_{e\tau}|\lesssim 3$, and
the latter is also stronger than those by
other on-going long-baseline experiments such as MINOS or OPERA:
If $\sin^22\theta_{13}\simeq0.07$,
then MINOS will give a bound $|\epsilon_{e\tau}|\lesssim
1$ at 90\%CL\,\cite{Yasuda:2007tx,Sugiyama:2007ub,Blennow:2007pu}, and
the combination of MINOS, OPERA and Double-CHOOZ
gives a bound
$-2.2 \,(-2.5) \lesssim\epsilon_{e\tau}\lesssim 0.6 \,(0.5)$ at 95\%CL
for $\sin^22\theta_{13}=0.05 \,(0.1)$\cite{EstebanPretel:2008qi}.

On the other hand, if the true values of the
new physics parameters lie much outside these bounds,
then T2KK can determine the values of $\epsilon_{ee}$,
$|\epsilon_{e\tau}|$, and arg($\epsilon_{e\tau}$).
In particular, if the values of $\theta_{13}$
and $|\epsilon_{e\tau}|$ are relatively large
($\sin^22\theta_{13}\gtrsim {\cal O}(0.01)$,
$|\epsilon_{e\tau}|\gtrsim 0.2$), then
we can determine the two phases 
$\delta$, arg($\epsilon_{e\tau}$) separately.
This determination is possible, because the oscillation probability
in Korea with the baseline $L$ = 1050 km
receives a non-negligible contribution from
the solar mass-squared difference $\Delta m^2_{21}$,
and it induces terms that approximately depend
only on arg($\epsilon_{e\tau}$).

Since our analysis is based on the ansatz (\ref{ansatz}),
all the results derived in this paper are approximate ones.
Nethertheless we believe that our conclusions are valid.

Long-baseline neutrino experiments with longer baselines
($L\gtrsim$ 1000 km) are sensitive to the matter effect
because of the matter effect contribution
appears in the form of $AL/2\sim L/4000$ km in the argument of
a sine function in the oscillation probability.
They are, therefore, also sensitive to a non-standard
interaction in propagation.
Among long-baseline experiments with longer baselines,
T2KK, for which
the matter effect and the
contribution from the solar mass-squared difference
are smaller than the
one from the atmospheric mass-squared difference
($|\Delta E_{21}|\sim A\ll |\Delta E_{31}|$),
is different from neutrino factories, for which
the contribution from the solar mass-squared difference
is smaller than the matter effect and the
one from the atmospheric mass-squared difference
($|\Delta E_{21}|\ll A\sim |\Delta E_{31}|$).
Consequently, the features of T2KK can complement
those of neutrino factories.
While T2KK is known to be powerful
in resolving parameter degeneracy
in the standard three-flavor scenario,
further studies on the
new physics potential of T2KK should be pursued.

\appendix

\section{Analytic treatment of the oscillation
probability in constant density matter
\label{appendixa}}

Throughout this paper, we assume that
the density of matter is constant.
In this appendix we derive analytically the neutrino oscillation
probability in constant density matter.
Let us start with the Dirac equation
\begin{eqnarray}
i{d\Psi \over dt}=
\left(U{\cal E}U^{-1}
+{\cal A}
\right)\Psi,
\label{dirac1}
\end{eqnarray}
where $\Psi\equiv(\nu_e,\nu_\mu,\nu_\tau)^T$ is the flavor eigenstate,
\begin{eqnarray}
{\cal E}&\equiv&{\mbox{\rm diag}}\left(E_1,E_2,E_3\right),
\label{energy1}
\end{eqnarray}
is a diagonal matrix with the energy eigenvalues in vacuum,
and we assume the nonstandard matter potential ${\cal A}$ defined
in Eq.\,(\ref{matter-np}).
In practical calculations, it is easier to
subtract the mass matrix $U{\cal E}U^{-1}+{\cal A}$
by $E_1{\bf 1}$,
which only affects the phase of the oscillation amplitude.
In the following, therefore, by Eq.\,(\ref{energy1})
we mean
\begin{eqnarray}
{\cal E}&\equiv&{\mbox{\rm diag}}\left(E_1,E_2,E_3\right)
-E_1\mbox{\bf 1}={\mbox{\rm diag}}\left(0,\Delta E_{21},\Delta E_{31}\right),
\label{energy2}
\end{eqnarray}
where $\Delta E_{jk}\equiv E_j-E_k\simeq \Delta m^2_{jk}/2E$.

The $3\times3$ matrix on the
right hand side of the Dirac equation (\ref{dirac1})
can be formally
diagonalized as:
\begin{eqnarray}
U{\cal E}U^{-1}+{\cal A}
=\tilde{U}\tilde{{\cal E}}\tilde{U}^{-1},
\label{sch3}
\end{eqnarray}
where
\begin{eqnarray}
\tilde{{\cal E}}&\equiv&{\mbox{\rm diag}}\left(
\tilde{E}_1,\tilde{E}_2,\tilde{E}_3\right)
\nonumber
\end{eqnarray}
is a diagonal matrix with the energy eigenvalues
in matter.

As in the case of the oscillation probability in vacuum,
Eq.\,(\ref{dirac1}) can be integrated as
\begin{eqnarray}
\Psi(L)=\tilde{U}\exp\left(-i\tilde{{\cal E}}L\right)\tilde{U}^{-1}\Psi(0)
\nonumber
\end{eqnarray}
Thus the oscillation probability
$P(\nu_\alpha\rightarrow\nu_\beta)$ is
given by
\begin{eqnarray}
P(\nu_\alpha\rightarrow\nu_\beta)&=&
\left|\left[\tilde{U}\exp\left(-i{\cal E}L\right)\tilde{U}^{-1}
\right]_{\beta\alpha}\right|^2\nonumber\\
&=&\delta_{\alpha\beta}-4\sum_{j<k}\mbox{\rm Re}\left(\tilde{X}^{\alpha\beta}_j
\tilde{X}^{\alpha\beta\ast}_k\right)
\sin^2\left({\Delta \tilde{E}_{jk}L \over 2}\right)\nonumber\\
&{\ }&-2\sum_{j<k}\mbox{\rm Im}\left(\tilde{X}^{\alpha\beta}_j
\tilde{X}^{\alpha\beta\ast}_k\right)
\sin\left(\Delta \tilde{E}_{jk}L\right),
\label{probv}
\end{eqnarray}
where we have defined
\begin{eqnarray}
\tilde{X}^{\alpha\beta}_j&\equiv&\tilde{U}_{\alpha j}\tilde{U}^\ast_{\beta j},
\label{xtilde}\\
\Delta \tilde{E}_{jk}&\equiv&\tilde{E}_j-\tilde{E}_k,
\nonumber
\end{eqnarray}
and $\alpha, \beta = (e, \mu, \tau)$
and $j, k = (1, 2, 3)$ stand for the indices of the flavor and mass eigenstates,
respectively.
Eq.\,(\ref{probv}) shows that we can obtain the oscillation probability
if we know the energy eigenvalues $\tilde{E}_j$ and
$\tilde{X}^{\alpha\beta}_j$.  The former can be obtained by
the eigenvalue equation $|U{\cal E}U^{-1}+{\cal A}-t\mbox{\bf 1}|=0$,
while the latter can be obtained by the formalism of
Kimura, Takamura and Yokomakura\,\cite{Kimura:2002hb,Kimura:2002wd}.

Let us briefly see how we get
$\tilde{X}^{\alpha\beta}_j$\,\cite{Kimura:2002hb,Kimura:2002wd,Xing:2005gk,Yasuda:2007jp,Meloni:2009ia}.
From the unitarity condition of the matrix $\tilde{U}$,
we have
\begin{eqnarray}
\delta_{\alpha\beta}=\left[\tilde{U}\tilde{U}^{-1}\right]_{\alpha\beta}
=\sum_j\tilde{U}_{\alpha j}\tilde{U}^\ast_{\beta j}
=\sum_j\tilde{X}^{\alpha\beta}_j.
\label{const1}
\end{eqnarray}
Note that the quantity $\tilde{X}^{\alpha\beta}_j$ was defined
in Eq.\,(\ref{xtilde}).
Furthermore we take the $(\alpha,\beta)$ component of the both hand sides
in Eq.\,(\ref{sch3}) and its square:
\begin{eqnarray}
%&{\ }&
\left[U{\cal E}U^{-1}+{\cal A}\right]_{\alpha\beta}
=\left[\tilde{U}\tilde{{\cal E}}\tilde{U}^{-1}\right]_{\alpha\beta}
=\sum_j\tilde{U}_{\alpha j}\tilde{E}_j\tilde{U}^\ast_{\beta j}
=\sum_j\tilde{E}_j\tilde{X}^{\alpha\beta}_j
\nonumber\\
%&=&
%&{\ }&
\left[\left(U{\cal E}U^{-1}+{\cal A}\right)^2\right]_{\alpha\beta}
=\left[\tilde{U}\tilde{{\cal E}}^2\tilde{U}^{-1}\right]_{\alpha\beta}
=\sum_j\tilde{U}_{\alpha j}\tilde{E}^2_j\tilde{U}^\ast_{\beta j}
=\sum_j\tilde{E}^2_j\tilde{X}^{\alpha\beta}_j.
%\nonumber\\
%&=&
\label{const2}
\end{eqnarray}
From Eqs.\,(\ref{const1}) and (\ref{const2}),
we have
\begin{eqnarray}
\left(\begin{array}{ccc}
1&1&1\cr
\tilde{E}_1&\tilde{E}_2&\tilde{E}_3\cr
\tilde{E}^2_1&\tilde{E}^2_2&\tilde{E}^2_3
\end{array}\right)
\left(\begin{array}{c}
\tilde{X}^{\alpha\beta}_1\cr
\tilde{X}^{\alpha\beta}_2\cr
\tilde{X}^{\alpha\beta}_3
\end{array}\right)
=\left(\begin{array}{r}
Y^{\alpha\beta}_1\cr
Y^{\alpha\beta}_2\cr
Y^{\alpha\beta}_3
\end{array}\right),
\label{simultaneous}
\end{eqnarray}
where we have introduced notations
($Y^{\alpha\beta}_1=\delta_{\alpha\beta}$):
\begin{eqnarray}
Y^{\alpha\beta}_\ell\equiv\left[\left(U{\cal E}U^{-1}+{\cal A}
\right)^{\ell-1}\right]_{\alpha\beta}.
\label{Y}
\end{eqnarray}
Eq.\,(\ref{simultaneous}) can be easily solved by inverting the
Vandermonde matrix:
\begin{eqnarray}
\left(\begin{array}{c}
\tilde{X}^{\alpha\beta}_1\cr
\tilde{X}^{\alpha\beta}_2\cr
\tilde{X}^{\alpha\beta}_3
\end{array}\right)
=\left(\begin{array}{ccc}
\displaystyle
\frac{{\ }1}{\Delta \tilde{E}_{21} \Delta \tilde{E}_{31}}
(\tilde{E}_2\tilde{E}_3, & -(\tilde{E}_2+\tilde{E}_3),&
1)\cr
\displaystyle
\frac{-1}{\Delta \tilde{E}_{21} \Delta \tilde{E}_{32}}
(\tilde{E}_3\tilde{E}_1, & -(\tilde{E}_3+\tilde{E}_1),&
1)\cr
\displaystyle
\frac{{\ }1}{\Delta \tilde{E}_{31} \Delta \tilde{E}_{32}}
(\tilde{E}_1\tilde{E}_2, & -(\tilde{E}_1+\tilde{E}_2),&
1)\cr
\end{array}\right)
\left(\begin{array}{r}
Y^{\alpha\beta}_1\cr
Y^{\alpha\beta}_2\cr
Y^{\alpha\beta}_3
\end{array}\right).
\label{solx}
\end{eqnarray}
Note that
$Y^{\alpha\beta}_\ell$ $(\ell=1,2,3)$ in Eq.\,(\ref{Y})
can be expressed by the known quantities, so that
the oscillation probability (\ref{probv}) can
be expressed analytically in terms of the known quantities.

To summarize this appendix, the following is a set of the procedures
to obtain the analytic form of the oscillation probability
in the presence of constant generic matter potential:

(i) Obtain the roots $\tilde{E}_j$ of the eigenvalue equation
$|U{\cal E}U^{-1}+{\cal A}-t{\bf 1}|=0$.

(ii) Obtain the coefficients $\tilde{X}^{\alpha\beta}_j$
in Eq.\,(\ref{solx})
by evaluating $Y^{\alpha\beta}_\ell\equiv\left[\left(U{\cal E}U^{-1}+{\cal A}
\right)^{\ell-1}\right]_{\alpha\beta}$.

(iii) Substitute $\tilde{E}_j$ and $\tilde{X}^{\alpha\beta}_j$
in Eq.\,(\ref{probv}) to obtain $P(\nu_\alpha\rightarrow\nu_\beta)$.

\section{The disappearance oscillation probability of
$\nu_\mu$ at high-energy in the standard three-flavor
scenario in matter
\label{appendixb}}
In this appendix
we derive the high-energy behavior of the the
oscillation probability $P(\nu_\mu\rightarrow\nu_\mu)$
in the standard three-flavor scenario in matter.
At high-energy we can safely ignore the contribution
from the solar neutrino mass-squared difference
$\Delta m^2_{21}$.  In this case the three eigenvalues
can be easily obtained as
$\tilde{E}_1, \tilde{E}_2, \tilde{E}_3 = \lambda_-, 0,
\lambda_+$\,\cite{Yasuda:1998sf}, where
\begin{eqnarray}
\lambda_\pm\equiv\frac{\Delta E_{31}+A}{2}
\pm\frac{1}{2}\sqrt{(\Delta E_{31}\cos2\theta_{13}-A)^2
+(\Delta E_{31}\sin2\theta_{13})^2}.
\nonumber
\end{eqnarray}
They have the following behavior at high-energy
($|\Delta E_{31}|=|\Delta m^2_{31}|/2E \ll A$):
\begin{eqnarray}
\tilde{E}_1&\simeq&\Delta E_{31}c^2_{13}
\nonumber\\
\tilde{E}_2&=&0
\nonumber\\
\tilde{E}_3&\simeq&A,
\nonumber
\end{eqnarray}
where $c_{jk}\equiv\cos\theta_{jk}$.
From these and the fact $Y^{\mu\mu}_2=\Delta E_{31}|U_{\mu3}|^2$,
$Y^{\mu\mu}_3=\Delta E_{31}^2|U_{\mu3}|^2$
we get the coefficients
\begin{eqnarray}
X^{\mu\mu}_1&\simeq&\frac{|U_{\mu3}|^2}{c^2_{13}},
~X^{\mu\mu}_2\simeq1-\frac{|U_{\mu3}|^2}{c^2_{13}},
~X^{\mu\mu}_3=\left(\frac{\Delta E_{31}}{A}\right)^2
|U_{\mu3}|^2s^2_{13},
\nonumber
\end{eqnarray}
where $s_{jk}\equiv\sin\theta_{jk}$.
Hence we obtain the following behavior of the disappearance probability
at high-energy:
\begin{eqnarray}
%&{\ }&
\lim_{E\to\infty}
\frac{1-P(\nu_\mu\rightarrow\nu_\mu)}{(\Delta E_{31}/A)^2}
%\nonumber\\
&\simeq&
\lim_{E\to\infty}
\frac{1}{(\Delta E_{31}/A)^2}
\left[
4\frac{|U_{\mu3}|^2}{c^2_{13}}
\left(1-\frac{|U_{\mu3}|^2}{c^2_{13}}\right)
\sin^2\left(\frac{c^2_{13}\Delta E_{31}L}{2}\right)
\right.\nonumber\\
&{\ }&\left.
\qquad\qquad\qquad~~~~
+4\left(\frac{\Delta E_{31}}{A}\right)^2
|U_{\mu3}|^2s^2_{13}\sin^2\left(
\frac{AL}{2}\right)
\right]
\nonumber\\
&\simeq&
\sin^22\theta_{23}
\left(\frac{c^2_{13}AL}{2}\right)^2
+s^2_{23}\sin^22\theta_{13}\sin^2\left(
\frac{AL}{2}\right),
\nonumber
\end{eqnarray}
or in other words, we have
\begin{eqnarray}
1-P(\nu_\mu\rightarrow\nu_\mu)\propto \frac{1}{E^2}.
\label{he0}
\end{eqnarray}
Eq.\,(\ref{he-std}) is the prediction
for the high-energy behavior of the
disappearance probability from the
standard three-flavor oscillations, and
it is perfectly consistent with the
atmospheric neutrino data\,\cite{Ashie:2005ik}.

\section{The disappearance oscillation probability of
$\nu_\mu$ at high-energy with new physics:
%\hglue 1.3cm 
(i) $\epsilon_{e\mu}$ and $\epsilon_{\mu\tau}$
\label{appendixc}}

In this appendix
we consider the disappearance oscillation probability
with new physics in propagation given by
Eq.\,(\ref{matter-np}), particularly with
nonvanishing $\epsilon_{e\mu}$ and $\epsilon_{\mu\tau}$.
To simplify calculations,
let us normalize $U{\cal E}U^{-1}+{\cal A}$
by $A(1+\epsilon_{ee})$:
\begin{eqnarray}
U{\cal E}U^{-1}+{\cal A}
=A'\left[U\mbox{\rm diag}\left(0,
\frac{\Delta E_{21}}{A'}, \frac{\Delta E_{31}}{A'}\right)
U^{-1}
+\left(
\begin{array}{ccc}
 1 & \epsilon'_{e\mu} & \epsilon'_{e\tau}\\
 (\epsilon'_{e\mu})^\ast & \epsilon'_{\mu\mu} & \epsilon'_{\mu\tau}\\
 (\epsilon'_{e\tau})^\ast & (\epsilon'_{\mu\tau})^\ast & \epsilon'_{\tau\tau}
\end{array}
\right)\right],
\nonumber
\end{eqnarray}
where we have defined\footnote{
Here, we have assumed that $|1+\epsilon_{ee}|$
is not very small.  This assumption is
reasonable, because otherwise the solar neutrino
would not feel the matter effect very much and
it would contradict with the solar neutrino data.}
\begin{eqnarray}
A'&\equiv& A(1+\epsilon_{ee})
\label{aprime}\\
\epsilon'_{\alpha\beta}&\equiv&\frac{\epsilon_{\alpha\beta}}{1+\epsilon_{ee}}.
\label{epsprime}
\end{eqnarray}
From (\ref{epsilon-m}) we assume that
$|\epsilon'_{e\mu}|$, $|\epsilon'_{\mu\mu}|$, $|\epsilon'_{\mu\tau}|$
are relatively small compared with one, while
$|\epsilon'_{e\tau}|$ and $|\epsilon'_{\tau\tau}|$
could be of order ${\cal O}(1)$.

First, we evaluate the energy eigenvalues for
$|\Delta E_{31}/A'|\ll 1$,
$|\epsilon'_{\alpha\mu}|\ll 1~(\alpha=e, \mu, \tau)$,
$|\epsilon'_{\tau\tau}-|\epsilon'_{e\tau}|^2|\ll 1$,
while $|\epsilon'_{\tau\tau}|$ and $|\epsilon'_{e\tau}|$
are not necessarily expected to be small.
Furthermore, we will ignore small corrections
due to $\Delta E_{21}/A'$ and $\sin\theta_{13}$.
Then the mass matrix becomes
\begin{eqnarray}
&{\ }&\frac{1}{A'}\left(U{\cal E}U^{-1}+{\cal A}\right)
\nonumber\\
&\simeq&
\left(
\begin{array}{ccc}
1 & \epsilon'_{e\mu} & \epsilon'_{e\tau}\\
(\epsilon'_{\mu e})^\ast 
& \epsilon'_{\mu\mu} +s^2_{23}\eta{\ }
& \epsilon'_{\mu\tau}+c_{23}s_{23}\eta\\
(\epsilon'_{\tau e})^\ast 
& (\epsilon'_{\mu\tau})^\ast +c_{23}s_{23}\eta
& \epsilon'_{\tau\tau}+c^2_{23}\eta{\ }{\ }
\end{array}
\right)
\equiv
\left(
\begin{array}{ccc}
1 & \epsilon'_{e\mu} & \epsilon'_{e\tau}\\
(\epsilon'_{e\mu})^\ast 
& \epsilon''_{\mu\mu}
& \epsilon''_{\mu\tau}\\
(\epsilon'_{e\tau})^\ast 
& (\epsilon''_{\mu\tau})^\ast
& \epsilon''_{\tau\tau}
\end{array}
\right),
\label{doubleprime}
\end{eqnarray}
where \begin{eqnarray}
\eta\equiv\Delta E_{31}/A'
\label{eta}
\end{eqnarray}
has been introduced and the difference between
$\epsilon''_{\alpha\beta}$ and
$\epsilon'_{\alpha\beta}$ is the term proportional
to $\eta$.
Expanding the eigenvalue equation to first order in
$\epsilon'_{\tau\tau}-|\epsilon'_{e\tau}|^2$,
and to second order in $\eta$,
$\epsilon'_{e\mu}$, $\epsilon'_{\mu\mu}$ and $\epsilon'_{\mu\tau}$,
we have
\begin{eqnarray}
0&=&\left|\frac{1}{A'}\left(U{\cal E}U^{-1}+{\cal A}\right)-t{\bf 1}\right|
\nonumber\\
&=&-t^3+t^2(1+\epsilon'_{\tau\tau}+\eta)
\nonumber\\
&{\ }&+t\left\{|\epsilon'_{e\mu}|^2+
|\epsilon''_{\mu\tau}|^2+|\epsilon'_{e\tau}|^2-\epsilon'_{\tau\tau}
-(1+s^2_{23}\epsilon'_{\tau\tau})\eta-c^2_{23}s^2_{23}\eta^2\right\}
\nonumber\\
&{\ }&-\left[|\epsilon''_{\mu\tau}|^2+\epsilon'_{\tau\tau}|\epsilon'_{e\mu}|^2
-2\mbox{\rm Re}\{\epsilon'_{e\mu}\epsilon''_{\mu\tau}
(\epsilon'_{e\tau})^\ast\}+s^2_{23}(|\epsilon'_{e\tau}|^2
-\epsilon'_{\tau\tau})\eta-c^2_{23}s^2_{23}\eta^2
\right].
\label{eigenvalueeq}
\end{eqnarray}
In the limit where all small parameters $\eta$,
$\epsilon'_{\tau\tau}-|\epsilon'_{e\tau}|^2$,
$\epsilon'_{e\mu}$, $\epsilon'_{\mu\mu}$, $\epsilon'_{\mu\tau}$,
are zero, Eq.\,(\ref{eigenvalueeq}) gives the roots
$t=0, 0, 1+\epsilon'_{\tau\tau}$.
If we include the corrections to first order in the small
parameters, then we get
\begin{eqnarray}
\left\{
\begin{array}{c}
t_2\cr
t_1
\end{array}
\right\}&=&
\frac{1+s^2_{23}\epsilon'_{\tau\tau}}{1+\epsilon'_{\tau\tau}}\frac{\eta}{2}
\nonumber\\
&{\ }&\pm\frac{1}{2(1+\epsilon'_{\tau\tau})}
\left[
4(1+\epsilon'_{\tau\tau})
\left\{|\epsilon''_{\mu\tau}|^2+\epsilon'_{\tau\tau}|\epsilon'_{e\mu}|^2
-2\mbox{\rm Re}(\epsilon'_{e\mu}\epsilon''_{\mu\tau}
(\epsilon'_{e\tau})^\ast)
\right.\right.
\nonumber\\
&{\ }&
~~\left.\left.
+s^2_{23}(|\epsilon'_{e\tau}|^2
-\epsilon'_{\tau\tau})\eta-c^2_{23}s^2_{23}\eta^2
\right\}
+(1+s^2_{23}\epsilon'_{\tau\tau})^2\eta^2
\right]^{1/2}
\label{t123}\\
t_3&=&1+\epsilon'_{\tau\tau}+{\cal O}(\eta, \epsilon'_{\alpha\mu}).
\nonumber
\end{eqnarray}

Next, we evaluate the quantities $Y^{\mu\mu}_j$.
They are given by
\begin{eqnarray}
Y^{\mu\mu}_2&=&\frac{1}{A'}\left(U{\cal E}U^{-1}+{\cal A}\right)_{\mu\mu}
=\epsilon'_{\mu\mu}+s^2_{23}\eta=\epsilon''_{\mu\mu}\,,
\nonumber\\
Y^{\mu\mu}_3&=&\frac{1}{(A')^2}
\left[\left(U{\cal E}U^{-1}+{\cal A}\right)^2\right]_{\mu\mu}
=|\epsilon'_{e\mu}|^2+|\epsilon''_{\mu\mu}|^2+|\epsilon''_{\mu\tau}|^2,
\nonumber
\end{eqnarray}
where $\epsilon''_{\alpha\beta}$ are defined in Eq.\,(\ref{doubleprime}).
Plugging these results into Eq.\,(\ref{solx}), we obtain
\begin{eqnarray}
\tilde{X}^{\mu\mu}_1&=&\frac{1}{\Delta t_{21}(1+\epsilon'_{\tau\tau})}
\left[
(t_2-\epsilon'_{\mu\mu}-s^2_{23}\eta)(1+\epsilon'_{\tau\tau})
+|\epsilon'_{e\mu}|^2+|\epsilon''_{\mu\mu}|^2+|\epsilon''_{\mu\tau}|^2
\right]
\nonumber\\
\tilde{X}^{\mu\mu}_2&=&\frac{-1}{\Delta t_{21}(1+\epsilon'_{\tau\tau})}
\left[
(t_1-\epsilon'_{\mu\mu}-s^2_{23}\eta)(1+\epsilon'_{\tau\tau})
+|\epsilon'_{e\mu}|^2+|\epsilon''_{\mu\mu}|^2+|\epsilon''_{\mu\tau}|^2
\right]
\nonumber\\
\tilde{X}^{\mu\mu}_3&=&\frac{1}{(1+\epsilon'_{\tau\tau})^2}
\left[t_1\,t_2-\frac{1+s^2_{23}\epsilon'_{\tau\tau}}{1+\epsilon'_{\tau\tau}}
(\epsilon'_{\mu\mu}+s^2_{23}\eta)\eta
+|\epsilon'_{e\mu}|^2+|\epsilon''_{\mu\mu}|^2+|\epsilon''_{\mu\tau}|^2
\right],
\label{x123}
\end{eqnarray}
where $\Delta t_{21}\equiv t_2-t_1$ and $t_j~(j=1,2,3)$
are given by Eq.\,(\ref{t123}).

Let us now consider the situation in which
the limit $E\to\infty$ or $\eta=\Delta m^2_{31}/(2A'E)\to 0$
is taken
while $\epsilon'_{e\mu}\ne0$ and $\epsilon'_{\mu\tau}\ne0$
are kept.  In this case, from Eq.\,(\ref{epsilon-m}),
which implies $|\epsilon'_{e\mu}|^2 \lesssim 0.1$ and
$|\epsilon'_{\mu\tau}|^2 \lesssim 0.1$,
and from Eqs.\,(\ref{x123}), we get
\begin{eqnarray}
&{\ }&1-P(\nu_\mu\rightarrow\nu_\mu)
\nonumber\\
&\simeq&4\tilde{X}^{\mu\mu}_1\tilde{X}^{\mu\mu}_2
\sin^2\left(\frac{\Delta t_{21}A'L}{2}\right)
+4\tilde{X}^{\mu\mu}_3(1-\tilde{X}^{\mu\mu}_3)
\sin^2\left(\frac{(1+\epsilon'_{\tau\tau})A'L}{2}\right),
\label{he1p}
\end{eqnarray}
where
\begin{eqnarray}
4\tilde{X}^{\mu\mu}_1\tilde{X}^{\mu\mu}_2
&\simeq&
1-\frac{1+\epsilon'_{\tau\tau}}
{|\epsilon'_{\mu\tau}|^2+\epsilon'_{\tau\tau}|\epsilon'_{e\mu}|^2
-2\mbox{\rm Re}(\epsilon'_{e\mu}\epsilon'_{\mu\tau}
(\epsilon'_{e\tau})^\ast)}
\left(\frac{|\epsilon'_{e\mu}|^2+|\epsilon'_{\mu\mu}|^2+|\epsilon'_{\mu\tau}|^2}
{1+\epsilon'_{\tau\tau}}-\epsilon'_{\mu\mu}
\right)^2,
\nonumber\\
4\tilde{X}^{\mu\mu}_3(1-\tilde{X}^{\mu\mu}_3)
&\simeq&4\tilde{X}^{\mu\mu}_3
\nonumber\\
&\simeq&4
\frac{|\epsilon'_{e\mu}|^2+|\epsilon'_{\mu\mu}|^2+|\epsilon'_{\mu\tau}|^2}
{(1+\epsilon'_{\tau\tau})^2}
-4\frac{|\epsilon'_{\mu\tau}|^2+\epsilon'_{\tau\tau}|\epsilon'_{e\mu}|^2
-2\mbox{\rm Re}(\epsilon'_{e\mu}\epsilon'_{\mu\tau}
(\epsilon'_{e\tau})^\ast)}{(1+\epsilon'_{\tau\tau})^3},
\nonumber\\
\Delta t_{21}&\simeq&2t_2
\simeq 2\left[
\frac{|\epsilon'_{\mu\tau}|^2+\epsilon'_{\tau\tau}|\epsilon'_{e\mu}|^2
-2\mbox{\rm Re}(\epsilon'_{e\mu}\epsilon'_{\mu\tau}
(\epsilon'_{e\tau})^\ast)}
{1+\epsilon'_{\tau\tau}}
\right].
\nonumber
\end{eqnarray}
Recovering the original notations, we obtain Eq.\,(\ref{he1}) from
Eq.\,(\ref{he1p}).
Since the argument
$(1+\epsilon'_{\tau\tau})A'L/2=(1+\epsilon_{ee}+\epsilon_{\tau\tau})AL/2$
of the sine function of the second term on the right hand side
of Eq.\,(\ref{he1}) is of the same order as $AL/2$,
in order for Eq.\,(\ref{he1}) to be consistent with the
atmospheric neutrino data at high-energy (cf. Eq.\,(\ref{he0})), at least
either of the followings has to be satisfied:
(i) $\Delta t_{21}=\tilde{X}^{\mu\mu}_3=0$ or
(ii) $4\tilde{X}^{\mu\mu}_1\tilde{X}^{\mu\mu}_2=\tilde{X}^{\mu\mu}_3=0$.
Let us introduce the two quantities:
\begin{eqnarray}
F&\equiv&|\epsilon'_{e\mu}|^2+|\epsilon'_{\mu\mu}|^2+|\epsilon'_{\mu\tau}|^2
\nonumber\\
G&\equiv&|\epsilon'_{\mu\tau}|^2+\epsilon'_{\tau\tau}|\epsilon'_{e\mu}|^2
-2\mbox{\rm Re}(\epsilon'_{e\mu}\epsilon'_{\mu\tau}
(\epsilon'_{e\tau})^\ast).
\nonumber
\end{eqnarray}
Then, in the case of (i), we have $F\simeq G\simeq 0$, which
implies $|\epsilon'_{e\mu}|\simeq|\epsilon'_{\mu\mu}|\simeq
|\epsilon'_{\mu\tau}|\simeq0$.
In the case of (ii), on the other hand, we get
\begin{eqnarray}
\left(\frac{F}{1+\epsilon'_{\tau\tau}}-\epsilon'_{\mu\mu}\right)^2
&\simeq&\frac{G}{1+\epsilon'_{\tau\tau}}
\nonumber\\
F&\simeq&\frac{G}{1+\epsilon'_{\tau\tau}}.
\nonumber
\end{eqnarray}
From these equations and the fact that $|\epsilon'_{e\mu}|^2 \lesssim 0.1$ and
$|\epsilon'_{\mu\tau}|^2 \lesssim 0.1$, it follows that
\begin{eqnarray}
|\epsilon'_{e\mu}|^2+|\epsilon'_{\mu\tau}|^2\simeq
\epsilon'_{\tau\tau}(\epsilon'_{\mu\mu})^2,
\nonumber
\end{eqnarray}
i.e., in the original notation,
\begin{eqnarray}
|\epsilon_{e\mu}|^2+|\epsilon_{\mu\tau}|^2\simeq
\frac{\epsilon_{\tau\tau}(\epsilon_{\mu\mu})^2}
{1+\epsilon_{ee}},
\nonumber
\end{eqnarray}
has to be satisfied at least,
if a consistent solution exists at all.
The bounds from (i) or (ii) are stronger that of Eq.\,(\ref{epsilon-m}),
and it is expected that the three parameters
$\epsilon_{e\mu}$, $\epsilon_{\mu\mu}$, $\epsilon_{\mu\tau}$
are negligibly small compared to
$\epsilon_{ee}$, $\epsilon_{e\tau}$ and $\epsilon_{\tau\tau}$,
although this speculation has to be verified by numerical
calculations.

\section{The disappearance oscillation probability of
$\nu_\mu$ at high-energy with new physics:
%\hglue 1.3cm 
(ii) $\epsilon_{\tau\tau}-|\epsilon_{e\tau}|^2/(1+\epsilon_{ee})$
\label{appendixd}}

In this appendix
we assume that $\epsilon_{e\mu}$ and $\epsilon_{\mu\tau}$
are zero and 
discuss the case in which $\epsilon_{ee}$, $\epsilon_{e\tau}$
and $\epsilon_{\tau\tau}$ are of order ${\cal O}(1)$.
In this case the energy eigenvalue equation (\ref{eigenvalueeq})
becomes
\begin{eqnarray}
0=-t^3+t^2(1+\epsilon'_{\tau\tau}+\eta)
-t\{\zeta'+(1+s^2_{23}\epsilon'_{\tau\tau})\eta\}
+s^2_{23}\zeta'\eta,
\nonumber
\end{eqnarray}
where $\epsilon'_{\alpha\beta}$ and $\eta$ are defined in
Eqs.\,(\ref{epsprime}) and (\ref{eta}), respectively, and
\begin{eqnarray}
\zeta'\equiv\epsilon'_{\tau\tau}-|\epsilon'_{e\tau}|^2.
\nonumber
\end{eqnarray}
In the case where $|\eta|\ll 1$, the three roots are given by
\begin{eqnarray}
t_1~~\,&\simeq&0
\nonumber\\
\left\{
\begin{array}{c}
t_2\cr
t_3
\end{array}
\right\}&\simeq&
\frac{1+\epsilon'_{\tau\tau}}
{2}
\mp\sqrt{\left(
\frac{1+\epsilon'_{\tau\tau}}
{2}\right)^2-\zeta'},
\nonumber
\end{eqnarray}
and $\tilde{X}_j^{\mu\mu}$ are given by
\begin{eqnarray}
\tilde{X}^{\mu\mu}_1&\simeq&
1-\frac{1+\epsilon'_{\tau\tau}}{\zeta'}\eta s^2_{23}
\nonumber\\
\tilde{X}^{\mu\mu}_2&\simeq&
\frac{t_3}{t_2\Delta t_{32}}\eta s^2_{23}
=\frac{t_3^2}{\zeta'\Delta t_{32}}\eta s^2_{23}
\nonumber\\
\tilde{X}^{\mu\mu}_3&\simeq&
-\frac{t_2}{t_3\Delta t_{32}}\eta s^2_{23}
=-\frac{\zeta'}{t_3^2\Delta t_{32}}\eta s^2_{23}.
\nonumber
\end{eqnarray}
The disappearance oscillation probability has
the following behavior:
\begin{eqnarray}
1-P(\nu_\mu\rightarrow\nu_\mu)
&\simeq&
-\eta\frac{4 s^2_{23}}{\Delta t_{32}}
\left[\frac{t_3^2}{\zeta'}
\sin^2\left(\frac{\zeta'A'L}{2t_3}\right)
+\frac{\zeta'}{t_3^2}
\sin^2\left(\frac{t_3A'L}{2}\right)
\right]
\nonumber\\
&{\ }&+\eta^2\frac{4s^4_{23}}{(\Delta t_{32})^2}
\sin^2\left(\frac{A'\Delta t_{32}L}{2}\right),
\nonumber
\end{eqnarray}
where 
$\Delta t_{32}\equiv t_3-t_2$.
By recovering the original notations,
we obtain Eq.\,(\ref{he2}).

\section{The disappearance oscillation probability of
$\nu_\mu$ at high-energy with new physics:
%\hglue 1.3cm 
(iii) $\epsilon_{ee}$, $\epsilon_{e\tau}$ and
$\epsilon_{\tau\tau}=|\epsilon_{e\tau}|^2/(1+\epsilon_{ee})$
\label{appendixe}}

In this appendix we assume
$\zeta'=\epsilon'_{\tau\tau}-|\epsilon'_{e\tau}|^2=0$
and derive the high-energy behavior (\ref{he3}).
In this case, introducing the new variable
$t_\beta\equiv|\epsilon_{e\tau}|/(1+\epsilon_{ee})$
(cf. Eq.\,(\ref{tanbeta})),
we have
\begin{eqnarray}
|\epsilon'_{e\tau}|&=&\frac{|\epsilon_{e\tau}|}{1+\epsilon_{ee}}
=t_\beta,
\nonumber\\
\epsilon'_{\tau\tau}&=&|\epsilon'_{e\tau}|^2=t^2_\beta.
\nonumber
\end{eqnarray}
Recovering the small corrections due to $\theta_{13}$,
we have the energy eigenvalue equation:
\begin{eqnarray}
0&=&-t^3+t^2(1+t^2_\beta+\eta)
-t\eta\left(1+t^2_\beta
-\left|U_{e3}+U_{\tau3}\epsilon'_{e\tau}\right|^2\right),
\nonumber\\
&=&-t\left\{t^2-(1/c^2_\beta+\eta)t+\eta(c''_{13})^2/c^2_\beta
\right\},
\nonumber
\end{eqnarray}
where $c''_{13}\equiv \cos\theta''_{13}$ and
$\theta''_{13}$ is defined by Eq.\,(\ref{doubleprime13})
The three roots in this case are
$t_1=0$, $t_2\simeq \eta(c''_{13})^2$, $t_3\simeq 1/c^2_\beta$.
Then we have the following high-energy behavior (\ref{he3})
in the limit $E\to\infty$:
\begin{eqnarray}
\frac{1-P(\nu_\mu\rightarrow\nu_\mu)}{(\Delta E_{31}/A)^2}
&=&
\frac{1-P(\nu_\mu\rightarrow\nu_\mu)}
{(1+\epsilon_{ee})^2\eta^2}
\nonumber\\
&\simeq&
4\frac{s^2_{23}}
{(c''_{13})^2}
\left\{1-\frac{s^2_{23}}
{(c''_{13})^2}
\right\}
\left\{\frac{(c''_{13})^2AL}{2}
\right\}^2
\nonumber\\
&{\ }&+\frac{s^2_{23}}{(c''_{13})^2}
\sin^22\theta''_{13}
\left(\frac{c^2_\beta}{1+\epsilon_{ee}}\right)^2
\sin^2\left(\frac{(1+\epsilon_{ee})AL}{2c^2_\beta}\right).
\nonumber
\end{eqnarray}

\section{The analytic expression of
$P(\nu_\mu\to\nu_e)$ with new physics in propagation
\label{appendixf}}

In this appendix we discuss the analytic form of
the oscillation probability
$P(\nu_\mu\to\nu_e)$.
From the formula (\ref{probv}) we have
\begin{eqnarray}
P(\nu_\mu\to\nu_e)&=&
-4\mbox{\rm Re}\left(\tilde{X}^{\mu e}_1\tilde{X}^{\mu e\ast}_2\right)
\sin^2\left(\frac{\Delta\tilde{E}_{21}L}{2}\right)
-4\mbox{\rm Re}\left(\tilde{X}^{\mu e}_2\tilde{X}^{\mu e\ast}_3\right)
\sin^2\left(\frac{\Delta\tilde{E}_{32}L}{2}\right)
\nonumber\\
&{\ }&-4\mbox{\rm Re}\left(\tilde{X}^{\mu e}_1\tilde{X}^{\mu e\ast}_3\right)
\sin^2\left(\frac{\Delta\tilde{E}_{31}L}{2}\right)\nonumber\\
&{\ }&-
8\mbox{\rm Im}\left(\tilde{X}^{\mu e}_1\tilde{X}^{\mu e\ast}_2\right)
\sin\left(\frac{\Delta\tilde{E}_{21}L}{2}\right)
\sin\left(\frac{\Delta\tilde{E}_{32}L}{2}\right)
\sin\left(\frac{\Delta\tilde{E}_{31}L}{2}\right).
\label{probme}
\end{eqnarray}
From Eq.\,(\ref{solx}) $\tilde{X}^{\mu e}_{j}$ can be expressed as
\begin{eqnarray}
\tilde{X}^{\mu e}_1&=&
\frac{1}{\Delta\tilde{E}_{21}\Delta\tilde{E}_{31}}
\left\{Y^{\mu e}_3-(\tilde{E}_2+\tilde{E}_3)Y^{\mu e}_2\right\}
\nonumber\\
\tilde{X}^{\mu e}_2&=&
\frac{-1}{\Delta\tilde{E}_{21}\Delta\tilde{E}_{32}}
\left\{Y^{\mu e}_3-(\tilde{E}_3+\tilde{E}_1)Y^{\mu e}_2\right\}
\nonumber\\
\tilde{X}^{\mu e}_3&=&
\frac{1}{\Delta\tilde{E}_{31}\Delta\tilde{E}_{32}}
\left\{Y^{\mu e}_3-(\tilde{E}_1+\tilde{E}_2)Y^{\mu e}_2\right\},
\nonumber
\end{eqnarray}
where $Y^{\mu e}_j~(j=2,3)$ is defined by Eq.\,(\ref{Y}).
and are given by
\begin{eqnarray}
Y^{\mu e}_2&=&\Delta E_{31}X^{\mu e}_3
+\Delta E_{21}X^{\mu e}_2
\nonumber\\
Y^{\mu e}_3&=&(\Delta E_{31})^2X^{\mu e}_3
+(\Delta E_{21})^2X^{\mu e}_2
+A\Delta E_{31}\left\{
(1+\epsilon_{ee})X^{\mu e}_3
+\epsilon_{\tau e}X^{\mu\tau}_3
\right\}
\nonumber\\
&{\ }&+A\Delta E_{21}\left\{
(1+\epsilon_{ee})X^{\mu e}_2
+\epsilon_{\tau e}X^{\mu\tau}_2
\right\}.
\label{ymue}
\end{eqnarray}
Here we have introduced
the same quantity in vacuum:
\begin{eqnarray}
X^{\alpha\beta}_j\equiv U_{\alpha j}U_{\beta j}^\ast.
\nonumber
\end{eqnarray}

As shown in Ref.\,\cite{Yasuda:2007jp},
in the limit $\Delta m^2_{21}\to 0$,
$\tilde{E}_j$ can be expressed as roots of a quadratic equation.
First, let us review how to obtain them.
With the ansatz (\ref{ansatz}), we have
\begin{eqnarray}
{\cal A}=A''\,e^{i\gamma\lambda_9}e^{-i\beta\lambda_5}
\frac{\lambda_0+\lambda_9}{2}
e^{i\beta\lambda_5}e^{-i\gamma\lambda_9},
\label{np1}
\end{eqnarray}
where $\beta$ is defined by Eq.\,(\ref{tanbeta}),
\begin{eqnarray}
A''&\equiv& A(1+\epsilon_{ee})/c^2_\beta,
\label{a-doubleprime}\\
\gamma&\equiv&\frac{1}{2}\mbox{\rm arg}\,(\epsilon_{e\tau}),
\nonumber
\end{eqnarray}
and we have introduced notations for $3\times3$ hermitian matrices:
\begin{eqnarray}
\lambda_2&\equiv&\left(
\begin{array}{ccc}
0&-i&0\cr
i&0&0\cr
0&0&0
\end{array}\right),\quad
\lambda_5\equiv\left(
\begin{array}{ccc}
0&0&-i\cr
0&0&0\cr
i&0&0
\end{array}\right),\quad
\lambda_7\equiv\left(
\begin{array}{ccc}
0&0&0\cr
0&0&-i\cr
0&i&0
\end{array}\right),\nonumber\\
\lambda_0&\equiv&\left(
\begin{array}{ccc}
1&0&0\cr
0&0&0\cr
0&0&1
\end{array}\right),\quad
\lambda_9\equiv\left(
\begin{array}{ccc}
1&0&0\cr
0&0&0\cr
0&0&-1
\end{array}\right).
\nonumber
\end{eqnarray}
Here, $\lambda_2$, $\lambda_5$ and $\lambda_7$ are
the standard Gell-Mann matrices whereas
$\lambda_0$ and $\lambda_9$ are the notations which
are defined only in this paper.

The mass matrix can be written as
\begin{eqnarray}
%&{\ }&
U{\cal E}U^{-1}+{\cal A}
%\nonumber\\
%&=&
=e^{i\gamma\lambda_9}e^{-i\beta\lambda_5}
\left[
e^{i\beta\lambda_5}e^{-i\gamma\lambda_9}U{\cal E}U^{-1}
e^{i\gamma\lambda_9}e^{-i\beta\lambda_5}
+\mbox{\rm diag}\left(A'',0,0\right)
\right]
e^{i\beta\lambda_5}e^{-i\gamma\lambda_9}.
\nonumber
\end{eqnarray}
Here, we introduce the two unitary matrices:
\begin{eqnarray}
U'&\equiv& e^{i\beta\lambda_5}e^{-i\gamma\lambda_9}\,U\nonumber\\
&\equiv&\mbox{\rm diag}(1,1,e^{i\text{arg}\,U'_{\tau3}})\,
U''\,\mbox{\rm diag}
(e^{i\text{arg}\,U'_{e1}},e^{i\text{arg}\,U'_{e2}},1),
\nonumber
\end{eqnarray}
where $U$ is the $3\times3$ MNS matrix in the standard
parametrization\,\cite{Amsler:2008zz} and
$U''$ was defined in the second line in such a way
that the elements $U''_{e1}$, $U''_{e2}$, $U''_{\tau3}$
be real to be consistent with the standard parametrization
in Ref.\,\cite{Amsler:2008zz}~\footnote{
The element $U''_{\tau2}$ has to be also real, but it is
already satisfied because $U''_{\tau2}=U_{\tau2}$.}.
Then we have
\begin{eqnarray}
U{\cal E}U^{-1}+{\cal A}
&=&e^{i\gamma\lambda_9}e^{-i\beta\lambda_5}
\mbox{\rm diag}(1,1,e^{i\text{arg}\,U'_{\tau3}})\,
\left[U''{\cal E}U''^{-1}
+\mbox{\rm diag}\left(A'',0,0\right)
\right]\nonumber\\
&{\ }&\times\mbox{\rm diag}(1,1,e^{-i\text{arg}\,U'_{\tau3}})\,
e^{i\beta\lambda_5}e^{-i\gamma\lambda_9}.
\label{np3}
\end{eqnarray}
Before proceeding further, let us obtain the
expression for the three mixing angles $\theta''_{jk}$
and the Dirac phase $\delta''$ in $U''$.
Since
\begin{eqnarray}
U'=\left(\begin{array}{ccc}
c_\beta e^{-i\gamma}U_{e1}+s_\beta e^{i\gamma}U_{\tau1}&
c_\beta e^{-i\gamma}U_{e2}+s_\beta e^{i\gamma}U_{\tau2}&
c_\beta e^{-i\gamma}U_{e3}+s_\beta e^{i\gamma}U_{\tau3}\cr
U_{\mu1}&U_{\mu2}&U_{\mu3}\cr
c_\beta e^{-i\gamma}U_{\tau1}-s_\beta e^{i\gamma}U_{e1}&
c_\beta e^{-i\gamma}U_{\tau2}-s_\beta e^{i\gamma}U_{e2}&
c_\beta e^{-i\gamma}U_{\tau3}-s_\beta e^{i\gamma}U_{e3}
\end{array}
\right),
\nonumber
\end{eqnarray}
where $c_\beta\equiv\cos\beta$, $s_\beta\equiv\sin\beta$,
we get
\begin{eqnarray}
\theta''_{13}&=&\sin^{-1}|U''_{e3}|=
\sin^{-1}|c_\beta e^{-i\gamma}U_{e3}+s_\beta e^{i\gamma}U_{\tau3}|
\label{doubleprime13}\\
\theta''_{12}&=&\tan^{-1}(U''_{e2}/U''_{e1})=
\tan^{-1}\left(|c_\beta e^{-i\gamma}U_{e2}+s_\beta e^{i\gamma}U_{\tau2}|
/|c_\beta e^{-i\gamma}U_{e1}+s_\beta e^{i\gamma}U_{\tau1}|\right)
\label{doubleprime12}\\
\theta''_{23}&=&\tan^{-1}(U''_{\mu3}/U''_{\tau3})=
\tan^{-1}\left(U_{\mu3}/
|c_\beta e^{-i\gamma}U_{\tau3}-s_\beta e^{i\gamma}U_{e3}|\right)
\label{doubleprime23}\\
\delta''&=&-\mbox{\rm arg}\,\left[(U_{e3}')^{-1}U_{e1}'U_{e2}'
U_{\tau3}'\right].
\nonumber
\end{eqnarray}
As shown in Ref.\,\cite{Yasuda:1998sf}, in the limit
$\Delta m^2_{21}\rightarrow0$, the matrix on the right hand side
of Eq. (\ref{np3}) can be diagonalized as follows:
\begin{eqnarray}
&{\ }&U''{\cal E}U''^{-1}
+\mbox{\rm diag}\left(A'',0,0\right)-E_1{\bf 1}\nonumber\\
&=&e^{i\theta''_{23}\lambda_7}\Gamma_{\delta''}
e^{i\theta''_{13}\lambda_5}\Gamma_{\delta''}^{-1}
e^{i\theta''_{12}\lambda_2}
\mbox{\rm diag}\left(0,0,\Delta E_{31}\right)
e^{-i\theta''_{12}\lambda_2}\Gamma_{\delta''}
e^{-i\theta''_{13}\lambda_5}\Gamma_{\delta''}^{-1}
e^{-i\theta''_{23}\lambda_7}
+\mbox{\rm diag}\left(A'',0,0\right)\nonumber\\
&=&e^{i\theta''_{23}\lambda_7}\Gamma_{\delta''}
\left[e^{i\theta''_{13}\lambda_5}\mbox{\rm diag}\left(0,0,\Delta E_{31}\right)
e^{-i\theta''_{13}\lambda_5}
+\mbox{\rm diag}\left(A'',0,0\right)\right]
\Gamma_{\delta''}^{-1}e^{-i\theta''_{23}\lambda_7}\nonumber\\
&=&e^{i\theta''_{23}\lambda_7}\Gamma_{\delta''}
e^{i\tilde{\theta}''_{13}\lambda_5}
\mbox{\rm diag}\left(\Lambda_-,0,\Lambda_+\right)
e^{-i\tilde{\theta}''_{13}\lambda_5}
\Gamma_{\delta''}^{-1}e^{-i\theta''_{23}\lambda_7},
\nonumber
\end{eqnarray}
where $\Gamma_{\delta''}\equiv\mbox{\rm diag}(1,1,e^{-i\delta''})$,
$\Delta E_{31}\equiv\Delta m_{31}^2/2E$,
we have used the standard parametrization\,\cite{Amsler:2008zz}
$U''\equiv e^{i\theta''_{23}\lambda_7}\Gamma_{\delta''}
e^{i\theta''_{13}\lambda_5}\Gamma_{\delta''}^{-1}
e^{i\theta''_{12}\lambda_2}$, and the eigenvalues
$\Lambda_\pm$ and the effective mixing angle
$\tilde{\theta}''_{13}$ are defined by
\begin{eqnarray}
\Lambda_\pm&=&\frac{1}{2}\left(\Delta E_{31}+A''
\right)
\pm\frac{1}{2}
\sqrt{\left(\Delta E_{31}\cos2\theta_{13}''
-A''\right)^2
+(\Delta E_{31}\sin2\theta_{13}'')^2}\nonumber\\
\tan2\tilde{\theta}''_{13}
&=&\frac{\Delta E_{31}\sin2\theta_{13}''}
{\Delta E_{31}\cos2\theta_{13}''-A''}.
\nonumber
\end{eqnarray}

In the present case, since we consider the oscillation
probability $P(\nu_\mu\to\nu_e)$ in Korea at low energy,
i.e., at $L$ = 1050 km with $E$ = 0.7 GeV,
the mass-squared difference due to the solar neutrino oscillation
gives a non--negligible contribution $\Delta E_{21}L\sim 0.3$ and
the correction in
$|\Delta m^2_{21}|/|\Delta m^2_{31}|$ becomes important.

The discussions in Ref.\,\cite{Yasuda:2007jp}
can be generalized to the case with
non-vanishing $\Delta m^2_{21}$,
since discussions up to Eq.\,(\ref{np3}) are
valid for a generic value of $\Delta m^2_{21}\ne0$ and
all we have to do is to obtain the correction
to the energy eigenvalues due to small
$\Delta m^2_{21}$.
The energy eigenvalue
$\tilde{E}_j$ to first order in $|\Delta m^2_{21}|/|\Delta m^2_{31}|$
can be computed and are given by
\begin{eqnarray}
\tilde{E}_1&=&\Lambda_-+\frac{(s''_{12})^2}{\Lambda_+-\Lambda_-}
\left\{\Lambda_+-(c''_{13})^2A''\right\}\Delta E_{21}
\nonumber\\
\tilde{E}_2&=&(c''_{12})^2\Delta E_{21}
\nonumber\\
\tilde{E}_3&=&\Lambda_+-\frac{(s''_{12})^2}{\Lambda_+-\Lambda_-}
\left\{\Lambda_--(c''_{13})^2A''\right\}\Delta E_{21},
\label{tildeej}
\end{eqnarray}
where $c''_{jk}\equiv\cos\theta''_{jk}$ and
$s''_{jk}\equiv\sin\theta''_{jk}$.
As seen from these expressions,
in the limit $\Delta m^2_{21}\to 0$,
all the quantities depend on the phase
only through the combination $\delta+\mbox{\rm arg}(\epsilon_{e\tau})$,
since they depend only on $\theta''_{13}$ which
is a function of the combination $\delta+\mbox{\rm arg}(\epsilon_{e\tau})$.
On the other hand, the oscillation
probability $P(\nu_\mu\to\nu_e)$ in Korea has
the moderate dependence on $\delta$ and $\mbox{\rm arg}(\epsilon_{e\tau})$
separately, because the first order corrections in
$|\Delta m^2_{21}|/|\Delta m^2_{31}|$ to $\tilde{E}_j$
have dependence on $\theta''_{12}$,
which is approximately a function of $\mbox{\rm arg}(\epsilon_{e\tau})$ only.

\section{The T violating term in $P(\nu_\mu\to\nu_e)$
\label{appendixg}}

In this appendix,
to see the contribution of the two phases
$\delta$ and $\mbox{\rm arg}(\epsilon_{e\tau})$ in $P(\nu_\mu\to\nu_e)$,
we will study the T violating term, which is
the last line in Eq.\,(\ref{probme}).
This term contains the modified Jarlskog factor
$\mbox{\rm Im}(\tilde{X}^{\mu e}_1\tilde{X}^{\mu e\ast}_2)$,
and it can be rewritten as
\begin{eqnarray}
\mbox{\rm Im}(\tilde{X}^{\mu e}_1\tilde{X}^{\mu e\ast}_2)
&=&\frac{-1}
{(\Delta\tilde{E}_{21})^2\Delta\tilde{E}_{31}\Delta\tilde{E}_{32}}\,
\mbox{\rm Im}\left[
\left\{Y^{\mu e}_3-(\tilde{E}_2+\tilde{E}_3)Y^{\mu e}_2\right\}
\left\{Y^{\mu e}_3-(\tilde{E}_3+\tilde{E}_1)Y^{\mu e}_2\right\}^\ast
\right]
\nonumber\\
&=&\frac{-1}{\Delta\tilde{E}_{21}\Delta\tilde{E}_{31}\Delta\tilde{E}_{32}}\,
\mbox{\rm Im}\left\{
Y^{\mu e}_3(Y^{\mu e}_2)^\ast
\right\}.
\label{jarlgkog1}
\end{eqnarray}
In the present case with the ansatz (\ref{ansatz}),
instead of using the explicit expressions (\ref{ymue}),
it is convenient to work with the following form:
\begin{eqnarray}
Y^{\mu e}_2&=&(U{\cal E}U^{-1}+{\cal A})_{\mu e}
=(U{\cal E}U^{-1})_{\mu e}
\nonumber\\
Y^{\mu e}_3&=&\left[\left(U{\cal E}U^{-1}+{\cal A}
\right)^2\right]_{\mu e}=
(U{\cal E}^2U^{-1})_{\mu e}
+(U{\cal E}U^{-1})_{\mu e}{\cal A}_{ee}
+(U{\cal E}U^{-1})_{\mu\tau}{\cal A}_{\tau e}\,,
\nonumber
\end{eqnarray}
so that the factor $\mbox{\rm Im}\{Y^{\mu e}_3(Y^{\mu e}_2)^\ast\}$
in Eq.\,(\ref{jarlgkog1}) can be expressed as
\begin{eqnarray}
\mbox{\rm Im}\left\{
Y^{\mu e}_3(Y^{\mu e}_2)^\ast
\right\}
%\nonumber\\
&=&\mbox{\rm Im}\left[(U{\cal E}U^{-1})_{\mu e}^\ast
\left\{(U{\cal E}^2U^{-1})_{\mu e}
+(U{\cal E}U^{-1})_{\mu\tau}{\cal A}_{\tau e}\right\}\right]
\nonumber\\
&=&\mbox{\rm Im}\left\{
Y^{\mu e}_3(Y^{\mu e}_2)^\ast
\right\}_{\text{std}}
+\mbox{\rm Im}\left\{
Y^{\mu e}_3(Y^{\mu e}_2)^\ast
\right\}_{\text{NP}},
\label{jarlgkog2}
\end{eqnarray}
where the term
$(U{\cal E}U^{-1})_{\mu e}{\cal A}_{ee}$
in $Y^{\mu e}_3$ has dropped
as in the case of the standard scheme,
because $|(U{\cal E}U^{-1})_{\mu e}|^2{\cal A}_{ee}$
is real.
In Eq.\,(\ref{jarlgkog2}) we have introduced the notations:
\begin{eqnarray}
\mbox{\rm Im}
\left\{
Y^{\mu e}_3(Y^{\mu e}_2)^\ast
\right\}_{\text{std}}
&\equiv&
\mbox{\rm Im}\left[
(U{\cal E}U^{-1})_{\mu e}^\ast
(U{\cal E}^2U^{-1})_{\mu e}
\right]
\nonumber\\
&=&\mbox{\rm Im}\left[\{
(\Delta E_{31})^2X^{\mu e}_3+(\Delta E_{21})^2X^{\mu e}_2\}
\{\Delta E_{31}(X^{\mu e}_3)^\ast+\Delta E_{21}(X^{\mu e}_2)^\ast\}
\right]
\nonumber\\
&=&\Delta E_{21}\Delta E_{31}\Delta E_{32}\,
\mbox{\rm Im}\left\{
X^{\mu e}_3(X^{\mu e}_2)^\ast
\right\}
\label{jarlgkog-std}
\end{eqnarray}
is the Jarlskog factor in the standard three-flavor
scheme\,\cite{Naumov:1991ju}, and
\begin{eqnarray}
\mbox{\rm Im}\left\{
Y^{\mu e}_3(Y^{\mu e}_2)^\ast
\right\}_{\text{NP}}
&\equiv&
\mbox{\rm Im}
\left\{(U{\cal E}U^{-1})_{\mu e}^\ast
(U{\cal E}U^{-1})_{\mu\tau}{\cal A}_{\tau e}\right\}
\nonumber\\
&=&\mbox{\rm Im}\left[
A(\epsilon_{e\tau})^\ast
\{\Delta E_{31}(X^{\mu e}_3)^\ast+\Delta E_{21}(X^{\mu e}_2)^\ast\}
(\Delta E_{31}X^{\mu \tau}_3+\Delta E_{21}X^{\mu \tau}_2)
\right]
\nonumber\\
&=&A|\epsilon_{e\tau}|\left[
(\Delta E_{31})^2\,
\mbox{\rm Im}\{(X^{\mu e}_3)^\ast X^{\mu \tau}_3\,e^{2i\gamma}\}
\right.
\nonumber\\
&{\ }&
\qquad+\Delta E_{31}\Delta E_{21}\,
%\left\{
\mbox{\rm Im}\{(X^{\mu e}_3)^\ast X^{\mu \tau}_2\,e^{2i\gamma}
%\}
+
%\mbox{\rm Im}\{
(X^{\mu e}_2)^\ast X^{\mu \tau}_3\,e^{2i\gamma}\}
%\right\}
\nonumber\\
&{\ }&\left.
\qquad+(\Delta E_{21})^2\,
\mbox{\rm Im}\{(X^{\mu e}_2)^\ast X^{\mu \tau}_2\,e^{2i\gamma}\}
\right]
\label{jarlgkog-np}
\end{eqnarray}
is the extra contribution to the Jarlskog factor
due to new physics.
If $\theta_{13}$ is small, then
the dominant contribution in the new physics term
(\ref{jarlgkog-np}) comes from the middle one which is proportional to
$A|\epsilon_{e\tau}|\Delta E_{31}\Delta E_{21}
|X^{\mu e}_2| |X^{\mu \tau}_3|$, and this should
be compared with the standard factor 
(\ref{jarlgkog-std}) to examine which contribution
dominates the T violating term.

\section*{Acknowledgments}
The authors thank H. Minakata for useful comments and
T. Kajita and K. Kaneyuki for useful communications on the T2K and T2KK
experiments.
This research was partly supported by a Grant-in-Aid for Scientific
Research of the Ministry of Education, Science and Culture,
\#21540274, and the MEXT program "Support Program
for Improving Graduate School Education."

\end{document}